\documentclass[prd,onecolumn,showpacs,superscriptaddress,nofootinbib,floatfix,11pt]{revtex4-2}

\usepackage[utf8]{inputenc}
\usepackage{amsmath,amssymb}
\usepackage{slashed}
\usepackage{subfigure}
\usepackage[colorlinks=true, 
linkcolor=blue,
breaklinks=true,
urlcolor=magenta,
citecolor=blue]{hyperref}
\usepackage[usenames,dvipsnames]{color}
\usepackage{appendix}
\usepackage{braket,bm}
\usepackage{multirow}
\usepackage{enumitem}
\usepackage{color}
\usepackage{cancel}
\usepackage{orcidlink}
\usepackage{mathrsfs}

\newcommand{\R}\mathbf{}

\usepackage{graphicx}
\usepackage{tikz}
\usepackage[normalem]{ulem}
\usepackage{booktabs}
\usepackage{array}
\usepackage{overpic}
\usepackage{float}
\usepackage{threeparttable}
\allowdisplaybreaks[4]

\graphicspath{{figs/}}

\usetikzlibrary{calc}
\usetikzlibrary{intersections}
\usetikzlibrary{trees}
\usetikzlibrary{decorations.pathmorphing}
\usetikzlibrary{decorations.markings}
\usetikzlibrary{arrows.meta}
\usetikzlibrary{patterns}
\tikzset{
   global scale/.style={
      scale=#1,
      every node/.append style={scale=#1}},
   photon/.style={decorate, decoration={snake}, draw=red},
   nucleon/.style={draw=black, postaction={decorate},
      decoration={markings,mark=at position .65 with{\arrow[draw=black]{latex}}}},
   pion/.style={draw=blue, postaction={decorate},
      decoration={markings,mark=at position .55 with{\arrow[draw=blue]{}}}},
    nucleonstar/.style={draw=black, postaction={decorate},
      decoration={markings, mark=at position 0.7 with {\arrow[draw=black]{latex}}}},
    }

\newcommand{\itp}{\affiliation{ Institute of Theoretical Physics, Chinese Academy of Sciences, Beijing 100190, China}}

\newcommand{\su}{\affiliation{
Institute for Particle and Nuclear Physics, College of Physics,\\ Sichuan University, Chengdu, Sichuan 610065, China}}

\newcommand{\ucas}{\affiliation{School of Physical Sciences, University of Chinese Academy of Sciences, Beijing 100049, China}}

\newcommand{\phcc}{\affiliation{Peng Huanwu Collaborative Center for Research and Education, Beihang University, Beijing 100191, China}}

\newcommand{\hebtu}{\affiliation{Department of Physics and Hebei Key Laboratory of Photophysics Research and Application,\\
Hebei Normal University, Shijiazhuang 050024, China}}

\newcommand{\scnt}{\affiliation{Southern Center for Nuclear-Science Theory (SCNT), Institute of Modern Physics,\\
Chinese Academy of Sciences, Huizhou 516000, China}}

\newcommand\tbbint{{-\mkern -16mu\int}}

\newcommand\dbbint{{-\mkern -19mu\int}}

\newcommand\bbint{
{\mathchoice{\dbbint}{\tbbint}{\tbbint}{\tbbint}}
}
\newcommand{\md}{\mathrm{d}}

\begin{document}
\title{Revisiting Roy-Steiner-equation analysis of pion-kaon scattering from lattice QCD data}

\author{Xiong-Hui Cao\orcidlink{0000-0003-1365-7178}}
\email{xhcao@itp.ac.cn}
\su\itp

\author{Feng-Kun Guo\orcidlink{0000-0002-2919-2064}}\email{fkguo@itp.ac.cn}
\itp\ucas\phcc\scnt 

\author{Zhi-Hui Guo\orcidlink{0000-0003-0409-1504}}\email{zhguo@hebtu.edu.cn}
\hebtu

\author{Qu-Zhi Li\orcidlink{0009-0001-2640-1174}}
\email{liquzhi@scu.edu.cn}
\su

\begin{abstract}
A comprehensive analysis of $\pi K\rightarrow \pi K$ and $\pi\pi\rightarrow K\bar K$ amplitudes at large unphysical pion mass for all important partial waves is presented. 
A set of crossing-symmetric partial-wave hyperbolic dispersion relations is used to describe lattice QCD data at $m_\pi=391~$MeV. 
In the present analysis, the amplitudes for the $S$- and $P$-waves are formulated by combining the constraints of analyticity, unitarity, and crossing symmetry, fulfilling Roy-Steiner-type equations. 
We use these results to investigate the low-lying strange-meson resonances and resolve the instability problem tied to analytic continuation in prior lattice QCD studies based on the $K$-matrix formalism. 
At $m_\pi=391$~MeV, the rigorous Roy-Steiner-type equation approach allows us to determine the $S$-wave scattering lengths, $m_\pi a_0^{1/2}=\left(0.92_{-0.28}^{+0.06}\right)$, $m_\pi a_0^{3/2}=-\left(0.32_{-0.02}^{+0.05}\right)$, and the $\kappa$ (also known as $K_0^*(700)$) pole position, $\sqrt{s_\kappa}=\left(966_{-24}^{+41}-i 198_{-17}^{+38}\right)$~MeV. 
We also provide a detailed analysis of the complex validity domain of the Roy-Steiner-type equations.
\end{abstract}

\maketitle

\section{Introduction}\label{sec:intro}

As pseudo-Nambu-Goldstone bosons arising from spontaneous chiral symmetry breaking, pions and kaons are the two lightest hadrons in the quantum chromodynamics (QCD) spectrum. 
Pion-kaon ($\pi K$) scattering is a fundamental process for investigating strong interaction dynamics at low energies, particularly for exploring strangeness-related QCD phenomena. The low-energy parameters of this process, especially the scattering lengths, contain essential information about the spontaneous breaking of chiral symmetry in the strange-meson sector. 
Partial-wave analysis of $\pi K \rightarrow \pi K$ in the isospin $I=1/2$ channel shows that the $S$-wave $K_0^*(1430)$, $P$-wave $K^*(892)$, and $D$-wave $K_2^*(1430)$ resonances appear as distinct peaks in the partial-wave amplitudes.

However, hadronic resonances do not always manifest as distinct peaks in scattering amplitudes. 
The most rigorous way to characterize them is through pole singularities obtained by analytically continuing the scattering amplitudes into the complex energy plane. 
When these poles are located far from the physical region, they may only induce moderate variations in the physical phase shifts. 
A prime example is the $\kappa$ resonance (also known as $K_0^*(700)$), which is the strange counterpart to the $\sigma$ resonance (also known as $f_0(500)$) in $\pi\pi$ scattering. 
The smooth rise observed in the $\left(I,J\right)=\left(1/2,0\right)$ phase shift near the $\pi K$ threshold is primarily governed by the $\kappa$ resonance.

Lattice QCD (LQCD) stands as one of the most powerful tools for investigating nonperturbative aspects of low-energy strong interactions, particularly in the study of hadronic resonances. 
This systematic approach discretizes spacetime by placing quark and gluon fields on a finite grid. 
Through Monte Carlo importance sampling of field configurations within the path-integral framework, one can compute correlation functions, thereby allowing for the analysis of their spectral contents. 
The fundamental connection between scattering phase shifts and discrete spectra in a finite volume was first established by L\"uscher~\cite{Luscher:1985dn, Luscher:1986pf, Luscher:1990ux}; for a recent review see~\cite{Briceno:2017max}.

Before 2010, LQCD studies of $\pi K$ scattering focused mainly on determining the scattering lengths characterizing near-threshold dynamics, with early studies by Miao et al.~\cite{Miao:2004gy} and NPLQCD~\cite{Beane:2006gj} focusing on the $I=3/2$ channel, while Nagata et al.~\cite{Nagata:2008wk} and Fu et al.~\cite{Fu:2011wc} addressed the $I=1/2$ channel. 
Lang et al.~\cite{Lang:2012sv} and the PACS-CS Collaboration~\cite{Sasaki:2013vxa} extended these analyses to both isospin channels. 
The ETM Collaboration~\cite{Helmes:2018nug} has recently contributed further insight with their $2+1+1$ flavor calculations.

Early LQCD investigations of strange-meson resonances focused on analyzing energy spectra, searching for additional energy levels that might correspond to resonances~\cite{Prelovsek:2010kg, Alexandrou:2012rm}. 
However, a wrong spectrum would arise when the quark-disconnected contributions are neglected~\cite{Lang:2012sv, Guo:2013nja}. 
To date, there have been few comprehensive determinations of the energy-dependence of $\pi K$ scattering amplitudes with LQCD. 
The first study by Fu et al.~\cite{Fu:2012tj} examined $K^*(892)$ using the staggered action, with subsequent work~\cite{Fu:2013sua} extending the analysis to $\kappa$. 
Prelovsek et al.~\cite{Prelovsek:2013ela} determined the $P$-wave $\pi K$ phase shifts and the width of $K^*(892)$ without dynamical strange quarks. 
The RQCD Collaboration~\cite{Bali:2015gji} repeated the calculation for $K^*(892)$ with a pion mass close to the physical value. 
Brett et al.~\cite{Brett:2018jqw} calculated $S$- and $P$-wave $I=1/2$ amplitudes in $2+1$ flavors at $m_\pi=230~$MeV, and a possible virtual state (VS) $\kappa$ was suggested based on the effective-range expansion. Rendon et al.~\cite{Rendon:2020rtw} determined $S$- and $P$-wave $I=1/2$ $\pi K$ scattering amplitudes with $2+1$ flavors for two different pion masses around 317 and 176~MeV. 
At both pion masses, a broad $\kappa$ and a narrow $K^*(892)$ were found. 
They observed, however, that the pole positions of $\kappa$ remain stable only for parametrizations that include an Adler zero. 
A more comprehensive analysis by the Hadron Spectrum Collaboration (HSC)~\cite{Dudek:2014qha,Wilson:2014cna} determined resonance properties through calculations of $S$-, $P$-, and $D$-waves for both isospin channels. Their $2+1$ flavor LQCD study at $m_\pi=391~$MeV established the complete resonance spectrum in these partial waves. 
For the low-energy behavior of $I=1/2$ channels, the $S$-wave amplitude seems to contain a VS associated with $\kappa$ and an obvious shallow bound state (BS) $K^*(892)$ was found in the $P$-wave amplitude. 
They also calculated the $I=1/2$ $S$- and $P$-wave scattering amplitudes and the positions of $K^*(892)$ at four quark masses~\cite{Wilson:2019wfr}. 
Recently, Boyle et al.~\cite{Boyle:2024hvv,Boyle:2024grr} and Dawid et al.~\cite{Dawid:2025doq, Dawid:2025zxc} reported the first calculation of the $\left(I,J\right)=\left(1/2,1\right)$ channel and  $\left(I,J\right)=\left(3/2,0\right),\left(3/2,1\right)$ channels at physical quark masses, respectively. 
Furthermore, alternative approaches have been developed to extract $\pi K$ scattering amplitudes from finite-volume calculations using unitarized chiral perturbation theory (ChPT), as detailed in Refs.~\cite{Doring:2011nd, Doring:2012eu, Zhou:2014ana, Sadl:2024dbd, Yan:2023bwt, Guo:2018tjx, Guo:2016zep}.

In fact, a general consensus is yet to be reached on relating the current phase shift analysis of LQCD data to the pole contents. 
While LQCD is capable of providing accurate phase shifts, the extraction of resonances in the complex energy plane from these phase shifts generally requires the aid of phenomenological models. 
This represents one of the most significant challenges in the analysis of LQCD data. Particularly for non-ordinary resonances like $\kappa$, unconstrained $K$-matrix models lead to rather large uncertainties for the determinations of resonance pole positions~\cite{Wilson:2019wfr}; see also recent analysis of the higher $D_0^*$ pole~\cite{Asokan:2022usm}. 
The most intriguing and challenging examples emerge from the $m_\pi=391~$MeV case. 
In this case, not only does the $K^*(892)$ appear as a shallow BS~\cite{Dudek:2014qha, Wilson:2014cna, Wilson:2019wfr}, but also the $\sigma$ in $\pi\pi$ scattering becomes a BS~\cite{Briceno:2016mjc, Briceno:2017qmb}. 
For $\pi K\to \pi K$ scattering, the $\pi\pi\to K\bar{K}$ process acts as a cross channel. Properly accounting for the BS pole in the cross channel, when present, is essential for preserving crossing symmetry. 
Neglecting the contributions of these states from crossing symmetry makes it difficult to quantify the uncertainty of the extracted resonance poles. For a recent Roy equation study of the $\sigma$ resonance in $\pi\pi$ scattering at $m_\pi=391$~MeV, we refer to Ref.~\cite{Cao:2023ntr}. 
Consequently, the crossing symmetry is also expected to be crucial for the determination of the $\kappa$ resonance in the $\pi K$ scattering at unphysically large pion masses. In this context, we propose a rigorous reexamination of lattice $\pi K$ scattering phase shifts by utilizing the rigorous Roy-Steiner equation approach, with particular focus on the poorly constrained deep VS pole below threshold identified in HSC studies~\cite{Dudek:2014qha, Wilson:2014cna}.

For the case of physical quark masses, $\kappa$ also had remained a subject of considerable debates for several decades until it was established as the lowest-lying strange hadronic resonance in QCD in the past twenty years based on rigorous dispersive analyses of $\pi K$ scattering~\cite{Zheng:2003rw, Zhou:2006wm, Buettiker:2003pp, Descotes-Genon:2006sdr, Pelaez:2018qny, Pelaez:2020uiw}; see, e.g., Refs.~\cite{Yao:2020bxx, Pelaez:2020gnd} for reviews. 
The dispersive techniques such as crossing-symmetric partial-wave dispersion relations (DRs), have repeatedly been proved to be a powerful model-independent method. 
They are built upon general principles such as unitarity, analyticity and crossing symmetry. 
The resulting crossing-symmetric partial-wave DRs enable high-precision investigations of hadronic processes.
For $\pi \pi$ scattering, Roy equation~\cite{Roy:1971tc} is a coupled system of partial-wave DRs for the $\pi \pi$ partial waves. 
The Roy equation formalism has enabled the determination of low-energy $\pi \pi$ scattering amplitudes with unprecedented accuracy~\cite{Ananthanarayan:2000ht, Colangelo:2001df, Garcia-Martin:2011iqs}, leading to tightly constrained $\sigma$ pole parameters~\cite{Caprini:2005zr, Moussallam:2011zg, Garcia-Martin:2011nna, Pelaez:2015qba}. 
For the unequal mass case, Steiner equation~\cite{Hite:1973pm}, also known as Roy-Steiner (RS) equation, for $\pi N$ scattering is a set of partial-wave DRs that combine the $s/u$- ($\pi N\to \pi N$) and $t$-channel ($\pi\pi\to N\bar N$) amplitudes by means of hyperbolic DRs. 
In the last decade, $\pi N$ Steiner equation has been successfully applied to various fields~\cite{Ditsche:2012fv, Hoferichter:2012wf, Hoferichter:2015dsa, Hoferichter:2015tha, Hoferichter:2015hva,  Hoferichter:2016duk, RuizdeElvira:2017stg, Hoferichter:2018zwu} and confirmed the existence of the subthreshold pole singularity in the $\pi N$ $S_{11}$ channel~\cite{Wang:2017agd, Cao:2022zhn, Hoferichter:2023mgy}. 
The first complete construction and numerical solution of the $\pi K$ system of RS-type equations have been presented by B\"uttiker et al.~\cite{Buettiker:2003pp}. 

Thus, crossing-symmetrically constrained analysis of LQCD data with dispersive techniques is expected to be essential for the reliable determination of the $\kappa$ pole parameters. 
We follow the spirit of Ref.~\cite{Cao:2023ntr} on the Roy equation analysis of $\pi\pi$ scattering at unphysical pion masses; see also a similar analysis by HSC~\cite{Rodas:2023nec}.
Our analysis focuses on the heavy-pion mass with $m_\pi=391~$MeV, leveraging available HSC lattice results for $\pi K, \eta K$ scattering phase shifts~\cite{Dudek:2014qha, Wilson:2014cna, Wilson:2019wfr} and $\pi\pi, K\bar K$ scattering amplitudes~\cite{Dudek:2012gj, Dudek:2012xn, Briceno:2016mjc, Briceno:2017qmb}. 
Part of the results have been briefly presented in Ref.~\cite{Cao:2024zuy} with the emphasis on elaborating the phenomenological consequences, such as the outputs of phase shifts and pole contents, while this paper contains many additional technical details regarding the RS-type equation analysis. 

This work is organized as follows. 
We first briefly summarize the fixed-$t$, hyperbolic DRs and RS-type equations of $\pi K$ scattering in Sec.~\ref{sec:RS_eq}, where the $\sigma$ and $K^*(892)$ BS pole terms in the DRs are added to accommodate the lattice data. 
A key point is how crossing symmetry necessarily introduces new cut structures in the partial-wave amplitudes.
Then, a general framework for the complex validity domain of the hyperbolic DR is proposed in Sec.~\ref{sec:VD}. 
In Sec.~\ref{sec:Input}, we discuss the available lattice inputs and the asymptotic Regge amplitudes at $m_\pi=391~$MeV. 
The formalism of the $t$-channel input is discussed in Sec.~\ref{sec:t-channel}. 
We then proceed to the numerical solution in Sec.~\ref{sec:s-channel}. 
The phenomenological discussion on the $\kappa$ pole can be found in Sec.~\ref{sec:kappa}. 
We summarize our work in Sec.~\ref{sec:summary}. 
Various details about the kernels of the RS-type equation of $\pi K$ scattering and the subthreshold singularities are provided in Appendices~\ref{app:kernel} and \ref{app:VS & Adler0}, respectively.

%%%%%%%%%%%%%%%%%%%%%%%%%%%%%%%%%%%%%%%%%%%%%%%%%%%%%%%%%%%%%%%%%%%%%%%%%%%%

\section{Roy-Steiner-type equations of pion-kaon scattering}\label{sec:RS_eq}

\subsection{Notations and conventions}

We take the $s$-channel reaction to be $\pi(q_1)K(p_1) \rightarrow \pi(q_2)K(p_2)$ and the $t$-channel reaction to be $\pi(q_1)\pi\left(-q_2\right) \rightarrow \bar{K}(-p_1)K\left(p_2\right)$ with the usual Mandelstam variables:
\begin{align}
    s=(q_1+p_1)^2\ ,\quad t=(q_1-q_2)^2\ ,\quad u=(q_1-p_2)^2\ ,
\end{align}
which fulfill $s+t+u=2(m_\pi^2+m_K^2)$, where $m_\pi\equiv m_{\pi^{\pm,0}}$ and $ m_K\equiv m_{K^\pm}=m_{K^0,\bar{K}^0}$ denote the pion and kaon masses in the isospin limit, respectively. 
Additionally, it is convenient to define
\begin{align}
    m_{\pm}=m_K \pm m_\pi,\quad \Sigma=m_\pi^2+m_K^2,\quad\Delta=m_K^2-m_\pi^2,\quad t_\pi=4 m_\pi^2,\quad t_K=4 m_K^2\ .
\end{align}
Furthermore, we introduce the momentum in the $s$-channel center-of-mass (c.m.) system,
\begin{align}
    q\equiv q_{\pi K}=\frac{1}{2 \sqrt{s}} \sqrt{\lambda\left(s, m_\pi^2, m_K^2\right)}=\frac{1}{2 \sqrt{s}} \sqrt{s^2-2 s \Sigma+\Delta^2}\ ,
\end{align}
where the K\" all\' en function is given by $\lambda\left(s, m_\pi^2, m_K^2\right)=\left(s-m_{+}^2\right)\left(s-m_{-}^2\right)$, and the phase space factor of $\pi K$ system is defined as $\rho_{\pi K}(s)={2 q}/{\sqrt{s}}$. 
For the $t$-channel reaction, the c.m. momenta are $q_\pi={\sqrt{t-t_\pi}}/{2}$ and $q_K={\sqrt{t-t_K}}/{2}$, together with corresponding phase-space factors $\rho_\pi(t)={2 q_\pi}/{\sqrt{t}}$ and $\rho_K(t)={2 q_K}/{\sqrt{t}}$.

The amplitudes of physical $\pi K \to \pi K$ ($\pi\pi \to K\bar K$) scatterings can be written in terms of only two independent elements with total $s$-channel isospin $I_s=\frac{1}{2},\frac{3}{2}$ ($t$-channel isospin $I_t=0,1$). 
We assign the isospin multiplets of $\pi$ and $K$ using the usual Condon-Shortley phase convention for the Clebsch-Gordan coefficients,\footnote{Note that our conventions for $K$ multiplet differ from Ref.~\cite{Lang:1978fk}.} 
\begin{align}
    \begin{gathered}
        \left|\pi^{+}\right\rangle=-|1,+1\rangle_\pi, \quad\left|\pi^{-}\right\rangle=|1,-1\rangle_\pi, \quad\left|\pi^0\right\rangle=|1,0\rangle_\pi\ , \\
        \left|K^{+}\right\rangle=\left|\frac{1}{2},+\frac{1}{2}\right\rangle_K, \quad\left|K^0\right\rangle=\left|\frac{1}{2},-\frac{1}{2}\right\rangle_K, \quad\left|K^{-}\right\rangle=\left|\frac{1}{2},-\frac{1}{2}\right\rangle_{\bar{K}}, \quad\left|\bar{K}^0\right\rangle=-\left|\frac{1}{2},+\frac{1}{2}\right\rangle_{\bar{K}}\ .
    \end{gathered}
\end{align}
One gets the following relations between the physical amplitudes and isospin ones: 
\begin{align}\label{eq:F_isospin1}
    \begin{aligned}
        & F\left(K^{+} \pi^{-} \rightarrow K^{+} \pi^{-}\right)&&=\frac{1}{3}\!\left(2 F^{1 / 2}+F^{3 / 2}\right) , \\
        & F\left(K^{+} \pi^{-} \rightarrow K^0 \pi^0\right)=F\left(K^{-} \pi^0 \rightarrow \bar{K}^0 \pi^{-}\right)&&=\frac{\sqrt{2}}{3}\!\left(-F^{1 / 2}+F^{3 / 2}\right) , \\
        & F\left(K^{+} \pi^{+} \rightarrow K^{+} \pi^{+}\right)&&=F^{3 / 2}\ , \\
        & F\left(\pi^{+} \pi^{-} \rightarrow K^{+} K^{-}\right)&&=-\frac{1}{2} F^1-\sqrt{\frac{1}{6}} F^0\ , \\
        & F\left(\pi^{+} \pi^{-} \rightarrow K^0 \bar{K}^0\right)&&=-\frac{1}{2} F^1+\sqrt{\frac{1}{6}} F^0\ ,
    \end{aligned}
\end{align}
It is noticed that the amplitudes in the first and third lines in Eqs.~\eqref{eq:F_isospin1} are related by $s \leftrightarrow u$ crossing,
\begin{align}
    F^{1 / 2}(s, t, u)=\frac{3}{2} F^{3 / 2}(u, t, s)-\frac{1}{2} F^{3 / 2}(s, t, u)\ .
\end{align}

We also need the isospin-even and isospin-odd amplitudes $F^+$ and $F^-$, which are defined as 
\begin{equation}
    \hat F_{j i}=\delta_{j i} F^{+}+\frac{1}{2}\left[\tau_j, \tau_i\right] F^{-}
\end{equation}
with $i,j=1,2,3$ being the isospin indices of the pions. Sandwiching $\hat F_{j i}$ between the final and initial kaons, we get the corresponding scattering amplitude. 
The isospin-even and isospin-odd amplitudes $F^+$ and $F^-$ are symmetric and antisymmetric under $s\leftrightarrow u$ crossing, respectively. 
Using the isospin crossing matrix (see Refs.~\cite{Martin:1970Elementary,Lang:1978fk,Ditsche:2012fv} for detailed derivations), we can relate the $s$-channel isospin amplitudes $F^{1/2,3/2}(s,t,u)$ to $F^{\pm}(s,t,u)$ as follows,
\begin{align}\label{eq:s_isospin}
    \left(\begin{array}{l}
        F^{+} \\
        F^{-}
        \end{array}\right)=C_{\nu s}\!\left(\begin{array}{l}
        F^{1 / 2} \\
        F^{3 / 2}
        \end{array}\right), \quad\left(\begin{array}{l}
        F^{1 / 2} \\
        F^{3 / 2}
        \end{array}\right)=C_{s \nu}\!\left(\begin{array}{l}
        F^{+} \\
        F^{-}
        \end{array}\right), \quad C_{\nu s}=\frac{1}{3} C_{s \nu}=\frac{1}{3}\!\left(\begin{array}{cc}
        1 & 2 \\
        1 & -1
        \end{array}\right),
\end{align}
and we can also relate the $t$-channel isospin amplitudes $G^{0,1}(t,s,u)$ to amplitudes $F^{\pm}(s,t,u)$,
\begin{align}\label{eq:t_isospin}
    \left(\begin{array}{l}
        F^{+}(s,t,u) \\
        F^{-}(s,t,u)
        \end{array}\right)=C_{\nu t}\!\left(\begin{array}{l}
        G^0(t,s,u) \\
        G^1(t,s,u)
        \end{array}\right), \quad C_{\nu t}=C_{\nu s} C_{s t}=\left(\begin{array}{cc}
        \frac{1}{\sqrt{6}} & 0 \\
        0 & \frac{1}{2}
        \end{array}\right).
\end{align}
For later convenience, we rewrite the above relations as
\begin{align}\label{eq:G&F}
    G^0(t, s, u)=\sqrt{6} F^{+}(s, t, u)\ ,\quad G^1(t, s, u)=2 F^{-}(s, t, u)\ .
\end{align}

The partial-wave decompositions of both the $\pi K\to \pi K$ and $\pi\pi\to K\bar K$ scattering amplitudes are defined as,
\begin{align}
    & F^I(s, t, u)=16 \pi \sum_{J}(2 J+1) P_{J}\!\left(z_s\right) f_{J}^I(s)\ , \\
    & G^I(t, s, u)=16 \pi \sqrt{2} \sum_{J}(2 J+1)\left(q_\pi q_K\right)^{J} P_{J}\!\left(z_t\right) g_{J}^I(t)\ ,
\end{align}
where $P_J$ is the Legendre polynomial of the first kind. For the $t$-channel partial waves, only even/odd partial waves contribute for even/odd isospin states, respectively.
The scattering angles $\theta_s$ and $\theta_t$ in the $s$-channel and $t$-channel are defined as
\begin{align}
    z_s=\cos \theta_s=1+\frac{2 s t}{\lambda_s}\ , \quad z_t=\cos \theta_t=\frac{s-u}{4 q_\pi q_K}=\frac{\nu}{4 q_\pi q_K}\ ,
\end{align}
where $\lambda_s=\lambda\left(s, m_\pi^2, m_K^2\right)=s^2-2 s \Sigma+\Delta^2=4 s q^2(s)$.

Correspondingly, the partial-wave amplitudes are then given by
\begin{align}
    &f_{J}^I(s)=\frac{1}{32 \pi} \int_{-1}^1 \md z_s P_{J}\!\left(z_s\right) F^I\left(s, t\left(z_s\right)\right) ,\label{eq:s_PW}\\
    &g_{J}^I(t)=\frac{\sqrt{2}}{32 \pi\left(q_\pi q_K\right)^{J}} \int_0^1 \md z_t P_{J}\!\left(z_t\right) G^I\left(t, s\left(z_t\right)\right) .\label{eq:t_PW}
\end{align}
The $\pi K$ scattering lengths are defined as 
\begin{align} \label{eq:a0}
    a_J^I=\frac{2}{m_{+}} \lim_{s\to m_+^2}\frac{f_J^I\left(s\right)}{q_{\pi K}(s)^{2 J}} .
\end{align}
The relations between the partial-wave $S$-matrix elements and partial-wave amplitudes can be written as
\begin{align}
    & S_{J}^I(s)_{\pi K \rightarrow \pi K}=1+i \frac{4 q}{\sqrt{s}} f_{J}^I(s) \,\theta\!\left(s-m_{+}^2\right) , \\
    & S_{J}^I(t)_{\pi \pi \rightarrow K \bar{K}}=i \frac{4\left(q_\pi q_K\right)^{J+1 / 2}}{\sqrt{t}} g_{J}^I(t) \,\theta\!\left(t-t_K\right) .
\end{align}
Furthermore, using the coupled-channel unitarity relation $\left|S^I_J(t)_{\pi \pi \rightarrow K \bar{K}}\right|^2= 1-\left|S^I_J(t)_{\pi \pi \rightarrow \pi\pi}\right|^2<1$, we can derive a unitarity bound for the cross-channel partial-wave amplitudes,
\begin{align}\label{eq:UB}
    \left|g_J^I(t)\right|<\frac{\sqrt{t}}{4\left(q_\pi q_K\right)^{J+1 / 2}}\ .
\end{align}

\subsection{Fixed-$t$ and hyperbolic dispersion relations}

For $\pi K\to \pi K$ and $\pi\pi\to K\bar K$ scatterings, numerous dispersive studies based on RS-type equations have been conducted. 
In the 1970s, partial-wave amplitudes were investigated using both fixed-$t$~\cite{Nielsen:1973au, Ader:1973qhy} and hyperbolic DRs~\cite{Johannesson:1974ma, Hedegaard-Jensen:1974plp, Johannesson:1976qp, Johannesson:1977zu} for these scattering processes. 
Nevertheless, certain approximations were employed in the treatment of these equations. 
Recent interest in RS-type equations for $\pi K$ scattering has been driven by high-precision experimental data, with primary objectives including the establishment of rigorous sum rules~\cite{Ananthanarayan:2000cp, Ananthanarayan:2001uy}, determination of low-energy phase shifts~\cite{Buettiker:2003pp}, and precise determination of the pole corresponding to the lowest-lying scalar strange meson $\kappa$~\cite{Descotes-Genon:2006sdr, Pelaez:2020uiw}. 
For a comprehensive overview, we direct the reader to the recent review~\cite{Pelaez:2020gnd}.

\subsubsection{Fixed-$t$ dispersion relations}

According to the Froissart-Martin bound~\cite{Froissart:1961ux, Martin:1962rt}, for fixed-$t$ DR, at most two subtractions are needed for $F^+$ and one subtraction for $F^-$, respectively, because $s-u$ can be factored out in the latter case. 
More detailed information on asymptotic behavior is provided by the Regge theory~\cite{Collins:1977jy}. 
According to Eq.~\eqref{eq:G&F}, $F^+$ and $F^-$ correspond to the exchange of isospin 0 and 1 states in the $t$-channel, respectively. 
At high energies, $F^+$ is dominated by the $t$-channel exchanges of the Pomeron, while $F^-$ receives its dominant contribution from the $t$-channel exchanges of the $\rho$-trajectory. 
Therefore, two subtractions are appropriate for $F^+$~\footnote{However, this is not strictly necessary, since the $F^\pm$ DRs contain integrals over the right-hand and left-hand cuts whose leading Regge trajectories cancel each other due to the $s\leftrightarrow u$ crossing properties~\cite{Pelaez:2020gnd}. Consequently, a once-subtracted DR could ensure the convergence of the fixed-$t$ DR.} while an unsubtracted fixed-$t$ DR is expected to converge for $F^-$. 
Previous studies~\cite{Johannesson:1976qp, Pelaez:2018qny, Pelaez:2020gnd} have partially used the so-called minimally subtracted fixed-$t$ DRs. 
Fewer subtractions are advantageous for reducing uncertainty propagation in subtraction terms, which may become excessive in the resonance region, while additional subtractions are beneficial when focusing on threshold or subthreshold regions. 
Therefore, we employ a twice-subtracted DR for $F^+$ and a once-subtracted DR for $F^-$ to improve convergence and reduce sensitivity to high-energy behavior.

Considering that $F^+$ is invariant under $s \leftrightarrow u$, we can set the subtraction point at $s=0$. 
Thus, we have~\cite{Johannesson:1976qp, Pelaez:2018qny, Pelaez:2020gnd}\footnote{The residue of a BS with spin-$J$ (for the full scattering amplitude) is $(-1)^{J+1}g^2 P_J(z_s(s_\text{pole},t))$~\cite{Martin:1970Elementary, Taylor1972}. Note that the result in~\cite{Martin:1970Elementary} misses a factor of $(-1)^J$.},
\begin{align}\label{eq:F+1}
    F^{+}(s, t)=&\, c^{+}(t)\mathbf{-16\pi g_{K^* \pi K}^2 P_1\left(z_s(s_{K^*},t)\right)\left(\frac{1}{s_{K^*}-s}+\frac{1}{s_{K^*}-u}\right)} \nonumber\\
    & +\frac{1}{\pi} \int_{m_{+}^2}^{\infty} \frac{\md s^{\prime}}{s^{\prime 2}}\left[\frac{s^2}{s^{\prime}-s}+\frac{u^2}{s^{\prime}-u}\right]\operatorname{Im} F^{+}\!\left(s^{\prime}, t\right) ,
\end{align}
where $c^+(t)$ is an unknown $t$-dependent subtraction term. The new BS pole term corresponding to $K^*$ (here and in the rest of the paper, the pole terms are in bold face for easy identification), which is only required when the light quark mass is sufficiently heavy, is also added to describe the lattice data. 
The fixed-$t$ DR for $F^-(s,t)$ is similar to that for $F^+(s,t)$. 
We take a once-subtracted DR for $F^-(s,t)/(s-u)$, yielding~\cite{Pelaez:2020gnd}
\begin{align}\label{eq:F-}
    F^{-}(s, t)=&\, c^{-}(t)(s-u)\mathbf{-16\pi g_{K^* \pi K}^2 P_1\left(z_s(s_{K^*},t)\right)\left(\frac{1}{s_{K^*}-s}-\frac{1}{s_{K^*}-u}\right)} \nonumber\\
    & +\frac{1}{\pi} \int_{m_{+}^2}^{\infty} \frac{\md s^{\prime}}{s^{\prime 2}} \left[\frac{s^2}{s^{\prime}-s}-\frac{u^2}{s^{\prime}-u}\right]\operatorname{Im} F^{-}\!\left(s^{\prime}, t\right) ,
\end{align}
where $c^-(t)$ is also an unknown $t$-dependent subtraction term.

\subsubsection{Hyperbolic dispersion relations}

In hyperbolic DRs, $s$ and $u$ are constrained on a hyperbola $(s-a)(u-a)=b$, where $a$ is a free parameter.
Here, it is assumed that the amplitude has $s \leftrightarrow u$ symmetry.\footnote{In fact, this requirement can be relaxed provided the kinematics satisfy $s \leftrightarrow u$ symmetry.} 
For $\pi K$ scattering, the $a=0$ case is investigated in Ref.~\cite{Buettiker:2003pp}, and Refs.~\cite{Pelaez:2018qny, Pelaez:2020gnd} extended RS-type equations to arbitrary $a$, although they partially used the minimally subtracted DRs. 
In this work, we do not employ the minimally subtracted DRs and consider the corresponding RS-type equations with $a\neq 0$. 
For detailed properties of hyperbolic DRs, we refer to the seminal work~\cite{Hite:1973pm}. 
For completeness, we present several key relations that will prove useful for subsequent discussions.
Using $s+t+u=2\Sigma$ and $(s-a)(u-a)=b$, we can solve
\begin{align}
    & s=s_b \equiv s_b(t)=\frac{1}{2}\!\left(2 \Sigma-t+\sqrt{(t+2 a-2 \Sigma)^2-4 b}\right) ,\nonumber\\
    & u=u_b \equiv u_b(t)=\frac{1}{2}\!\left(2 \Sigma-t-\sqrt{(t+2 a-2 \Sigma)^2-4 b}\right) ,
\end{align}
where $t$ and $b$ satisfy a linear relation:
\begin{align}\label{eq:HDR_t}
    t=-\frac{b}{s-a}+2 \Sigma-s-a.
\end{align}

We begin with $F^+$. 
Consider a twice-subtracted hyperbolic DR in the $t$-channel\footnote{In this scenario, the asymptotic behavior corresponds to the kinematics of backward scattering (for a detailed analysis, see Appendix D of Ref.~\cite{Ditsche:2012fv}). The most significant contributions arise from the $u$-channel $K^*$ and $K^*_2$ trajectories, thus a once-subtracted DR would suffice for convergence. However, to improve convergence and reduce dependence on high-energy behavior, we employ a twice-subtracted DR here.} with the subtraction point at $t=0$,
\begin{align}
    F^+(t ; a, b)=&\, f^+(b)+th^+(b)\mathbf{+F^+_\text{pole}(t;a,b)} \nonumber\\
    & +\frac{t^2}{\pi} \int_{-\infty}^{t_\text{L}} \mathrm{d} t^{\prime} \frac{\operatorname{Im} F^+\left(t^{\prime} ; a, b\right)}{t^{\prime 2}(t^{\prime}-t)}+\frac{t^2}{\pi} \int_{t_\pi}^{\infty} \mathrm{d} t^{\prime} \frac{\operatorname{Im} F^+\left(t^{\prime} ; a, b\right)}{t^{\prime 2}(t^{\prime}-t)}\ ,
\end{align}
where $f^+(b)$ and $h^+(b)$ are unknown $b$-dependent subtraction terms, $F^+_\text{pole}(t;a,b)$ is a pole term, and $t_\text{L}$ denotes the branch point of the left-hand cut (LHC). 
By transforming the $t^\prime$-integral into the $s^\prime$-integral, a twice-subtracted hyperbolic DR is obtained after some algebraic simplifications\footnote{In general, the pole term would include a ``constant term'' proportional to $(s_{K^*}-a)^{-1}$ (which may depend on $t$), explicitly dependent on $a$. 
In principle, such $a$-dependence would cancel with contributions from the dispersion integral at infinity, rendering the final result $a$-independent. However, numerically, since the dispersion integral must be truncated at a finite energy, this contribution would depend on $a$, introducing unnecessary numerical complications. 
Since we effectively neglect integral contributions from infinity, we can simply ignore this constant term (or consider it absorbed into the subtraction term) to avoid these numerical difficulties.}
\begin{align}\label{eq:HDR_F+}
    F^{+}\!\left(s_b, t\right)=&\, f^{+}(b)+t h^{+}(b) \nonumber\\
    &\mathbf{-16\pi g_{K^* \pi K}^2 P_1\left(z_s(s_{K^*},t)\right)\left(\frac{1}{s_{K^*}-s_b}+\frac{1}{s_{K^*}-u_b}\right)+16\pi g_{\sigma \pi\pi}g_{\sigma K\bar{K}}\frac{1}{\sqrt{3}\!\left(s_\sigma-t\right)}} \nonumber\\
    &+\frac{t^2}{\sqrt{6} \pi} \int_{t_\pi}^{\infty} \frac{\mathrm{d} t^{\prime}}{t^{\prime 2}\!\left(t^{\prime}-t\right)} \operatorname{Im} G^0\left(t^{\prime},s_b^\prime\right) \nonumber\\
    &+\frac{1}{\pi} \int_{m_{+}^2}^{\infty} \mathrm{d} s^{\prime}\left[\frac{2 s^{\prime}-2 \Sigma+t}{s^{\prime 2}-2\Sigma s^\prime+s_b u_b-a(s_b+u_b-2\Sigma)+s^{\prime}t-at} \right.\nonumber\\
    &\left.-\frac{2 s^{\prime}-2 \Sigma-t}{s^{\prime 2}-2\Sigma s^\prime+s_b u_b-a(s_b+u_b-2\Sigma)}\right] \operatorname{Im} F^{+}\!\left(s^{\prime}, t_b^{\prime}\right) ,
\end{align}
where
\begin{align}
    s_b^{\prime} \equiv s_b\!\left(t^{\prime}\right) , \quad u_b^{\prime} \equiv u_b\!\left(t^{\prime}\right) , \quad t_b^\prime= 2 \Sigma-s^{\prime}-\frac{b}{s^{\prime}-a}-a\ .
\end{align}
Equation~\eqref{eq:HDR_F+} contains two unknown functions, $f^+(b)$ and $h^+(b)$, along with two pole terms corresponding to $K^*$ and $\sigma$. 
When the pole terms are omitted and $a=0$ is taken, this equation reproduces the findings presented in Refs.~\cite{Buettiker:2003pp,Descotes-Genon:2006sdr}.

For $F^-$ (actually the DR is constructed for $F^-/(s-u)$), in principle, subtraction is not necessary.
However, we use once subtracted DR here, with the subtraction point at $t=0$, and obtain
\begin{align}\label{eq:HDR_F-}
    F^{-}\!\left(s_b, t\right)=&\, h^{-}(b)(s_b-u_b)\mathbf{-16\pi g_{K^* \pi K}^2 P_1\left(z_s(s_{K^*},t)\right)\left(\frac{1}{s_{K^*}-s_b}-\frac{1}{s_{K^*}-u_b}\right)} \nonumber\\
    &+\frac{t(s_b-u_b)}{2 \pi} \int_{t_\pi}^{\infty} \frac{\mathrm{d} t^{\prime}}{t^{\prime}\!\left(t^{\prime}-t\right)} \operatorname{Im}\left[\frac{G^1\left(t^{\prime}, s_b^{\prime}\right)}{\left(s_b^{\prime}-u_b^{\prime}\right)}\right] \nonumber\\
    &+\frac{(s_b-u_b)}{\pi} \int_{m_{+}^2}^{\infty} \mathrm{d} s^{\prime}\left[\frac{1}{s^{\prime 2}-2\Sigma s^\prime+s_b u_b-a(s_b+u_b-2\Sigma)+s^{\prime}t-at} \right.\nonumber\\
    &\left.-\frac{1}{s^{\prime 2}-2\Sigma s^\prime+s_b u_b-a(s_b+u_b-2\Sigma)}\right] \operatorname{Im} F^{-}\!\left(s^{\prime}, t_b^{\prime}\right) .
\end{align}
The formalism is equivalent to that used in Refs.~\cite{Ananthanarayan:2001uy, Pelaez:2018qny,Pelaez:2020gnd}. By taking $a=0$ and neglecting the pole term, it matches the result in Ref.~\cite{Buettiker:2003pp}. 

Next, we need to establish the relations between the $S$-wave scattering lengths and the corresponding hyperbolic subtraction terms. 
This can be achieved by matching the fixed-$t$ DR with the hyperbolic DR. 
In the literature, only partial derivations for the $a=0$ case~\cite{Buettiker:2003pp} have been presented. 
Therefore, in this section, we provide a pedagogical derivation with detailed steps to facilitate future applications. 
First, consider the hyperbolic DR for $F^-$ in Eq.~\eqref{eq:HDR_F-}. 
Our aim is to relate the subtraction term $h^-$ to the $\pi K$ scattering lengths through iterative application of the fixed-$t$ DR in Eq.~\eqref{eq:F-}.

The $\pi K$ scattering lengths are related to the $S$-wave scattering amplitudes at the $\pi K$ threshold, i.e., when $s = m_+^2, u = m_-^2$ and $t=0$, as in Eq.~\eqref{eq:a0}. 
By fixing $t = 0$ in Eqs.~\eqref{eq:F-} and~\eqref{eq:HDR_F-}, we obtain two relations,
\begin{align}
    F^-(s,0)=&\, c^-(0)(s-u)\mathbf{-16\pi g_{K^* \pi K}^2 \left(\frac{1}{s_{K^*}-s}-\frac{1}{s_{K^*}-u} \right)} \nonumber\\
    &+\frac{1}{\pi}\int_{m_+^2}^{\infty} \frac{\md s^\prime}{s^{\prime 2}}\!\left(\frac{s^2}{s^{\prime}-s}-\frac{u^2}{s^{\prime}-u}\right)\operatorname{Im}F^-(s^\prime,0)\ ,\label{eq:F-th1}\\
    F^-(s_b,0)=&\, h^-(b)(s_b-u_b)\mathbf{-16\pi g_{K^* \pi K}^2 \left(\frac{1}{s_{K^*}-s_b}-\frac{1}{s_{K^*}-u_b} \right)}\ .\label{eq:F-th2}
\end{align}
From Eq.~\eqref{eq:F-th1}, we can write 
\begin{align}\label{eq:c-_0}
    c^-(0)=&\, \frac{8\pi m_+ a^-_0}{m_+^2-m_-^2}\mathbf{+\frac{16\pi g_{K^* \pi K}^2}{(s_{K^*}-m_+^2)(s_{K^*}-m_-^2)}} \nonumber\\
    &-\frac{1}{m_+^2-m_-^2}\frac{1}{\pi} \int_{m_+^2}^{\infty}\frac{\md s^\prime}{s^{\prime 2}}\!\left(\frac{m_+^4}{s^{\prime}-m_+^2}-\frac{m_-^4}{s^{\prime}-m_-^2}\right)\operatorname{Im}F^-(s^\prime,0)\ .
\end{align}
By means of Eqs.~\eqref{eq:F-th1},~\eqref{eq:F-th2}, and~\eqref{eq:c-_0}, the expression for the subtraction term $h^-(b)$ is given by
\begin{align}\label{eq.h-_b}
    h^-(b)=&\, \frac{8\pi m_+ a^-_0}{m_+^2-m_-^2}\mathbf{+\frac{16\pi g_{K^* \pi K}^2}{(s_{K^*}-m_+^2)(s_{K^*}-m_-^2)}}\nonumber\\
    & +\frac{1}{\pi} \int_{m_{+}^2}^{\infty} \md s^{\prime} \left[\frac{1}{s^{\prime 2}-2 \Sigma s^{\prime}+s_b u_b-a(s_b+u_b-2 \Sigma)}-\frac{1}{s^{\prime 2}-2 \Sigma s^{\prime}+\Delta^2}\right]\operatorname{Im} F^{-}\!\left(s^{\prime}, 0\right) ,
\end{align}
where $t=0$ has been used.\footnote{Actually, we use the relation $b=s_b u_b-a(s_b+u_b)+a^2$. 
It is important to note that $b$ and $t$ are independent variables in the hyperbolic DR.} 
Finally, one can derive a compact expression for the DR of $F^-$ by involving the $S$-wave scattering lengths as the only parameters,
\begin{align}\label{eq:F-fin}
    \frac{F^-(s_b,t)}{s_b-u_b}=&\,\frac{8\pi m_+ a^-_0}{m_+^2-m_-^2}\mathbf{+\frac{16\pi g_{K^* \pi K}^2}{(s_{K^*}-m_+^2)(s_{K^*}-m_-^2)}-\frac{16\pi g_{K^* \pi K}^2 P_1\left(z_s(s_{K^*},t)\right)}{(s_{K^*}-s_b)(s_{K^*}-u_b)}}\nonumber\\
    &+\frac{t}{2 \pi} \int_{t_\pi}^{\infty} \frac{\mathrm{d} t^{\prime}}{t^{\prime}\!\left(t^{\prime}-t\right)} \operatorname{Im}\left[\frac{G^1\left(t^{\prime}, s_b^{\prime}\right)}{\left(s_b^{\prime}-u_b^{\prime}\right)}\right] \nonumber\\
    &+\frac{1}{\pi} \int_{m_{+}^2}^{\infty} \md s^{\prime} \left[\frac{1}{s^{\prime 2}-2 \Sigma s^{\prime}+s_b u_b-a(s_b+u_b-2 \Sigma)}-\frac{1}{s^{\prime 2}-2 \Sigma s^{\prime}+\Delta^2}\right]\operatorname{Im} F^{-}\!\left(s^{\prime}, 0\right) \nonumber\\
    &+\frac{1}{\pi} \int_{m_{+}^2}^{\infty} \md s^{\prime} \left[\frac{1}{s^{\prime 2}-2 \Sigma s^{\prime}+s_b u_b-a\left(s_b+u_b-2 \Sigma\right)+s^\prime t-at} \right.\nonumber\\
    & \left.-\frac{1}{s^{\prime 2}-2 \Sigma s^{\prime}+s_b u_b-a(s_b+u_b-2\Sigma)}\right]\operatorname{Im} F^{-}\!\left(s^{\prime}, t^\prime_b\right) ,
\end{align}
which is consistent with the result of Ref.~\cite{Pelaez:2020gnd} when the pole terms are omitted.

$F^+$ is more complicated than $F^-$, because the subtraction term is a linear function of $t$. 
First, by setting $t=0$ and using Eqs.~\eqref{eq:F+1} and~\eqref{eq:HDR_F+}, one can follow the derivation of $F^-$ to obtain an explicit expression for the subtraction term 
\begin{align}\label{eq:f+b}
    f^+(b)=&\, 8\pi m_+ a^+_0\mathbf{+16\pi g_{K^* \pi K}^2 \left(\frac{1}{s_{K^*}-m_+^2}+\frac{1}{s_{K^*}-m_-^2} \right)-\frac{16\pi  g_{\sigma \pi\pi}g_{\sigma K\bar{K}}}{\sqrt{3} s_\sigma}} \nonumber\\
    &+\frac{1}{\pi}\int_{m_+^2}^{\infty} \frac{\md s^\prime}{s^{\prime 2}}\!\left(\frac{s_b^2}{s^{\prime}-s}+\frac{u_b^2}{s^{\prime}-u}\right)\operatorname{Im}F^+(s^\prime,0) \nonumber\\
    &-\frac{1}{\pi} \int_{m_+^2}^{\infty}\frac{\md s^\prime}{s^{\prime 2}}\!\left(\frac{m_+^4}{s^{\prime}-m_+^2}+\frac{m_-^4}{s^{\prime}-m_-^2}\right)\operatorname{Im}F^+(s^\prime,0)\ .
\end{align}
To derive the expression for the other linear subtraction term $h^+(b)$, we employ some rapidly convergent sum rules. 
By dividing both sides of Eq.~\eqref{eq:HDR_F+} by $t$ and taking the limit $t \to \infty$, one obtains 
\begin{align}\label{eq:sum1}
    0=&\, h^{+}(b)\mathbf{-32\pi g^2_{K^* \pi K}\frac{s_{K^*}}{(s_{K^*}-a)\lambda_{s_{K^*}}}} -\frac{1}{\sqrt{6} \pi} \int_{t_\pi}^{\infty} \frac{\mathrm{d} t^{\prime}}{t^{\prime 2}} \operatorname{Im} G^0\left(t^{\prime}, s_b^{\prime}\right)\nonumber\\
    & +\frac{1}{\pi} \int_{m_{+}^2}^{\infty} \mathrm{d} s^{\prime}\left[\frac{1}{s^{\prime 2}-2 \Sigma s^{\prime}+s_b u_b-a\left(s_b+u_b-2 \Sigma\right)}\right] \operatorname{Im} F^{+}\!\left(s^{\prime}, t_b^{\prime}\right) .
\end{align}
However, this sum rule for $h^+(b)$ exhibits slow convergence. 
Using the DR of $F^-$ from Eq.~\eqref{eq:F-fin}, we can derive an additional sum rule by taking the limit $t \to \infty$:
\begin{align}
    0= & \frac{8 \pi m_{+} a_0^{-}}{m_{+}^2-m_{-}^2}\mathbf{+\frac{16 \pi g_{K^* \pi K}^2}{\left(s_{K^*}-m_{+}^2\right)\left(s_{K^*}-m_{-}^2\right)}-32 \pi g_{K^* \pi K}^2 \frac{s_{K^*}}{\left(s_{K^*}-a\right) \lambda_{S_{K^*}}}}-\frac{1}{2 \pi} \int_{t_\pi}^{\infty} \frac{\mathrm{d} t^{\prime}}{t^{\prime}} \operatorname{Im}\left[\frac{G^1\left(t^{\prime}, s_b^{\prime}\right)}{\left(s_b^{\prime}-u_b^{\prime}\right)}\right]\nonumber\\
    & -\frac{1}{\pi} \int_{m_{+}^2}^{\infty} \mathrm{d} s^{\prime} \frac{1}{s^{\prime 2}-2 \Sigma s^{\prime}+s_b u_b-a(s_b+u_b-2 \Sigma)} \operatorname{Im} F^{-}\left(s^{\prime}, t_b^{\prime}\right) \nonumber\\
    & +\frac{1}{\pi} \int_{m_{+}^2}^{\infty} \mathrm{d} s^{\prime}\left[\frac{1}{s^{\prime 2}-2 \Sigma s^{\prime}+s_b u_b-a(s_b+u_b-2 \Sigma)}-\frac{1}{s^{\prime 2}-2 \Sigma s^{\prime}+\Delta^2}\right] \operatorname{Im} F^{-}\left(s^{\prime}, 0\right).
\end{align}
Combining these two produces a rapidly converging sum rule~\cite{Buettiker:2003pp},
\begin{align}\label{eq:h+b_fin}
    h^+(b)=&\, \frac{8 \pi m_{+} a_0^{-}}{m_{+}^2-m_{-}^2}\mathbf{+\frac{16\pi g_{K^* \pi K}^2}{(s_{K^*}-m_+^2)(s_{K^*}-m_-^2)}}+\frac{1}{\pi} \int_{t_\pi}^{\infty} \frac{\mathrm{d} t^{\prime}}{t^{\prime}} \operatorname{Im}\left[\frac{ G^0\left(t^{\prime}, s_b^{\prime}\right)}{\sqrt{6} t^\prime}-\frac{G^1\left(t^{\prime}, s_b^{\prime}\right)}{2\left(s_b^{\prime}-u_b^{\prime}\right)}\right] \nonumber\\
    & +\frac{1}{\pi} \int_{m_{+}^2}^{\infty} \mathrm{d} s^{\prime}\left[\frac{1}{s^{\prime 2}-2 \Sigma s^{\prime}+s_b u_b-a(s_b+u_b-2 \Sigma)}-\frac{1}{s^{\prime 2}-2 \Sigma s^{\prime}+\Delta^2}\right] \operatorname{Im} F^{-}\!\left(s^{\prime}, 0\right) \nonumber\\
    & -\frac{1}{\pi} \int_{m_{+}^2}^{\infty} \mathrm{d} s^{\prime}\left[\frac{1}{s^{\prime 2}-2 \Sigma s^{\prime}+s_b u_b-a\left(s_b+u_b-2 \Sigma\right)}\right] \operatorname{Im} \left[F^{+}\!\left(s^{\prime}, t_b^{\prime}\right)+F^{-}\!\left(s^{\prime}, t_b^{\prime}\right)\right] .
\end{align}
Finally, from Eqs.~\eqref{eq:HDR_F+}, \eqref{eq:f+b}, and~\eqref{eq:h+b_fin}, a compact expression for the hyperbolic DR of $F^+$ is obtained with the $S$-wave $\pi K$ scattering lengths $a_0^{\pm}$ as the only parameters,
\begin{align}\label{eq:F+_fin}
    F^+(s_b,t)=&\, 8 \pi m_{+}\!\left(a_0^{+}+t \frac{a_0^{-}}{m_{+}^2-m_{-}^2}\right)+\frac{t}{\pi} \int_{t_\pi}^{\infty} \frac{\mathrm{d} t^{\prime}}{t^{\prime}}\operatorname{Im} \left[\frac{G^0\left(t^{\prime}, s_b^{\prime}\right)}{\sqrt{6}\!\left(t^{\prime}-t\right)}-\frac{G^1\left(t^{\prime}, s_b^{\prime}\right)}{2\left(s_b^{\prime}-u_b^{\prime}\right)}\right]\mathbf{+\frac{16 \pi g_{\sigma \pi \pi} g_{\sigma K \bar{K}}t}{\sqrt{3}s_\sigma(s_\sigma-t)}}\nonumber\\
    & \mathbf{+16\pi g_{K^* \pi K}^2\frac{2s_{K^*}+t-m_+^2-m_-^2}{\left(s_{K^*}-m_+^2\right)\left(s_{K^*}-m_-^2\right)}-16\pi g_{K^* \pi K}^2 P_1\left(z_s(s_{K^*},t)\right)\left(\frac{1}{s_{K^*}-s_b}+\frac{1}{s_{K^*}-u_b}\right)} \nonumber\\
    & +\frac{1}{\pi} \int_{m_{+}^2}^{\infty} \md s^{\prime} \frac{1}{s^{\prime 2}}\left[\frac{s^\prime\left((2\Sigma)^2-2s_b u_b+2a(s_b+u_b-2 \Sigma)\right)-2\Sigma \left(s_b u_b-a(s_b+u_b-2 \Sigma)\right)}{s^{\prime 2}-2\Sigma s^\prime+s_b u_b-a(s_b+u_b-2 \Sigma)} \right.\nonumber\\
    & \left.-\frac{s^\prime\left((2\Sigma)^2-2\Delta^2\right)-2\Sigma \Delta^2}{s^{\prime 2}-2\Sigma s^\prime+\Delta^2}\right]\operatorname{Im} F^{+}\!\left(s^{\prime}, 0\right)\nonumber\\
    & +\frac{t}{\pi} \int_{m_{+}^2}^{\infty} \mathrm{d} s^{\prime}\left[\frac{1}{s^{\prime 2}-2 \Sigma s^{\prime}+s_b u_b-a(s_b+u_b-2 \Sigma)}-\frac{1}{s^{\prime 2}-2 \Sigma s^{\prime}+\Delta^2}\right] \operatorname{Im} F^{-}\!\left(s^{\prime}, 0\right) \nonumber\\
    & -\frac{t}{\pi} \int_{m_{+}^2}^{\infty} \mathrm{d} s^{\prime}\left[\frac{1}{s^{\prime 2}-2 \Sigma s^{\prime}+s_b u_b-a\left(s_b+u_b-2 \Sigma\right)}\right] \operatorname{Im} F^{-}\!\left(s^{\prime}, t_b^{\prime}\right)\nonumber\\
    & +\frac{1}{\pi} \int_{m_{+}^2}^{\infty} \mathrm{d} s^{\prime}\left[\frac{2 s^{\prime}-2 \Sigma+t}{s^{\prime 2}-2\Sigma s^\prime+s_b u_b-a(s_b+u_b-2\Sigma)+s^{\prime}t-at} \right.\nonumber\\
    &\left.-\frac{2 s^{\prime}-2 \Sigma}{s^{\prime 2}-2\Sigma s^\prime+s_b u_b-a(s_b+u_b-2\Sigma)}\right] \operatorname{Im} F^{+}\!\left(s^{\prime}, t_b^{\prime}\right) ,
\end{align}
which reduces to the result in Ref.~\cite{Buettiker:2003pp} as long as $a \to 0$ and all pole terms are omitted.

\subsection{$t$-channel Roy-Steiner-type equations with partial-wave amplitudes}

In order to combine analyticity with unitarity, the RS-type equations have to be cast into a form that permits diagonal $s$-channel partial-wave unitarity relations. 
Therefore, we work in the $s$-channel isospin basis $I_s=1/2,3/2$ instead of the $I_s=\pm$ basis. 
By utilizing the relations~\eqref{eq:s_isospin} and~\eqref{eq:t_isospin}, then performing the partial-wave expansion of Eqs.~\eqref{eq:F-fin} and~\eqref{eq:F+_fin} and projecting onto the $t$-channel partial waves, the complete $t$-channel RS-type equations become
\begin{align}
    g_0^0(t)=&\, \text{ST}^0_0(t)+\text{PT}^0_0(t) \nonumber\\
    &+\frac{t}{\pi} \int_{t_\pi}^{\infty} \frac{\mathrm{d} t^{\prime}}{t^{\prime}} \frac{\operatorname{Im} g_0^0\left(t^{\prime}\right)}{t^{\prime}-t}-\frac{3 \sqrt{6}}{8} \frac{t}{\pi} \int_{t_\pi}^{\infty} \frac{\mathrm{d} t^{\prime}}{t^{\prime}} \operatorname{Im} g_1^1\left(t^{\prime}\right) \nonumber\\
    &+\sum_{J^{\prime}=2}^{\infty}\frac{t}{\pi} \int_{t_\pi}^{\infty} \frac{\mathrm{d} t^{\prime}}{t^{\prime}}G_{0,2 J^{\prime}-2}^0\left(t, t^{\prime}\right) \operatorname{Im} g_{2 J^{\prime}-2}^0\left(t^{\prime}\right)+\sum_{J^{\prime}=2}^{\infty}\frac{t}{\pi} \int_{t_\pi}^{\infty} \frac{\mathrm{d} t^{\prime}}{t^{\prime}}G_{0,2 J^{\prime}-1}^1\left(t, t^{\prime}\right) \operatorname{Im} g_{2 J^{\prime}-1}^1\left(t^{\prime}\right) \nonumber\\
    &+\sum_{J^{\prime}=0}^\infty \frac{1}{\pi} \int_{m_{+}^2}^{\infty} \mathrm{d} s^{\prime}\left[G_{0, J^{\prime}}^{1/2}\!\left(t, s^{\prime}\right) \operatorname{Im} f_{J^{\prime}}^{1/2}\!\left(s^{\prime}\right)+G_{0, J^{\prime}}^{3/2}\!\left(t, s^{\prime}\right) \operatorname{Im} f_{J^{\prime}}^{3/2}\!\left(s^{\prime}\right)\right] ,\label{eq:g00}\\
    g_1^1(t)=&\,\text{ST}^1_1(t)+\text{PT}^1_1(t)+\frac{t}{\pi} \int_{t_\pi}^{\infty} \frac{\mathrm{d} t^{\prime}}{t^{\prime}} \frac{\operatorname{Im} g_1^1\left(t^{\prime}\right)}{\left(t^{\prime}-t\right)}+\sum_{J^{\prime}=2}^{\infty}\frac{t}{\pi} \int_{t_\pi}^{\infty} \frac{\mathrm{d} t^{\prime}}{t^{\prime}} G_{1,2 J^{\prime}-1}^1\left(t, t^{\prime}\right) \operatorname{Im} g_{2 J^{\prime}-1}^1\left(t^{\prime}\right) \nonumber\\
    &+\sum_{J^{\prime}=0}^{\infty}\frac{1}{\pi} \int_{m_{+}^2}^{\infty} \mathrm{d} s^{\prime} \left[G_{1, J^{\prime}}^{1/2}\!\left(t, s^{\prime}\right) \operatorname{Im} f_{J^{\prime}}^{1/2}\!\left(s^{\prime}\right)+G_{1, J^{\prime}}^{3/2}\!\left(t, s^{\prime}\right) \operatorname{Im} f_{J^{\prime}}^{3/2}\!\left(s^{\prime}\right)\right] ,\label{eq:g11}
\end{align}
where $\text{ST}^{I_t}_J$ and $\text{PT}^{I_t}_J$ correspond to the projections of the subtraction and pole terms, respectively. Each partial-wave $g^{I_t}_J$ is coupled to all other higher ($J^\prime>J$) $t$-channel partial waves through the kernel functions $G^{I_t}_{J,J^\prime}$. 
Moreover, the equations involve the $s$-channel partial waves $f^{I_s}_J$ via the kernels $G^{I_s}_{J,J^\prime}$. 
The $s$-channel kernels in the isospin basis are related to the expressions in the $I_s=\pm$ basis,
\begin{align}
    &G^{1/2}_{J,J^\prime}(t,s^\prime)=\frac{1}{3}\!\left(G^{+}_{J,J^\prime}(t,s^\prime)+tG^{-}_{J,J^\prime}(t,s^\prime)\right),\nonumber\\
    &G^{3/2}_{J,J^\prime}(t,s^\prime)=\frac{1}{3}\!\left(2 G^{+}_{J,J^\prime}(t,s^\prime)-tG^{-}_{J,J^\prime}(t,s^\prime)\right),\quad J=0,2,4\cdots\ ,\\
    &G^{1/2}_{J,J^\prime}(t,s^\prime)=\frac{1}{3}G^{-}_{J,J^\prime}(t,s^\prime),\quad G^{3/2}_{J,J^\prime}(t,s^\prime)=-\frac{1}{3}G^{-}_{J,J^\prime}(t,s^\prime),\quad J=1,3,5\cdots\ .
\end{align}
In Appendix~\ref{app:kernel_t}, the derivation and explicit expressions of different kernel functions are discussed following Refs.~\cite{Buettiker:2003pp, Pelaez:2020gnd}, after having corrected several typographical errors (see the Appendix for details) and extended the results therein.

\subsection{$s$-channel Roy-Steiner-type equations with partial-wave amplitudes}

After performing the partial-wave expansion of Eqs.~\eqref{eq:F-fin} and~\eqref{eq:F+_fin} and the subsequent projection onto the $s$-channel partial waves, one obtains the $s$-channel RS-type equations
\begin{align}\label{eq:fIl}
    f_J^{I}(s)= & \text{ST}^{I}_J(s)+\text{PT}^{I}_J(s) \nonumber\\
    & +\frac{1}{\pi} \int_{m_{+}^2}^{\infty} \mathrm{d} s^{\prime} \sum_{J^\prime=0}^{\infty} \left\{K_{J, J^\prime}^{I,1/2}\!\left(s, s^{\prime}\right) \operatorname{Im} f_{J^\prime}^{1/2}\!\left(s^{\prime}\right)+K_{J,J^\prime}^{I,3/2}\!\left(s, s^{\prime}\right) \operatorname{Im} f_{J^\prime}^{3/2}\!\left(s^{\prime}\right)\right\} \nonumber\\
    & +\frac{1}{\pi} \int_{t_\pi}^{\infty} \mathrm{d} t^{\prime} \sum_{J^\prime=0}^{\infty} \left\{K_{J,2 J^\prime}^{I,0}\!\left(s, t^{\prime}\right) \operatorname{Im} g_{2 J^\prime}^0\left(t^{\prime}\right)+K_{J,2 J^\prime+1}^{I,1}\!\left(s, t^{\prime}\right) \operatorname{Im} g_{2 J^\prime+1}^1\left(t^{\prime}\right)\right\}\ .
\end{align}
This work focuses on three channels at low energies, namely $(I,J)=\left(\frac{1}{2},0\right), \left(\frac{1}{2},1\right)$, and $\left(\frac{3}{2},0\right)$. 
The corresponding subtraction terms can be written as
\begin{align}
    \text{ST}^{1/2}_0(s) &=\text{ST}^+_0(s)+2\text{ST}^-_0(s)=\frac{1}{2} m_{+} a_0^{1/2}+\frac{1}{12} m_{+}\!\left(a_0^{1/2}-a_0^{3/2}\right)\frac{\left(s-m_{+}^2\right)\left(5 s+3 m_{-}^2\right)}{\left(m_{+}^2-m_{-}^2\right) s}\ ,\nonumber\\
    \text{ST}^{1/2}_1(s) &=\text{ST}^+_1(s)+2\text{ST}^-_1(s),\quad \text{ST}^{3/2}_J(s)=\text{ST}^+_J(s)-\text{ST}^-_J(s)\ ,
\end{align}
where $\text{ST}^{1/2}_0(s)$ is consistent with the result of Ref.~\cite{Descotes-Genon:2006sdr}. 
The pole terms read
\begin{align}
    \text{PT}^{1/2}_J(s)=\text{PT}^+_J(s)+2\text{PT}^-_J(s),\quad \text{PT}^{3/2}_J(s)=\text{PT}^+_J(s)-\text{PT}^-_J(s)\ .
\end{align}
The kernels can be written as
\begin{align}
    K^{\frac{1}{2},0}_{J,J^\prime}(s, t^{\prime})=&\, K^{+,0}_{J,J^\prime}(s, t^{\prime})\ ,\quad K^{\frac{3}{2},0}_{J,J^\prime}(s, t^{\prime})=K^{+,0}_{J,J^\prime}(s, t^{\prime})\ ,\\
    K^{\frac{1}{2},1}_{J,J^\prime}(s, t^{\prime})=&\, K^{+,1}_{J,J^\prime}(s, t^{\prime})+2 K^{-,1}_{J,J^\prime}(s, t^{\prime})\ ,\quad K^{\frac{3}{2},1}_{J,J^\prime}(s, t^{\prime})=K^{+,1}_{J,J^\prime}(s, t^{\prime})-K^{-,1}_{J,J^\prime}(s, t^{\prime})\ ,\\
    K^{\frac{1}{2},\frac{1}{2}}_{J,J^\prime}(s, s^{\prime})=&\, \frac{1}{3}\!\left(K^{+,+}_{J,J^\prime}(s, s^{\prime})+K^{+,-}_{J,J^\prime}(s)+2 K^{-,-}_{J,J^\prime}(s, s^{\prime})\right) ,\\
    K^{\frac{3}{2},\frac{1}{2}}_{J,J^\prime}(s, s^{\prime})=&\, \frac{1}{3}\!\left(K^{+,+}_{J,J^\prime}(s, s^{\prime})+K^{+,-}_{J,J^\prime}(s, s^{\prime})-K^{-,-}_{J,J^\prime}(s, s^{\prime})\right) ,\\
    K^{\frac{1}{2},\frac{3}{2}}_{J,J^\prime}(s, s^{\prime})=&\, \frac{1}{3}\!\left(2 K^{+,+}_{J,J^\prime}(s, s^{\prime})-K^{+,-}_{J,J^\prime}(s, s^{\prime})-2 K^{-,-}_{J,J^\prime}(s, s^{\prime})\right) ,\\
    K^{\frac{3}{2},\frac{3}{2}}_{J,J^\prime}(s, s^{\prime})=&\, \frac{1}{3}\!\left(2 K^{+,+}_{J,J^\prime}(s, s^{\prime})-K^{+,-}_{J,J^\prime}(s, s^{\prime})+K^{-,-}_{J,J^\prime}(s, s^{\prime})\right) .
\end{align}
The detailed derivation and explicit expressions for the $s$-channel kernel functions, with a nonzero $a$, are presented in Appendix~\ref{app:kernel_s}.

\subsection{Singularity structure of the partial-wave amplitudes}\label{subsec:singularity}

\begin{figure}[t]
    \centering
    \includegraphics[width=1\textwidth,angle=-0]{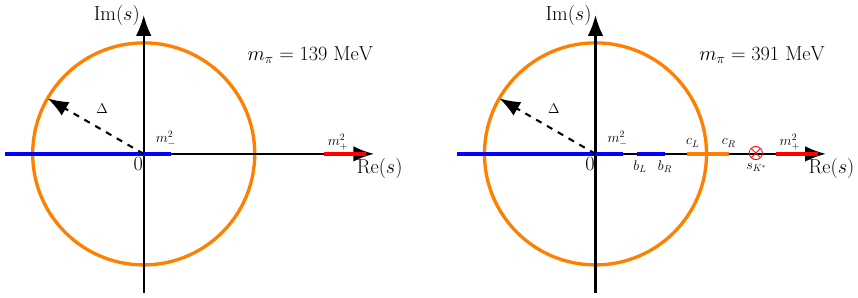}
    \caption{The cut structure of the $\pi K$ partial-wave amplitudes in the complex $s$ plane for the physical $m_\pi=139~$MeV case (left) and the unphysical $m_\pi=391~$MeV case (right). The cuts located in the intervals of $[b_L,b_R]$ and $[c_L,c_R]$ in the right panel are caused by the BS states $K^*$ and $\sigma$ at large pion masses, respectively.}\label{fig:singularity}
\end{figure}

At the physical pion mass, i.e., $m_\pi=139~$MeV, the singularity structure of the $\pi K$ partial-wave amplitudes has been systematically studied; see Ref.~\cite{Lang:1978fk} and references therein.
For the sake of self-consistency, we enumerate the related cut singularities as shown in the left panel of Fig.~\ref{fig:singularity}:
\begin{itemize}
    \item[a)] the physical cut (right-hand cut) $s \geq m_+^2$;
    \item[b)] the crossed physical cut or LHC $u \geq m_+^2$, i.e. $s \leq m_-^2$;
    \item[c)] the LHC from $t$-channel unitarity ($t \geq t_\pi$) leads to
    \begin{itemize}
        \item[$\text{I}$)] a circular cut with radius $\Delta=m_K^2-m_\pi^2$ corresponding to $t \in\left[t_\pi, t_K\right]$;
        \item[$\text{II}$)] a LHC for $s \leq 0$ corresponding to $t \geq t_K$;
    \end{itemize}
    \item[d)] an easily overlooked singularity of the type $|s|^{-\text{const}}$ at $s=0$ depending on the high energy behaviour in the $t$- and $u$-channels.
\end{itemize}

As the masses of the $u$ and $d$ quarks are gradually increased (while keeping the $s$ quark mass fixed) until $m_\pi \simeq 391~$MeV (and $m_K\simeq 549~$MeV), the scattering amplitude exhibits two new features. 
First, the $I=1/2$ $P$-wave $\pi K$ scattering shows the emergence of a shallow BS, $K^*(892)$. 
Second, in the $t$-channel $\pi\pi \to K \bar{K}$ process, the $I=0$ $S$-wave develops a BS pole of $\sigma$. These two BSs introduce new LHCs. 
For general methods of computing the branch points of the LHC, we refer to~\cite{Kennedy:1962ovz}; the results are presented in the right panel of Fig.~\ref{fig:singularity}:
\begin{itemize}
    \item[e)] The $s$-channel vector BS $K^*(892)$ first introduces a BS pole below the threshold in $f^{1/2}_1(s)$. 
    Additionally, crossing symmetry implies that the exchange of a $u$-channel vector BS will introduce a short cut on the real axis, with branch points located at
    \begin{align}
        b_L=\frac{\Delta^2}{m_{K^*}^2}\ ,\quad b_R=2\Sigma-m_{K^*}^2\ ,
    \end{align}
    which can be easily derived from, i.e., $B(s,s_{K^*})\propto \ln \left(s_{K^*}+s-2 \Sigma\right)-\ln \left(s_{K^*}-\frac{\Delta^2}{s}\right)$ in Eqs.~\eqref{eq:PT-0} and~\eqref{eq:PT-1}.
    This cut lies entirely within the domain surrounded by the circular cut and is close to the pseudo-threshold $s=m_-^2$.
    \item[f)] For the $t$-channel $\sigma$ BS pole, the partial-wave projection onto the $s$-channel introduces a somewhat subtle cut, with branch points located at
    \begin{align}
        c_{L,R}=\frac{1}{2}\!\left(2\Sigma-s_\sigma\mp\sqrt{(t_K-s_\sigma)(t_\pi-s_\sigma)}\right) .
    \end{align}
    This cut actually consists of a short cut that crosses the circular cut and an additional circular cut.\footnote{Clearly, this circular cut arises from a different physical cause than the previously mentioned circular cut corresponding to $t \in\left[t_\pi, t_K\right]$, but they coincide in position. 
    Hence, we refer to both as the circular cut.} 
    A mathematical understanding can be obtained by analyzing the cut of the expression $\ln \left(1+\frac{\lambda_s}{s_\sigma s}\right)$ in Eqs.~\eqref{eq:PT+0} and~\eqref{eq:PT+1}. 
    Let us denote $s=x+iy$, then the cut $1+\frac{\lambda_s}{s_\sigma s}\leq 0$ can be written as
    \begin{align} \label{eq:xy}
        \frac{x \Delta^2}{x^2+y^2}+x+s_\sigma-2\Sigma\leq 0\ ,\quad y\left(1-\frac{\Delta^2}{x^2+y^2}\right)=0\ .
    \end{align}
    There exist two solutions, with one real solution corresponding to
    \begin{align}
        y=0\ ,\quad x\in\left[c_L, c_R\right] ,
    \end{align}
    and the other complex solution associated with a circle, 
    \begin{align}
        x^2+y^2=\Delta^2\ ,
    \end{align}
    and the inequality in Eq.~\eqref{eq:xy}, $x\leq 2\Delta +(4m_\pi^2-s_\sigma)$, is trivially satisfied along the circle when the $\sigma$ is a BS pole.
\end{itemize}

To provide an intuitive understanding, we show the relevant branch points below in ascending order (all in units of MeV):
\begin{align} 
    &m_- = 158 \ ,\quad \sqrt{b_L} = 159 \ ,\quad \sqrt{b_R} = 190 \ ,\quad \sqrt{c_L} = 303 \ ,\nonumber\\
    &\sqrt{\Delta} = 385 \ ,\quad \sqrt{c_R} = 491 \ ,\quad \sqrt{s_{K^*}} = 934\ ,\quad m_+ = 940 \ , 
    \label{eq:values}
\end{align}
where the value of $\sqrt{s_{K^*}}$ is taken from Ref.~\cite{Wilson:2019wfr}, and $\sqrt{s_\sigma}=759~$MeV~\cite{Cao:2023ntr} has been used. 
Although these structures are somewhat distant from the physical region, they have a significant impact on the location of the $\kappa$ pole and the Adler zero. 
In particular, the study in Ref.~\cite{Dudek:2014qha} used various $K$-matrix models for analytic continuation and found the $\kappa$ to be a VS pole, which is approximately located in the range $600 \sim 750$~MeV. 
Although the pole lies deep below the physical threshold, it is positioned at similar distances from both the unitarity cut and the LHCs (particularly the points $s=c_R$ and $s=\Delta$). 
Consequently, the $\kappa$ pole position is sensitive to the various LHC singularities, and a reliable and precise determination of the $\kappa$ properties in this case can only be achieved by properly accounting for all LHC contributions. The RS-type equations with strict crossing symmetry discussed above provide a valuable tool for this purpose.
%%%%%%%%%%%%%%%%%%%%%%%%%%%%%%%%%%%%%%%%%%%%%%%%%%%%%%%%%%%%%%%%%%%%%%%%%%%%

\section{Validity domain of the $s$-channel partial-wave hyperbolic dispersion relations}\label{sec:VD}

Fixed-$t$ and hyperbolic DRs are valid only within a finite region as constrained by axioms (microcausality, unitarity and crossing symmetry) in quantum field theory~\cite{Martin:1969ina, Sommer:1970mr}. 
Within the validity domain, the scattering amplitude is analytic and the partial-wave expansion converges. 
Lehmann was the first to show that this analytic domain is bounded by an ellipse~\cite{Lehmann:1958ita}, and Martin subsequently enlarged this domain by incorporating unitarity constraints and crossing symmetry~\cite{Martin:1965jj,Martin:1966zsy}.
The enlarged domain is bounded by an ellipse called the Lehmann-Martin (LM) ellipse.

In this section, we describe how to derive the complex validity domain of hyperbolic DRs. 
The derivation of the real validity domain for partial-wave hyperbolic DRs, particularly the optimization of the parameter $a=0$ to maximize this domain, has been thoroughly investigated in Refs.~\cite{Ditsche:2012fv, Pelaez:2020gnd}.
However, the complex validity domain for general values of $a$ remains poorly explored.
Only a few works~\cite{ Descotes-Genon:2006sdr, Hoferichter:2011wk, Hoferichter:2012pm, Pelaez:2020gnd, Cao:2022zhn, Hoferichter:2023mgy} have discussed relevant derivations, but they provide only specific examples accompanied with certain deficiencies (to be discussed below).  
In fact, a comprehensive framework for calculating the complex validity domain for arbitrary parameters $a$ has not yet been established. 
In the following, we present such a framework and provide two examples to illustrate this issue.

\subsection{Generic approach}

We now turn to describing the LM ellipse~\cite{Lehmann:1958ita, Martin:1965jj, Martin:1966zsy} and the Mandelstam double spectral region~\cite{Mandelstam:1958xc}. 
These frameworks have been recently reviewed in Refs.~\cite{Ditsche:2012fv, Pelaez:2020gnd}, but for completeness, we briefly recapitulate the setup here. 

The partial-wave expansions of $\operatorname{Im}F^\pm(s,t)$ in terms of Legendre polynomials $P_J(z)$, with $z=1+\frac{2st}{\lambda_s}$ or $z=\frac{s-u}{4q_\pi q_K}$, converge in the complex $z$ plane within the LM ellipse
\begin{align}
    \frac{(\operatorname{Re} z)^2}{A^2}+\frac{(\operatorname{Im} z)^2}{B^2}=1\ ,
\end{align}
with foci at $z= \pm 1$. 
Without loss of generality, we assume $A^2-B^2=1$, which implies that the semi-major axis $A$ (along the real axis) exceeds the semi-minor axis $B$ (along the imaginary axis). 
The size of this convergence ellipse is determined by the maximum real value $z^{\max}=A$ at which it first encounters a singularity of $\operatorname{Im}F(s, t)$, i.e., when it reaches the Mandelstam double spectral region where the double spectral function~\cite{Mandelstam:1958xc} $\rho_{st}$ (or $\rho_{su}$, $\rho_{tu}$) is non-zero.\footnote{In the double spectral region, the considered channel and its crossed channel are simultaneously in the physical region. For instance, in the case of $\rho_{st}$, both $s$ and $t$ channels are simultaneously in the physical region.}  
For unequal-mass elastic scattering such as $\pi K$ scattering, the hyperbolic DRs~\eqref{eq:F-fin} and~\eqref{eq:F+_fin} involve both the $s$-channel amplitudes $F^\pm(s^\prime,t^\prime)$ and the $t$-channel amplitudes $G^{0,1}(t^\prime,s^\prime)$ simultaneously. 
To obtain RS-type equations, we need to expand $\operatorname{Im} F^{+}\!\left(s^{\prime}, t^{\prime}\right)$ in terms of $z_{s^{\prime}}$, i.e., the partial-wave expansion in the $s$-channel, and expand $\operatorname{Im} G^0\left(t^{\prime}, s^{\prime}\right)$ in terms of $z_{t^{\prime}}$, i.e., the partial-wave expansion in the $t$-channel. 
We now have two LM ellipses: one with respect to $z_{s^{\prime}}$ and the other with respect to $z_{t^{\prime}}$.

\begin{figure}[t]
    \centering
    \includegraphics[width=.5\textwidth,angle=-0]{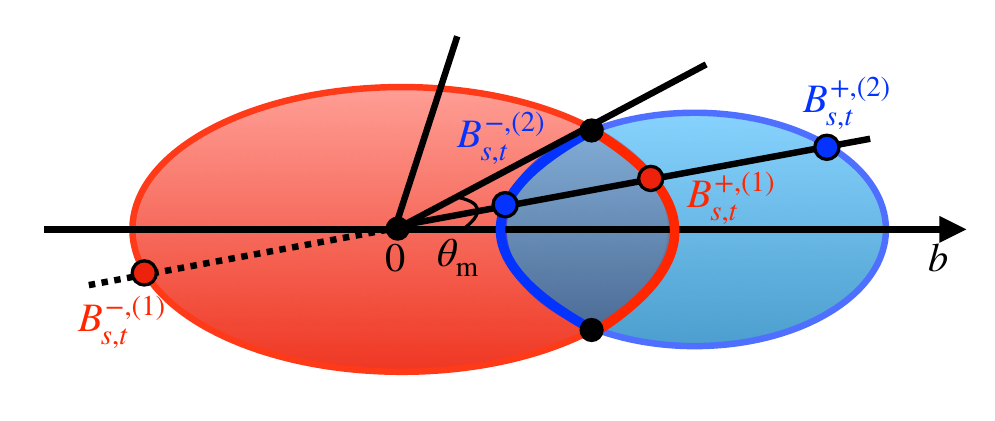}    
    \caption{Two LM ellipses: the red one encompasses the origin, while the blue one does not. 
    Lines extending from the origin at different angles intersect the two ellipses at different points. 
    The superscripts $(1)$ and $(2)$ correspond to the intersection points of these lines with the red and blue ellipses, respectively. The straight line connecting $B_{s,t}^{-,(1)}$ and $B_{s,t}^{+,(2)}$ also intersects the red ellipse at $B_{s,t}^{+,(1)}$ and the blue ellipse at $B_{s,t}^{-,(2)}$, where the $\pm$ superscripts indicate the two solutions of Eq.\eqref{eq:ellipsev1}.} \label{fig:ellipse_overlap}
\end{figure}

First, we discuss the partial-wave expansion corresponding to the internal integral variable $s^\prime$ in the hyperbolic DRs. 
The convergence region of the partial-wave expansion in the $s$-channel is an ellipse corresponding to $z_{s^{\prime}}$,
\begin{align}\label{eq:ellipse_s_v1}
\frac{\left(\operatorname{Re} z_{s^{\prime}}\right)^2}{A_{s^{\prime}}^2}+\frac{\left(\operatorname{Im} z_{s^{\prime}}\right)^2}{B_{s^{\prime}}^2}=1\ ,
\end{align}
where the foci are located at $z_{s^{\prime}}= \pm 1$, and $A_{s^{\prime}}^2-B_{s^{\prime}}^2=1$. 
For a fixed $s^{\prime}$, the maximum value of $z_{s^{\prime}}$ on the real axis, i.e., the semi-major axis $A_{s^{\prime}}$, should be at the boundary of the corresponding double spectral region at $t^{\prime}=T_{st}\!\left(s^{\prime}\right)$, i.e.,
\begin{align}
    z_{s^{\prime}}^{\max } = A_{s^{\prime}} =1+\frac{2 s^{\prime} T_{st}\!\left(s^{\prime}\right)}{\lambda_{s^{\prime}}}, \quad \forall s^{\prime} \geq m_{+}^2\ .
\end{align}
For a fixed $s^{\prime}$, the convergence ellipse of $z_{s^{\prime}}$ determines the LM ellipse of $t^{\prime}$ since $t^{\prime}$ is a linear function of $z_{s^{\prime}}$. 
According to the linear relation between $t'$ and $b$, viz. $b=\left(s^{\prime}-a\right)(2\Sigma-s^{\prime}-t^{\prime}-a)$, it further determines the LM ellipse of $b$. 
The foci of this ellipse of $b$ are at $\left(s^{\prime}-a\right)\left(2\Sigma-s^{\prime}-a\right)$ and $\left(s^{\prime}-a\right)\left(2\Sigma-s^{\prime}+\lambda_{s^{\prime}} / s^{\prime}-a\right)$. 
To begin with, let us assume that the origin $b=0$ lies within this ellipse of $b$. 
Consequently, the boundary of the ellipse can be conveniently represented in polar coordinates as $\left(B_s\!\left(s^{\prime}, \theta\right), \theta\right)$ of the complex $b$ plane. 
According to Eq.~\eqref{eq:ellipse_s_v1}, $B_s\!\left(s^{\prime}, \theta\right)$ must satisfy
\begin{align}\label{eq:ellipsev1}
    \frac{\left[1+\frac{2 s^{\prime}}{\lambda_{s^{\prime}}}\!\left(2\Sigma-s^{\prime}-a-\frac{B_s\!\left(s^{\prime}, \theta\right) \cos \theta}{s^{\prime}-a}\right)\right]^2}{A_{s^{\prime}}^2}+\frac{\left[\frac{2 s^{\prime}}{\lambda_{s^{\prime}}}\!\left(\frac{B_s\!\left(s^{\prime}, \theta\right) \sin \theta}{s^{\prime}-a}\right)\right]^2}{B_{s^{\prime}}^2}=1\ .
\end{align}
By solving this quadratic equation for the radial coordinate $B_s$, we obtain two roots. 
When $\theta$ is set to 0, this yields two solutions,
\begin{align}
B_s^{-}\!\left(s^{\prime},0\right) &=\left(s^{\prime}-a\right)\left(2\Sigma-s^{\prime}-T_{st}\!\left(s^{\prime}\right)-a\right) ,\nonumber\\
B_s^{+}\!\left(s^{\prime},0\right) &=\left(s^{\prime}-a\right)\left(2\Sigma-s^{\prime}+\frac{\lambda_{s^{\prime}}}{s^{\prime}}+T_{st}\!\left(s^{\prime}\right)-a\right) .
\end{align}
The root corresponding to $B_s^{+}$ represents the maximum allowable value of $b$ on the real axis. Furthermore, setting $\theta$ to $\pi$ yields,
\begin{align}
B_s^{-}\!\left(s^{\prime}, \pi\right) &=-\left(s^{\prime}-a\right)\left(2\Sigma-s^{\prime}+\frac{\lambda_{s^{\prime}}}{s^{\prime}}+T_{st}\!\left(s^{\prime}\right)-a\right) ,\nonumber\\
B_s^{+}\!\left(s^{\prime}, \pi\right) &=-\left(s^{\prime}-a\right)\left(2\Sigma-s^{\prime}-T_{st}\!\left(s^{\prime}\right)-a\right) .
\end{align}
One can readily observe that $B_s^{\pm}(s',0)$ and $B_s^{\mp}(s',\pi)$ are essentially the same functions, differing only by a phase difference of $\pi$, which implies that only one of the solutions is independent. 
The physical condition requires the modulus to be greater than 0, so it suffices to consider $B_s^+$ alone, corresponding to the red ellipse in Fig.~\ref{fig:ellipse_overlap}. 
Unfortunately, the ellipse of $b$ does not always encompass the origin $b=0$, necessitating modifications to the preceding discussion. 
A naive approach is to constrain the parameter $a$ to ensure that the origin is always within the ellipse~\cite{Cao:2022zhn}. 
As noted in Ref.~\cite{Hoferichter:2023mgy}, this restriction appears to be unnecessary, despite the lack of a detailed explanation. 
In the following, we present a thorough description.

As the angle $\theta$ changes and $s^\prime$ takes values in the interval $\left[m_{+}^2, \infty\right)$, there exist infinite LM ellipses. 
Our goal is to obtain the overlap of these LM ellipses. 
It is noted that there is a special class of ellipses which do not encompass the origin. 
Consequently, the overlap is analogous to the one illustrated in Fig.~\ref{fig:ellipse_overlap}. 
Clearly, in this case, the angle $\theta$ has a bound, and the maximum angle $\theta_\mathrm{m}$ can be obtained by setting the discriminant of Eq.~\eqref{eq:ellipsev1} to zero, i.e., when $B_s^+=B_s^-$. 
The boundary of the overlap is given by the following conditions (see Fig.~\ref{fig:ellipse_overlap} for illustration of $B_s^\pm$), 
\begin{align}\label{eq:Bst}
    B_s(\theta) & = B_s^{(1)}(\theta) \cup B_s^{(2)}(\theta),\nonumber\\
    B_s^{(1)}(\theta) & =\min _{s^{\prime} \geq m_{+}^2} B_s^{+}\!\left(s^{\prime}, \theta\right) ,\quad \theta\in [-\theta_\mathrm{m},\theta_\mathrm{m}]\ ,\nonumber\\
    B_s^{(2)}(\theta) & =\max _{s^{\prime} \geq m_{+}^2,\ \left(s^{\prime}-a\right)\left(2\Sigma-s^{\prime}-T_{st}\!\left(s^{\prime}\right)-a\right)\geq 0} B_s^{-}\!\left(s^{\prime}, \theta\right) ,\quad \theta\in [-\theta_\mathrm{m},\theta_\mathrm{m}]\ ,
\end{align}
where the condition $\ \left(s^{\prime}-a\right)\left(2\Sigma-s^{\prime}-T_{st}\!\left(s^{\prime}\right)-a\right)\geq 0$ ensures that the ellipses do not encompass the origin. 
Then, only if all the ellipses encompass the origin ($\ \left(s^{\prime}-a\right)\left(2\Sigma-s^{\prime}-T_{st}\!\left(s^{\prime}\right)-a\right)<0$), we obtain
\begin{align}
    B_s(\theta)=\min _{s^{\prime} \geq m_{+}^2} B_s^{+}\!\left(s^{\prime}, \theta\right) ,\quad \theta\in (-\pi,\pi]\ ,
\end{align}
which is consistent with the condition from Ref.~\cite{Pelaez:2020gnd}. 

Next, we analyze the partial-wave expansion for the integral variable $t^\prime$ in the hyperbolic DRs. 
The convergence region is within an ellipse for $z_{t^{\prime}}$,
\begin{align}\label{eq:ellipsev2}
\frac{\left(\operatorname{Re} z_{t^{\prime}}\right)^2}{A_{t^{\prime}}^2}+\frac{\left(\operatorname{Im} z_{t^{\prime}}\right)^2}{B_{t^{\prime}}^2}=1\ ,
\end{align}
with foci at $z_{t^{\prime}}= \pm 1$, and $A_{t^{\prime}}^2 - B_{t^{\prime}}^2 = 1$. 
According to the definition, $z_{t^{\prime}} = \frac{\nu^{\prime}}{4 q_\pi\left(t^{\prime}\right) q_K\left(t^{\prime}\right)}$ and $\left(s^{\prime} - a\right)\left(u^{\prime} - a\right) = b, s^{\prime} + u^{\prime} + t^{\prime} = \Sigma$, we obtain $\nu^{\prime 2} = \left(t^{\prime} - 2\Sigma + 2 a\right)^2 - 4 b$. 
Consequently, we find that $z_{t^{\prime}}^2$ and $b$ satisfy a linear relation: $z_{t^{\prime}}^2 = \frac{\left(t^{\prime} - 2\Sigma + 2 a\right)^2 - 4 b}{16 q_\pi\left(t^{\prime}\right)^2 q_K\left(t^{\prime}\right)^2}$~\cite{Ditsche:2012fv}. 
Based on Eq.~\eqref{eq:ellipsev2}, we can derive that $z_{t^{\prime}}^2$ must also satisfy a new elliptical equation,
\begin{align}
\frac{\left(\operatorname{Re} z_{t^{\prime}}^2 - \frac{1}{2}\right)^2}{\hat{A}_{t^{\prime}}^2}+\frac{\left(\operatorname{Im} z_{t^{\prime}}^2\right)^2}{\hat{B}_{t^{\prime}}^2}=1\ ,
\end{align}
where $\hat{A}_{t^{\prime}} = \left(A_{t^{\prime}}^2 + B_{t^{\prime}}^2\right) / 2 = A_{t^{\prime}}^2 - 1 / 2$ and $\hat{B}_{t^{\prime}} = A_{t^{\prime}} B_{t^{\prime}} = A_{t^{\prime}} \sqrt{A_{t^{\prime}}^2 - 1}$. 
The center of the new ellipse is at $(1 / 2, 0)$, with two foci at $1 / 2 \pm 1 / 2$. 
We first express the boundary of the double spectral function $\rho_{st}$ using $\nu_{st}(t)$~\cite{Ditsche:2012fv, Pelaez:2020gnd}. 
The final boundary is determined by the function $N_{st}(t) \equiv \min \nu_{st}(t)$. 
When $t^{\prime} > t_K$, for a fixed $t^{\prime}$, the maximum value of $z_{t^{\prime}}$ on the real axis is
\begin{align}
z_{t^{\prime}}^{\max} = A_{t^{\prime}} = \frac{\nu_{\max}^{\prime}}{4q_\pi\left(t^{\prime}\right) q_K\left(t^{\prime}\right)} = \frac{N_{st}\!\left(t^{\prime}\right)}{4q_\pi\left(t^{\prime}\right) q_K\left(t^{\prime}\right)}, \quad \forall t^{\prime} > t_K\ ,
\end{align}
whereas when $t_\pi < t^{\prime} < t_K$, $z_{t^{\prime}} = \frac{\nu^{\prime}}{4q_\pi\left(t^{\prime}\right) q_K\left(t^{\prime}\right)}$ is purely imaginary, because $q_K\left(t^{\prime}\right) = \sqrt{t^{\prime} - t_K} / 2 = i \sqrt{t_K - t^{\prime}} / 2 = i q_K^{-}\!\left(t^{\prime}\right)$. 
In this case, we can determine the maximum value of $z_{t^{\prime}}$ on the imaginary axis,
\begin{align}
\operatorname{Im} z_{t^{\prime}}^{\max} = B_{t^{\prime}} = \frac{N_{st}\!\left(t^{\prime}\right)}{4q_\pi\left(t^{\prime}\right) q_K^{-}\!\left(t^{\prime}\right)}, \quad t_\pi < t^{\prime} < t_K\ .
\end{align}
Once we have obtained the semi-major or semi-minor axis of the LM ellipse corresponding to $z_{t^{\prime}}$, we can further determine the semi-major axis $\hat{A}_{t^{\prime}}$ of the LM ellipse for $z_{t^{\prime}}^2$. 
Given the linear relation between $z_{t^{\prime}}^2$ and $b$, we can conclude that the convergence region for $b$ is also an ellipse. 
Similarly, expressing its boundary in polar coordinates as $\left(B_t\!\left(t^{\prime}, \theta\right), \theta\right)$ and using Eq.~\eqref{eq:ellipsev2}, $B_t\!\left(t^{\prime}, \theta\right)$ should satisfy
\begin{align}
\frac{\left(\frac{\left(t^{\prime}-\Sigma+2 a\right)^2-4 B_t\!\left(t^{\prime}, \theta\right) \cos \theta}{16 q_\pi\left(t^{\prime}\right)^2 q_K\left(t^{\prime}\right)^2}-\frac{1}{2}\right)^2}{\hat{A}_{t^{\prime}}^2}+\frac{\left(\frac{4 B_t\!\left(t^{\prime}, \theta\right) \sin \theta}{16 q_\pi\left(t^{\prime}\right)^2 q_K\left(t^{\prime}\right)^2}\right)^2}{\hat{B}_{t^{\prime}}^2}=1\,. 
\end{align}
The quadratic equation for $B_t$ yields two roots. 
Setting $\theta$ to 0 in the root expressions yields
\begin{align}
& B_t^{-}\!\left(t^{\prime}, 0\right)=\frac{\left(t^{\prime}-2 \Sigma+2 a\right)^2-N_{s t}\!\left(t^{\prime}\right)^2}{4}\ , \nonumber\\
& B_t^{+}\!\left(t^{\prime}, 0\right)=\frac{\left(t^{\prime}-2 \Sigma+2 a\right)^2-16\left[q_\pi\left(t^{\prime}\right) q_K\left(t^{\prime}\right)\right]^2+N_{s t}\!\left(t^{\prime}\right)^2}{4}\ ,
\end{align}
where the root corresponding to $B_t^{+}$ is the maximum allowable value of $b$ on the real axis. 
Analogous to the restrictions~\eqref{eq:Bst}, the boundaries of the overlap can be given by
\begin{align}
    B_t(\theta) & = B_t^{(1)}(\theta) \cup B_t^{(2)}(\theta),\nonumber\\
    B_t^{(1)}(\theta) & =\min _{t^{\prime} \geq t_\pi} B_t^{+}\!\left(s^{\prime}, \theta\right) ,\quad \theta\in [-\theta_\mathrm{m}^\prime,\theta_\mathrm{m}^\prime]\ ,\\
    B_t^{(2)}(\theta) & =\max _{t^{\prime} \geq t_\pi, \left(t^{\prime}-2 \Sigma+2 a\right)^2-N_{s t}\!\left(t^{\prime}\right)^2\geq 0} B_t^{-}\!\left(s^{\prime}, \theta\right) ,\quad \theta\in [-\theta_\mathrm{m}^\prime,\theta_\mathrm{m}^\prime]\ .
\end{align}

The LM ellipses for the internal variables $s^{\prime}$ and $t^{\prime}$ jointly constrain the allowed range of the parameter $b$ in the DRs. 
Next, we proceed with the partial-wave expansion by integrating over the variable $t$, which can be transformed into an integration over $b$, $\int_{(s-a)\left(\Delta^2 / s-a\right)}^{(s-a)(2\Sigma-s-a)} \mathrm{d} b(\cdots)$. 
The integration contour for $b$ must lie entirely within the convergence domain.
This condition determines the validity boundary of the partial-wave amplitude in the complex $s$ plane:
\begin{align}
(s-a)(2\Sigma-s-a)-B_{s, t}(\theta) \exp (i \theta) & =0\ , \label{eq:boundaryv1}\\
(s-a)\left(\Delta^2 / s-a\right)-B_{s, t}(\theta) \exp (i \theta) & =0\ ,\label{eq:boundaryv2}
\end{align}
where the constraint in Eq.~\eqref{eq:boundaryv1} is generally the stronger one. 
Although the constraint~\eqref{eq:boundaryv2} is generally weaker, it is subtle and requires special attention when $a \neq 0$. 
When $a \neq 0$, the expression $(s-a)\left(\Delta^2 / s-a\right)$ diverges as $s \rightarrow 0$. 
Although this constraint is weaker than Eq.~\eqref{eq:boundaryv1} in most cases, it becomes crucial near $s=0$. 
In other words, to ensure that the integration limits for $b$ lie within the convergence region of $b$, the vicinity of $s=0$ must be excluded from the validity region of the RS-type equations through the boundary constraint~\eqref{eq:boundaryv2}. 
Specifically, for the case $a=0$, the expression $(s-a)\left(\Delta^2 / s-a\right)$ simplifies to $\Delta^2$, making it independent of $s$ and leaving only condition~\eqref{eq:boundaryv1} as the relevant constraint.
From a mathematical perspective, for the RS-type equation, the partial-wave amplitude can be safely continued to the neighborhood of $s=0$ only if the hyperbolic parameter $a=0$. 
Previous works~\cite{Pelaez:2020gnd, Hoferichter:2023mgy} with the situation when $a\neq 0$ did not take this factor into consideration. 
This is also why, in what follows, we prefer to solve the RS-type equation with $a=0$.

The above results only cover the validity domain of the double spectral function $\rho_{st}$. 
The method is also useful for calculating other regions, including those corresponding to $\rho_{su}$. 
Moreover, the methodology presented here for $\pi K$ scattering can be directly extended to other hadronic systems, such as $\pi N$ scattering. 
In the following, we use two specific physical examples to clarify some deficiencies in the previous works~\cite{Pelaez:2020gnd, Hoferichter:2023mgy}.

\subsection{Examples for $\pi K$ and $\pi N$ scattering with physical masses}

\begin{figure}[t]
    \centering
    \includegraphics[width=1\textwidth,angle=-0]{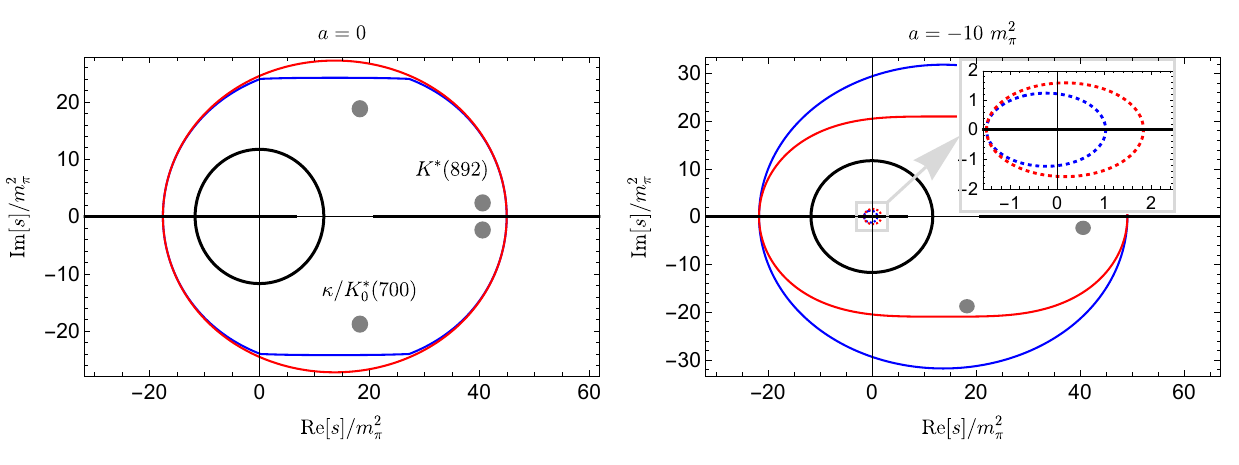}    
    \caption{Complex validity domain for the $s$-channel RS-type equations with physical meson masses. 
    The red and blue lines denote the constraints from the $s$- and $t$-channel LM ellipses associated with $\rho_{st}$, and the black lines the various cuts of the partial-wave amplitudes. 
    The left panel corresponds to the $a=0$ case, while the right panel corresponds to the $a=-10m_\pi^2$ case, where the dashed lines are derived from Eq.~\eqref{eq:boundaryv2}.
    The dashed ellipses encircle the $s=0$ point, and the region enclosed by these ellipses must be excluded from the complex validity domain.
    The gray dots indicate the positions of known resonance poles.} \label{fig:piK_139}
\end{figure}

The Mandelstam double spectral region for physical $\pi K$ scattering is presented in the seminal work~\cite{Descotes-Genon:2006sdr}. 
Using the results from Ref.~\cite{Descotes-Genon:2006sdr}, the complex validity domain corresponding to both $s$-channel and $t$-channel LM ellipses is shown in Fig.~\ref{fig:piK_139}. 
We observe that the boundary for the $a=0$ case is consistent with those of Ref.~\cite{Descotes-Genon:2006sdr}. 
Furthermore, we have completed the boundary for the $a=-10 m_\pi^2$ case used in Ref.~\cite{Pelaez:2020gnd}, including the previously neglected small boundary region near $s=0$. The latter is shown as dashed ellipses in the right panel of Fig.~\ref{fig:piK_139}, and the region enclosed by these ellipses must be excluded from the complex validity domain.

\begin{figure}[tt]
    \centering
    \includegraphics[width=1\textwidth,angle=-0]{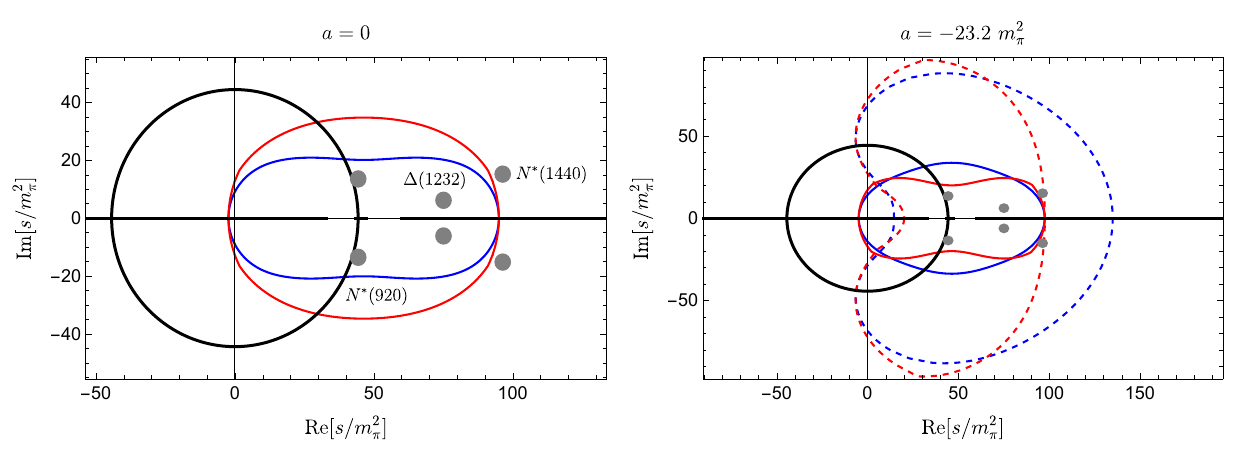}    \caption{Same as Fig.~\ref{fig:piK_139} but for the $\pi N$ scattering. In the right panel, the region outside the dashed ellipses contains the $s=0$ point and should be excluded from the validity domain. } \label{fig:piN_139}
\end{figure}

Recently, two groups used the $\pi N$ RS-type equation to explore the properties of low-lying nucleon resonances. 
Reference~\cite{Cao:2022zhn} first provided the complex validity domain of the RS-type equation with $a=0$, as shown in the left panel of Fig.~\ref{fig:piN_139}. 
Subsequently, Ref.~\cite{Hoferichter:2023mgy}, employing the optimized hyperbola parameter $a=-23.2~m_\pi^2$ from Ref.~\cite{Ditsche:2012fv}, presented the corresponding complex validity domain, which, however, overlooked the constraint of Eq.~\eqref{eq:boundaryv2}. 
The complete result is shown in the right panel of Fig.~\ref{fig:piN_139}.

\subsection{Validity domain at $m_\pi=391$~MeV}

\begin{figure}[t]
    \centering
    \includegraphics[width=.8\textwidth,angle=-0]{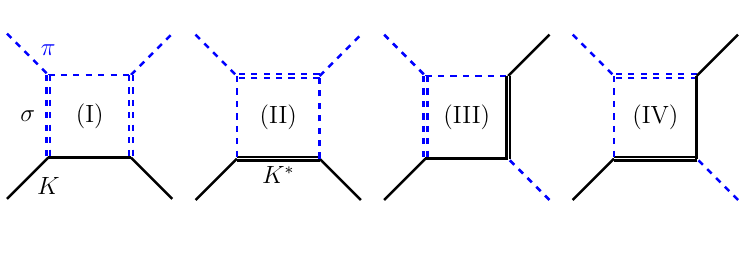}\caption{Unitarity box diagrams that are used to calculate the double spectral regions of $\pi K$ scattering at $m_\pi=391~$MeV. At this pion mass, both the $\sigma$ and the $K^*$ become BS poles. }\label{fig:piK_uni}
\end{figure}
\begin{figure}[t]
    \centering
    \includegraphics[width=.5\textwidth,angle=-0]{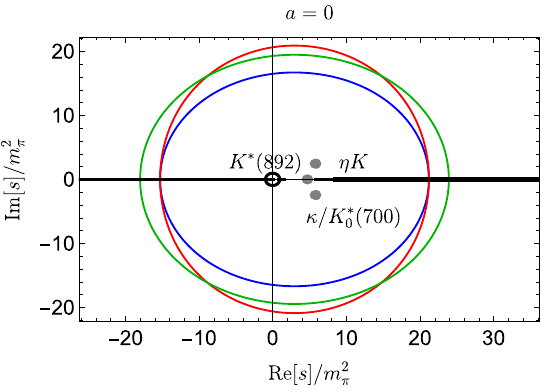} 
    \caption{Complex validity domain of the RS-type equation for $\pi K$ scattering with $a=0$ at $m_\pi=391~$MeV. 
    The blue and red lines correspond to the boundaries in the $s^{\prime}$ and $t^{\prime}$ integrals associated with $\rho_{s t}$, respectively. 
    The green line corresponds to the boundary in the $s^{\prime}$ integral associated with $\rho_{s u}$. 
    The unitarity $\pi K$ and $\eta K$ cuts along the real axis, together with the LHCs of the partial-wave amplitude, are also depicted as black lines. 
    In addition, the gray dots indicate the positions of known resonance poles below the $\eta K$ threshold.} \label{fig:piK_391}
\end{figure}
We assume that the scattering amplitudes satisfy Mandelstam double spectral representation~\cite{Mandelstam:1958xc}, so the semi-major axis of the ellipse is limited by the singularities of double spectral functions $\rho_{st}$ and $\rho_{su}$. 
These boundaries are generated by considering exchanging the lightest particles in the unitarity relations in the $s$-, $t$- and $u$-channels. 
In the case of $\pi K$ scattering at $m_\pi=391~$MeV, the boundaries come from the contributions illustrated in Fig.~\ref{fig:piK_uni}. 

The equations that describe the boundary of the spectral function $\rho_{st}$ are
\begin{align}
    b_\text{I}=&\,(t-4s_\sigma)\lambda_s-4s_\sigma^2 s=0\ ,\\
    b_\text{II}=&\, s_{K^*}\!\left(4 \Sigma s_\sigma-4 s\left(s_\sigma-2 m_\pi^2\right)-2 t\left(s+s_\sigma\right)\right)+4\Sigma s_\sigma s-4 \Delta^2 s_\sigma \nonumber\\
    &-4 m_K^2 s_\sigma^2+s_{K^*}^2\left(t-4 m_\pi^2\right)-4 m_\pi^2 s^2+t\left(s-s_\sigma\right)^2\ ,
\end{align}
where the subscripts I and II indicate the diagrams in Fig.~\ref{fig:piK_uni} that yield the corresponding boundary equations.
Through $s\leftrightarrow u$ exchange, similar equations are obtained for $\rho_{tu}$. The equations that describe the boundary of the nonzero $\rho_{su}$ are
\begin{align}
    b_\text{III}=& \left(\left(2s_{K^*}s+\lambda_s-2\Delta^2\right)\left(2s_\sigma s+\lambda_s\right)+\lambda_s\left(2st+\lambda_s\right)\right)^2 \nonumber\\
    &-16s_\sigma s^2\left(s-2\Sigma+s_{K^*}\right)\left(s_{K^*}s-\Delta^2\right)\left(s_\sigma s+\lambda_s\right) ,\\
    b_\text{IV}= &\,\lambda_s\left(s-s_\sigma-s_{K^*}\right)^2-2m_K^2 t\left(\widetilde{\lambda_s}+2m_K^2s_\sigma-4m_\pi^2 s_\sigma\right) \nonumber\\
    &+2t\left(-m_\pi^2\left(2m_\pi^2+s_\sigma\right)+s\left(2s_\sigma s_{K^*}+\widetilde{\lambda_s}\right)\right)+t^2\widetilde{\lambda_s}\ ,
\end{align}
where $\widetilde{\lambda_s}\equiv \lambda(s,s_\sigma,s_{K^*})$. 
The result is displayed in Fig.~\ref{fig:piK_391}. 
As explained above, we will always set $a=0$ in the subsequent analysis.

%%%%%%%%%%%%%%%%%%%%%%%%%%%%%%%%%%%%%%%%%%%%%%%%%%%%%%%%%%%%%%%%%%%%%%%%%%%%

\section{Inputs to solve the equations}\label{sec:Input}

In this section, our primary focus is to present the inputs used in the subsequent numerical calculations. The RS-type equations~\eqref{eq:g00},~\eqref{eq:g11} and~\eqref{eq:fIl} form a set of highly coupled nonlinear integral equations when we parameterize or formally express the scattering amplitudes in these equations using phase shifts and inelasticities. 
Numerically solving such equations requires truncation at low energies and low partial waves, since the contributions from low energies and low partial-waves in RS-type equations dominate the entire system. 
Here, we primarily aim to determine the low-energy behavior of the $t$-channel $S$- and $P$-wave $\pi\pi\to K\bar K$ amplitudes, the $s$-channel $I=1/2$ $S$- and $P$-wave $\pi K$ scattering amplitudes, and the $I=3/2$ $S$-wave $\pi K$ scattering amplitude. 
All remaining contributions are incorporated as inputs to determine the boundary conditions for the RS-type equations.

For convenience, we recast the RS-type equations into the following form:
\begin{align}
    g_{J}^I(t)= &\, \Delta_{J}^I(t)+\frac{t}{\pi} \int_{t_\pi}^{t_\mathrm{m}} \frac{\md t^{\prime}}{t^{\prime}} \frac{\operatorname{Im} g_{J}^I(t^\prime)}{t^{\prime}-t}\ , \quad \mbox{for }(I,J)=(0,0),(1,1)\ ,\label{eq:RS_t}\\
    f_J^{I}(s)= &\, \text{ST}^{I}_J(s)+\text{PT}^{I}_J(s)+s\text{DT}^I_J(s)+t\text{KT}^I_J(s) \nonumber\\
    & +\frac{1}{\pi} \int_{m_{+}^2}^{s_\mathrm{m}} \mathrm{d} s^{\prime} \left\{K_{J, 0}^{I,1/2}\!\left(s, s^{\prime}\right) \operatorname{Im} f_{0}^{1/2}\!\left(s^{\prime}\right)+K_{J, 1}^{I,1/2}\!\left(s, s^{\prime}\right) \operatorname{Im} f_{1}^{1/2}\!\left(s^{\prime}\right)\right. \nonumber\\
    &\left.+K_{J,0}^{I,3/2}\!\left(s, s^{\prime}\right) \operatorname{Im} f_{0}^{3/2}\!\left(s^{\prime}\right)\right\}\ ,\quad \mbox{for }(I,J)=\left(\frac{1}{2},0\right),\left(\frac{1}{2},1\right),\left(\frac{3}{2},0\right) ,\label{eq:RS_s}
\end{align}
where the driving terms (DTs), i.e., $\Delta^I_J$ and $s\text{DT}^I_J$, and the $t$-channel kernel terms $t\text{KT}^I_J$ can be expressed as
\begin{align}
    \Delta^0_0(t)\simeq &\, \text{ST}^0_0(t)+\text{PT}^0_0(t)-\frac{3 \sqrt{6}}{8} \frac{t}{\pi} \int_{t_\pi}^{t_\mathrm{m}} \frac{\mathrm{d} t^{\prime}}{t^{\prime}} \operatorname{Im} g_1^1\left(t^{\prime}\right) \nonumber\\
    &+\frac{1}{\pi} \int_{m_{+}^2}^{s_c} \mathrm{d} s^{\prime}\left\{G_{0, 0}^{1/2}\!\left(t, s^{\prime}\right) \operatorname{Im} f_{0}^{1/2}\!\left(s^{\prime}\right)+G_{0, 1}^{1/2}\!\left(t, s^{\prime}\right) \operatorname{Im} f_{1}^{1/2}\!\left(s^{\prime}\right)\right.\nonumber\\
    & \left.+G_{0, 0}^{3/2}\!\left(t, s^{\prime}\right) \operatorname{Im} f_{0}^{3/2}\!\left(s^{\prime}\right)+G_{0, 1}^{3/2}\!\left(t, s^{\prime}\right) \operatorname{Im} f_{1}^{3/2}\!\left(s^{\prime}\right)+G_{0, 2}^{1/2}\!\left(t, s^{\prime}\right) \operatorname{Im} f_{2}^{1/2}\!\left(s^{\prime}\right)\right\} \\
    \Delta^1_1(t)\simeq &\, \text{ST}^1_1(t)+\text{PT}^1_1(t)+\frac{1}{\pi} \int_{m_{+}^2}^{s_c} \mathrm{d} s^{\prime}\left\{G_{1, 0}^{1/2}\!\left(t, s^{\prime}\right) \operatorname{Im} f_{0}^{1/2}\!\left(s^{\prime}\right)+G_{1, 1}^{1/2}\!\left(t, s^{\prime}\right) \operatorname{Im} f_{1}^{1/2}\!\left(s^{\prime}\right)\right.\nonumber\\
    & \left.+G_{1, 0}^{3/2}\!\left(t, s^{\prime}\right) \operatorname{Im} f_{0}^{3/2}\!\left(s^{\prime}\right)+G_{1, 1}^{3/2}\!\left(t, s^{\prime}\right) \operatorname{Im} f_{1}^{3/2}\!\left(s^{\prime}\right)+G_{1, 2}^{1/2}\!\left(t, s^{\prime}\right) \operatorname{Im} f_{2}^{1/2}\!\left(s^{\prime}\right)\right\} \\
    s\text{DT}^I_J(s)\simeq &\, \frac{1}{\pi} \int_{s_\mathrm{m}}^{s_\mathrm{c}} \mathrm{d} s^{\prime} \left\{K_{J, 0}^{I,1/2}\!\left(s, s^{\prime}\right) \operatorname{Im} f_{0}^{1/2}\!\left(s^{\prime}\right)+K_{J, 1}^{I,1/2}\!\left(s, s^{\prime}\right) \operatorname{Im} f_{1}^{1/2}\!\left(s^{\prime}\right)\right. \nonumber\\
    &\left.+K_{J,0}^{I,3/2}\!\left(s, s^{\prime}\right) \operatorname{Im} f_{0}^{3/2}\!\left(s^{\prime}\right)\right\}+\frac{1}{\pi} \int_{m_+^2}^{s_\mathrm{c}} \mathrm{d} s^{\prime} K_{J, 1}^{I,3/2}\!\left(s, s^{\prime}\right) \operatorname{Im} f_{1}^{3/2}\!\left(s^{\prime}\right) \nonumber\\
    & +\frac{1}{\pi} \int_{m_+^2}^{s_\mathrm{c}} \mathrm{d} s^{\prime} K_{J, 2}^{I,1/2}\!\left(s, s^{\prime}\right) \operatorname{Im} f_{2}^{1/2}\!\left(s^{\prime}\right)+\frac{1}{32\pi}\int_{-1}^{1}\md z~P_J(z)F_\text{Regge}^{I}(s,t(s,z))\ ,\label{eq:DTs}\\
    t\text{KT}^I_J(s)\simeq &\, \frac{1}{\pi} \int_{t_\pi}^{t_\mathrm{c}} \mathrm{d} t^{\prime} \left\{K_{J,0}^{I,0}\!\left(s, t^{\prime}\right) \operatorname{Im} g_{0}^0\left(t^{\prime}\right)+K_{J,1}^{I,1}\!\left(s, t^{\prime}\right) \operatorname{Im} g_{1}^1\left(t^{\prime}\right)\right\}\ .
\end{align}
In the above equations, the $\simeq$ symbol indicates that high-energy contributions and higher partial waves are either approximated using Regge theory inputs or truncated.
The primary objective of this section is to justify the approximations for the aforementioned DTs and to provide detailed explanations of the relevant expressions.

\subsection{Lattice QCD input}

First, we need to incorporate results from LQCD as inputs for the RS-type equations. 
These come from HSC~\cite{Dudek:2012gj, Dudek:2012xn, Wilson:2014cna, Briceno:2016mjc, Briceno:2017qmb} and recent Roy equation analyses of $\pi\pi$ scattering at unphysical light quark masses~\cite{Cao:2023ntr}. 

The LQCD input can be divided into two categories. The first concerns the $t$-channel, i.e., the $\pi\pi \to K\bar{K}$ process, involving the phases $\phi^I_J$ and moduli $|g^I_J|$. The second concerns the $s$-channel $\pi K \to \pi K$ scattering in the energy region between 1.4~GeV and 1.8~GeV. 
These include the $S$- and $P$-wave phase shifts for isospin $I=1/2$ and $I=3/2$, as well as the $D$-wave phase shift for $I=1/2$~\cite{Dudek:2014qha, Wilson:2014cna, Wilson:2019wfr}. The $D$-wave is included to account for the effect of the narrow resonance $K_2^*(1430)$. 

For the $(I,J)=(0,0)$ input in the $t$ channel when $t_\pi < t < t_K$, i.e., in the pseudo-physical region, Watson's final-state interaction theorem~\cite{Watson:1952ji} ensures that $\phi^0_0 = \delta^0_0$, where $\delta^0_0$ is the phase shift of $\pi\pi$ elastic scattering in the $(I,J)=(0,0)$ channel; when $t \geq t_K$, the phase $\phi^0_0$ can be obtained through coupled-channel unitarity, i.e., $\phi^0_0=\delta^0_{0, \pi}+\delta^0_{0, K}$, where $\delta^0_{0, K}$ is the phase shift of elastic $K\bar K$ scattering. 
The available lattice phase shift data extend approximately up to 1.44~GeV~\cite{Briceno:2016mjc}. 
Despite the presence of some uncertainties, fortunately, the inelastic phase value remains relatively small in the region spanning from $t_K$ to $(1.44~{\rm GeV})^2$ where $\phi^0_0<20^\circ$, resulting in negligible contributions to the $s$-channel partial waves through the dispersion integral. 
Thus, the most significant contributions arise from the region $t_\pi < t < t_K$, where the phase shifts are provided by reliable Roy equation analysis~\cite{Cao:2023ntr}. 
In the $(I,J)=(1,1)$ channel, the phase $\phi^1_1=\delta^1_1$ is primarily determined by the elastic contribution from the near-threshold $\rho$ meson~\cite{Wilson:2014cna, Cao:2023ntr}.

For the $s$-channel $\pi K \to \pi K$ scattering, we need LQCD data in the range 1.4~GeV$<\sqrt{s}<1.8$~GeV. Let us discuss the possible coupled-channel effects. 
Inelastic effects in the $\left(I,J\right)=\left({1}/{2}, 0\right)$ channel emerge when the $\eta K$ channel opens ($M_\eta = 589$~MeV). 
Fortunately, the coupling between the $\eta K$ and $\pi K$ channels is weak, exhibiting only a slightly repulsive interaction~\cite{Wilson:2014cna}.
Therefore, the inelasticity $\eta^{1/2}_0$ deviates only slightly from the elastic value of unity.
Detailed lattice data analyses find no evidence for a significant phase shift cusp at the $\eta K$ threshold. Rather, the phase shift exhibits smooth behavior across this threshold~\cite{Wilson:2014cna}. 

The situation for the $\left(I,J\right)=\left({1}/{2}, 2\right)$ channel is relatively simple. 
When the $\pi\pi K$ channel is neglected, weak inelastic effects appear around $K_2^*(1430)$~\cite{Wilson:2014cna}. 
In contrast, the $\left(I,J\right)=\left({1}/{2}, 1\right)$ channel is more subtle. 
Recall that at the physical pion mass, the low-energy region ($\sqrt{s} < 1.2$~GeV) is dominated by the narrow resonance $K^*(892)$. According to the Review of Particle Physics~\cite{ParticleDataGroup:2024cfk}, the next excited vector resonance is $K^*(1410)$, which couples very weakly to $\pi K$, with a branching fraction below $7\%$. 
When $m_\pi = 391$~MeV, LQCD results show that the elastic approximation is well satisfied up to at least around 1.4~GeV~\cite{Wilson:2014cna}, but the behavior in the range from 1.4~GeV to 1.8~GeV remains unknown in the lattice calculation. 
Due to the weak coupling of $K^*(1410)$ to the $K \pi$ channel, we can neglect its contribution and assume that the elastic approximation works up to 1.8~GeV. 
Thus, we adopt a linear extrapolation to obtain $\delta^{1/2}_1$ in the energy region from 1.4 to 1.8~GeV. Regarding the effects from higher excited resonances, we collectively include their contributions via the Regge amplitudes in the following subsection.  

For the $I=3/2$ channel, repulsive interactions dominate in the low-energy region. At the physical pion mass, the $S$-, $P$-, and $D$-waves are weakly repulsive and do not exhibit any resonant behavior below 1.7~GeV. For $m_\pi = 391$~MeV, these conclusions remain valid~\cite{Wilson:2014cna}. The $S$-wave phase shift is negative and remains approximately elastic up to 1.8~GeV. The $P$-wave appears to exhibit a weak attractive interaction, but the uncertainties are too large to draw definitive conclusions. The $D$-wave phase shift is nearly zero, so its contribution can be safely neglected.

\subsection{Asymptotic Regge amplitudes}

In addition to the LQCD inputs (both direct and extrapolated), the high-energy (above 1.8~GeV) and higher partial-wave inputs of the DTs in Eq.~\eqref{eq:RS_s} are also needed. 
These can be estimated using Regge theory~\cite{Regge:1959mz} (see Refs.~\cite{Collins:1977jy, Donnachie:2002en, Gribov:2003nw} for pedagogical introductions), in which Reggeons and resonances are approximately dual representations of the same physics. 
Due to the lack of relevant scattering cross-section information for $m_\pi=391$~MeV, we need to employ models based on Regge theory, such as the modified Veneziano-Lovelace-Shapiro model (VLS model, also known as the Veneziano model) that includes the Pomeron contribution~\cite{Veneziano:1968yb, Lovelace:1968kjy, Shapiro:1969km}. 
For $\pi K$ scattering, we follow the approach of Refs.~\cite{Kawarabayashi:1969yd, Kawarabayashi:1969hqo}. 

In the asymptotic region where $s^\prime \to +\infty$ with $t$ fixed, based on the assumptions of the VLS model, the $\rho$- and $f_2$-Regge trajectories are linear and degenerate, as are the $K^*$- and $K_2^*$-Regge trajectories.
Therefore, the asymptotic form of the imaginary part of the isospin-odd (receiving contribution from the $\rho$ trajectory) and isospin-even (receiving contribution from the $f_2$ trajectory) amplitudes can be written as
\begin{align}
    &\operatorname{Im}F^-(s^\prime,t)\overset{s^\prime\to\infty}{\simeq}\frac{\pi\lambda}{\Gamma(\alpha_\rho+\alpha_1 t)}(\alpha_1 s^\prime)^{\alpha_\rho+\alpha_1 t}\ ,\\
    &\operatorname{Im}F^+(s^\prime,t)\overset{s^\prime\to\infty}{\simeq}\sigma_\text{P}s^\prime e^{\frac{b_\text{P} t}{2}}+\frac{\pi\lambda}{\Gamma(\alpha_\rho+\alpha_1 t)}(\alpha_1 s^\prime)^{\alpha_\rho+\alpha_1 t}\ ,
\end{align}
where the first term in the second line is the contribution from the Pomeron that has a zero isospin.
It is worth noting that the coupling $\lambda$ in $F^\pm$ is identical to ensure that there are no resonant structures in the repulsive $I=3/2$ channel. 
The parameters of the $\rho$-Regge trajectory can be roughly estimated through the Adler zero of $\pi\pi$ scattering, i.e., the soft pion theorem, and the mass of the $\rho$ meson (see, e.g., Refs.~\cite{Ananthanarayan:2000ht, Cao:2023ntr}). The first condition constains the $\rho$ trajectory at $t=m_\pi^2$ to be $\alpha_\rho + \alpha_1 m_\pi^2=1/2$, and the second condition leads to $\alpha_\rho + \alpha_1 m_\rho^2=1$ because the spin of $\rho$ is 1. Thus, we have 
\begin{align}
    \alpha_1=\frac{1}{2(m_\rho^2-m_\pi^2)}=0.87~\text{GeV}^{-2}\ ,\quad \alpha_\rho=\frac{1}{2}-\alpha_1 m_\pi^2=0.37\ ,
\end{align}
where $m_\rho=854.1~$MeV at $m_\pi=391$~MeV~\cite{Briceno:2017qmb} has been used. 
Unfortunately, the coupling $\lambda$ cannot be directly determined due to the inclusion of the Pomeron contribution. 
In the following, we determine the parameter $\lambda$ by matching our results to LQCD data in the intermediate energy region. 
The contribution of the Pomeron is very subtle. 
In principle, its contribution is directly related to the coupling of gluons and quarks. 
In the case of unphysically large $m_\pi$, the relevant parameters can be estimated using the additive quark rule as~\cite{Cao:2023ntr}
\begin{align}
    b_\text{P}\simeq b_\text{P}^\text{phy}=2.5~\text{GeV}^{-2}\ ,\quad \sigma_\text{P}\simeq 0.7 \sigma_\text{P}^\text{phy}\simeq 7~\text{mb}\ .
\end{align}

Due to crossing symmetry, we also need the asymptotic expression for the $t$-channel amplitudes $G^{0,1}$, i.e., the behavior of the amplitude in the region where $t' \to +\infty$, $u' \to -\infty$, and $s' \to a$~\cite{Buettiker:2003pp, Pelaez:2020gnd},
\begin{align}
    \frac{\operatorname{Im} G^0\left(t', s_b'\right)}{\sqrt{6}}=\frac{\operatorname{Im} G^1\left(t', s_b'\right)}{2}\overset{t'\to\infty}{\simeq}\frac{\pi \lambda\left(\alpha_1 t\right)^{\alpha_{K^*}+a \alpha_1}}{\Gamma\left(\alpha_{K^*}+a \alpha_1\right)}\left[1+\frac{\alpha_1 b}{t}\!\left(\psi\left(\alpha_{K^*}+a \alpha_1\right)-\ln \left(\alpha_1 t\right)\right)\right] ,
\end{align}
where the asymptotic region satisfies the relation $s - a \simeq -{b}/{t}$, and $\psi(z)$ is the polygamma function $\psi(z)=\frac{\mathrm{d}}{\mathrm{d} z}\ln \Gamma(z)$. There are two arguments for why the slope of the $K^*$ trajectory should be the same as that of the $\rho$ trajectory. 
First, the universal slope property of Regge trajectories implies that all slopes are approximately equal to unity; second, the soft theorem requires that the slopes of the two trajectories must be identical~\cite{Kawarabayashi:1969hqo}. 
The intercept can be determined from the soft $\pi$ theorem, i.e., $\alpha_{K^*}={1}/{2}-m_K^2 \alpha_1=0.24$. 
As a cross-check, it can also be derived using the soft kaon theorem, i.e., $\alpha_{K^*}=1-(m_\pi^2+m_K^2)\alpha_1=0.23$, and the results obtained from these two methods are consistent. 

\begin{figure}[tp]
    \centering
    \includegraphics[width=1\textwidth,angle=-0]{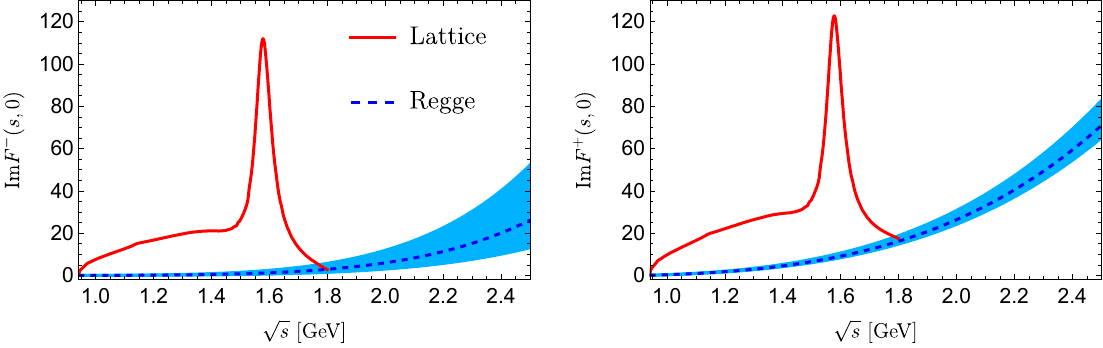} 
    \caption{Comparison of $\operatorname{Im}F^{\pm}(s,0)$ constructed from LQCD data (solid) and the Regge asymptotic amplitudes (dashed).} \label{fig:Regge_piK_391}
\end{figure}

All parameters in the Regge amplitude are known except for the coupling $\lambda$. 
We therefore constrain $\lambda$ by matching our amplitudes to the low-energy and low partial-wave LQCD data. The relevant results are shown in Fig.~\ref{fig:Regge_piK_391}, where the matching point is chosen to be $\sqrt{s_c}=1.8$~GeV\footnote{This matching pertains to the Regge amplitudes and should not be confused with $s_\text{m}$.} and $\lambda=10^{+10}_{-~5}$ is used. 
The high-energy contributions to the $t$-channel RS-type equation are neglected here due to their very weak impact on the low-energy part of the $s$-channel partial waves~\cite{Buettiker:2003pp, Ditsche:2012fv}.

\subsection{Driving terms}

\begin{figure}[tp]
    \centering
    \includegraphics[width=1\textwidth,angle=-0]{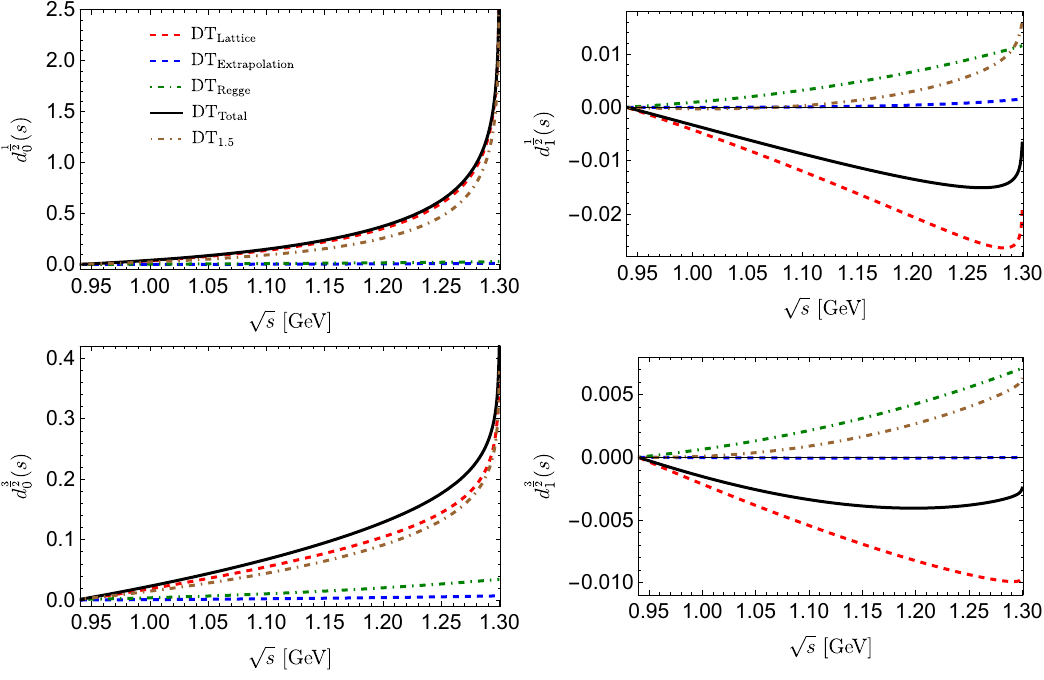} 
    \caption{Decomposition of the DTs in terms of LQCD input, extrapolated part between 1.4 and 1.8~GeV for $\delta_J^{I}$, and Regge theory above the matching point $s_c$. The black solid line labeled as $\mathrm{DT}_{\text {Total}}$ represents the sum of the aforementioned three components, where Regge contributions are introduced in the energy region above $\sqrt{s_c} = 1.8~\mathrm{GeV}$. The brown line labeled as $\mathrm{DT}_{1.5}$ represents the counterpart of $\mathrm{DT}_{\text {Total}}$ with the corresponding Regge contributions included in the energy region above $\sqrt{s_c} = 1.5$~GeV instead of $1.8$~GeV.} \label{fig:DT_Decomposition}
\end{figure}

We can now present the results of the various DTs, $s\text{DT}^I_J(s)\equiv d^I_J(s)$, in Eq.~\eqref{eq:DTs} for each partial wave, as shown in Fig.~\ref{fig:DT_Decomposition}.

Several points should be clarified in order. 
First, the influence of the DTs manifests in the intermediate energy region. 
The low partial-wave results are provided by LQCD, and the Regge contributions are very small for $S$-waves. 
This is consistent with the expected picture---the subtracted DR strongly suppresses the influence of high-energy behavior on low-energy physics. 
Second, for $P$-waves, since the phase shifts are very small in the low-energy region, the Regge contributions to the DTs are relatively large, being of the same order of magnitude as the low partial-wave contributions, but with opposite signs. 
Fortunately, the absolute values of the DTs for the $P$-waves are relatively small, so their impact on the low-energy region is also suppressed. 
In particular, for the $I=3/2$ $P$-wave, we also provide the DT for illustration purposes. 
As expected, the magnitude of this partial wave is very small.

To assess the influence of the matching point $s_c$ between the lattice data and the Regge contributions, we change $\sqrt{s_c}$ from $1.8~$GeV to $1.5~$GeV to estimate the contribution of the $K_2^*(1430)$ to the DTs; see Fig.~\ref{fig:DT_Decomposition}. 
When taking $\sqrt{s_c}=1.5~$GeV, the contribution of this $D$-wave narrow resonance is omitted. 
It can be clearly seen that for the $S$-waves, the contribution of $K_2^*(1430)$ is considerable compared to the Regge contribution, although it does not significantly affect the overall DTs. 
However, for the $P$-waves, their impact is substantial and can even change the sign of the DTs.
Fortunately, the overall magnitude of $P$-waves is small, so their effect on the low-energy region remains negligible.

%%%%%%%%%%%%%%%%%%%%%%%%%%%%%%%%%%%%%%%%%%%%%%%%%%%%%%%%%%%%%%%%%%%%%%%%%%%%

\section{The $t$-channel Muskhelishvili-Omn\` es problem}\label{sec:t-channel}

We are now in a position to analyze the $t$-channel RS-type equation, i.e., Eq.~\eqref{eq:RS_t}, which is a typical once-subtracted DR. 
This form of the equation is exactly a Muskhelishvili-Omn\`es (MO) problem~\cite{Muskhelishvili:1953, Omnes:1958hv} with a finite matching point $t_\mathrm{m}$ if the phase of $g^I_J$ is known below $t_\mathrm{m}$. 
The single-channel MO method is rigorously applied between the $\pi\pi$ and $K\bar{K}$ thresholds (elastic region), as Watson's final-state interaction theorem ensures unitarity. 
In this case, its solution depends on the amplitude value at $t_K$. 
In practice, this is inconvenient because there is an ambiguity of the threshold due to isospin breaking (the physical thresholds of $K^+K^-$ and $K^0\bar{K}^0$ are separated by $\sim8$~MeV), especially for the $I=0$ $S$-wave, and our isospin-conserving framework cannot account for it. 
Therefore, we choose the matching point $t_\text{m} \gtrsim t_K$. 
The optimal choice of $t_\text{m}$ is determined by LQCD data (in fact, different partial waves can have different matching points~\cite{Pelaez:2020gnd}), which will be discussed below. More extensive derivations can be found in Refs.~\cite{Buettiker:2003pp,Ditsche:2012fv}, and we include some necessary details for self-completeness.

\begin{figure}[tb]
    \centering
    \includegraphics[width=1\textwidth,angle=-0]{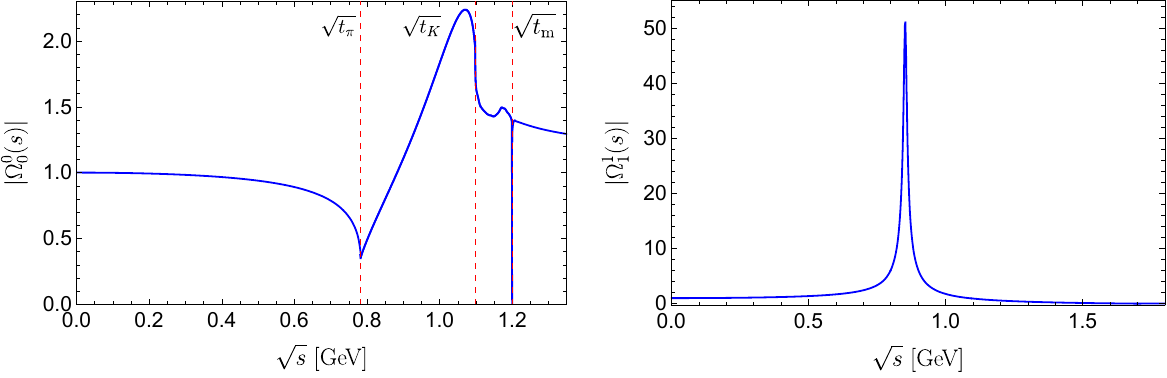} 
    \caption{The moduli of the Omn\`es functions for $\left(I,J\right)=\left(0,0\right)$ and $(1,1)$. The matching points are chosen as $\sqrt{t_\text{m}}=1.2$ and $1.7~$GeV, respectively.
} \label{fig:Omnes}
\end{figure}

We first introduce the Omn\`es function with a finite matching point $t_\mathrm{m}$,
\begin{align}
    \Omega_{J}^I(t)=\exp \left\{\frac{t}{\pi} \int_{t_\pi}^{t_\text{m}} \md t^{\prime}~\frac{\phi_{J}^I\left(t^{\prime}\right)}{t^{\prime}\!\left(t^{\prime}-t\right)}\right\}\ .
\end{align}
The $\Omega(t)$ function can be further written as 
\begin{align*}
        & \Omega(t)=|\Omega(t)| \exp \left\{i \phi(t) \,\theta\!\left(t-t_\pi\right) \,\theta\!\left(t_{\mathrm{m}}-t\right)\right\}\ , \quad \Omega(0)=1\ , \\
        & |\Omega(t)|=\exp \left\{\frac{t}{\pi} \bbint_{t_\pi}^{t_{\mathrm{m}}} \frac{\mathrm{d} t^{\prime}}{t^{\prime}} \frac{\phi\left(t^{\prime}\right)}{t^{\prime}-t}\right\}=|\bar{\Omega}(t)|\left|t_{\mathrm{m}}-t\right|^{x(t)}\ , \quad x(t)=\frac{\phi(t)}{\pi}\ ,\\
        & |\bar{\Omega}(t)|=\left|\frac{t_{\mathrm{m}}}{t_\pi}\!\left(t-t_\pi\right)\right|^{-x(t)} \exp \left\{\frac{t}{\pi} \int_{t_\pi}^{t_{\mathrm{m}}} \frac{\mathrm{d} t^{\prime}}{t^{\prime}} \frac{\phi\left(t^{\prime}\right)-\phi(t)}{t^{\prime}-t}\right\}\ .
\end{align*}
For brevity, the isospin and angular momentum labels are omitted for the moment. 
The symbol $\bbint$ represents the Cauchy principal value integral. 
In Fig.~\ref{fig:Omnes} we show the moduli $\left|\Omega^I_J\right|$ of the resulting once-subtracted Omn\`es functions for $J \in\{0,1\}$, where the choices $\sqrt{t_{\mathrm{m}}}=1.2~$GeV for $S$-wave and $\sqrt{t_{\mathrm{m}}}=1.7~$GeV for $P$-wave ensure $0<\left(\phi_0^0, \phi_1^1\right)\left(t_{\mathrm{m}}\right)<\pi$. 
Furthermore, for $J=0$ the Omn\`es function exhibits a strong cusp at the $\pi \pi$ threshold $\sqrt{t_\pi}$, while for $J=1$ it is fully dominated by the sharp $\rho$ peak.

To solve the MO problem with a finite matching point, a singularity may occur at $t = t_{\mathrm{m}}$. 
To ensure that the integral at the matching point is well-defined, we define an auxiliary function when $\phi(t_\text{m})>0$,\footnote{If $\phi(t_{\mathrm{m}}) < 0$, we can also define $F(t) = \frac{g(t) - \Delta(t)}{\Omega(t)t^l}(t_{\mathrm{m}} - t)^{-n}$.}
\begin{align}
	F(t)=\frac{g(t)-\Delta(t)}{\Omega(t)t^l} \left(t_{\mathrm{m}}-t\right)^n\ ,\quad n=\left[x\right] ,\quad x=\frac{\phi(t_\mathrm{m})}{\pi}\ ,
\end{align}
where $l=1$ for Eq.~\eqref{eq:RS_t}, and $[x]$ denotes the largest integer $\leq x$. 
The analytic function $F(t)$ has only a right-hand cut starting at $t_\text{m}$. 
Thus, $F(t)$ satisfies an $n$-th subtracted DR, which can be recast in terms of $g(t)$ as
\begin{align}
	g(t)= &\,\Delta(t)+\frac{t^l \Omega(t)}{(t_{\mathrm{m}}-t)^n}\left\{\mathcal{P}_{n-1}(t)+\frac{t^n}{\pi} \int_{t_{\pi}}^{t_{\mathrm{m}}} \mathrm{d} t^{\prime} \frac{(t_{\mathrm{m}}-t^\prime)^n\Delta\left(t^{\prime}\right) \sin \phi\left(t^{\prime}\right)}{t^{\prime n+l}\left|\Omega\left(t^{\prime}\right)\right|\left(t^{\prime}-t\right)}\right. \nonumber\\
    &\left.+\frac{t^n}{\pi} \int_{t_{\mathrm{m}}}^{\infty} \mathrm{d} t^{\prime} \frac{(t_{\mathrm{m}}-t^\prime)^n\operatorname{Im} g\left(t^{\prime}\right)}{t^{\prime n+l}\left|\Omega\left(t^{\prime}\right)\right|\left(t^{\prime}-t\right) \mid}\right\}\ .
\end{align}
It can be expressed in terms of the principal value integral as
\begin{align}\label{eq:Omnes_nl}
    |g(t)|=&\, \Delta(t) \cos \phi(t)+\frac{t^{l}|\Omega(t)|}{\left(t_{\mathrm{m}}-t\right)^{n}} \left\{\mathcal{P}_{n-1}(t)+\frac{t^{n}}{\pi} \bbint_{t_{\pi}}^{t_{\mathrm{m}}} \frac{\mathrm{d} t^{\prime}}{t^{\prime n+l}} \frac{\left(t_{\mathrm{m}}-t^{\prime}\right)^{n}}{\left|\Omega\left(t^{\prime}\right)\right|} \frac{\Delta\left(t^{\prime}\right) \sin \phi\left(t^{\prime}\right)}{t^{\prime}-t}\right.\nonumber\\
    &\left.+\frac{t^{n}}{\pi} \int_{t_{\mathrm{m}}}^{\infty} \frac{\mathrm{d} t^{\prime}}{t^{\prime n+l}} \frac{\left(t_{\mathrm{m}}-t^{\prime}\right)^{n}}{\left|\Omega\left(t^{\prime}\right)\right|} \frac{\mathrm{Im} g\left(t^{\prime}\right)}{t^{\prime}-t}\right\}\ .
\end{align}

We now proceed with the calculation by artificially selecting $0 < (\phi^0_0, \phi^1_1)(t_\text{m}) < \pi$, i.e., $n = 0$, which gives us
\begin{align}\label{eq:RS_MO}
    & g^I_J(t)=\Delta^I_J(t)+t \Omega^I_J(t)\left[\frac{1}{\pi} \int_{t_\pi}^{t_\text{m}} \md t^{\prime} \frac{\Delta^I_J\left(t^{\prime}\right) \sin \phi^I_J\left(t^{\prime}\right)}{|\Omega_{J}^I|\left(t^{\prime}\right) t^{\prime }\!\left(t^{\prime}-t\right)}+\frac{1}{\pi} \int_{t_\text{m}}^{\infty} \md t^{\prime} \frac{\left|g^I_J\left(t^{\prime}\right)\right| \sin \phi^I_J\left(t^{\prime}\right)}{|\Omega_{J}^I|\left(t^{\prime}\right) t^{\prime}\!\left(t^{\prime}-t\right)}\right] .
\end{align}
Numerical difficulties can arise in the integral appearing in the MO representation of $g^I_J$.
Although the singularity as $t^\prime \to t_{\mathrm{m}}$ is integrable by construction, the corresponding cusp at $\sqrt{t_m}$ (see the left panel in Fig.~\ref{fig:Omnes}) 
may contribute significantly to the integral, and a complete numerical treatment requires very careful handling of the discrete points in the integral to capture this effect. 
Our goal is to obtain numerical results away from the matching point while using analytical methods in a small region near the matching point to separate out this numerical cusp behavior. 
For a detailed discussion of this method, see Refs.~\cite{Buettiker:2003pp, Ditsche:2012fv}. 
Here, we only present the expressions for the relevant case $n=0, l=1$. First, near the matching point, we can approximate $|\Omega(t)|$ using its asymptotic form,
\begin{align}
    \left|\Omega\left(t \to t_{\mathrm{m}}\right)\right| \simeq\left|\bar{\Omega}\!\left(t_{\mathrm{m}}\right)\right|\left|t_{\mathrm{m}}-t\right|^{x}\ ,
\end{align}
with $x = \phi(t_\mathrm{m})/\pi$. 
For the case of $t<t_\mathrm{m}$, the integral over the region above the matching point can be rewritten as
\begin{align}
    \int_{t_{\mathrm{m}}}^{\infty} \mathrm{d} t^{\prime} \frac{\operatorname{Im} g\left(t^{\prime}\right)}{\left|\Omega\left(t^{\prime}\right)\right|t^\prime\left(t^{\prime}-t\right)} &=\int_{t_{\mathrm{m}}+\tau}^{\infty} \mathrm{d} t^{\prime} \frac{\operatorname{Im} g\left(t^{\prime}\right)}{\left|\Omega\left(t^{\prime}\right)\right|t^\prime\left(t^{\prime}-t\right)}+\frac{\operatorname{Im} g\left(t_{\mathrm{m}}\right)}{\left|\bar{\Omega}\!\left(t_{\mathrm{m}}\right)\right|t_\mathrm{m}} \int_{t_{\mathrm{m}}}^{t_{\mathrm{m}}+\tau} \frac{\mathrm{d} t^{\prime}}{\left|t_{\mathrm{m}}-t^{\prime}\right|^x\left(t^{\prime}-t\right)} \nonumber\\
    &=\int_{t_{\mathrm{m}}+\tau}^{\infty} \mathrm{d} t^{\prime} \frac{\operatorname{Im} g\left(t^{\prime}\right)}{\left|\Omega\left(t^{\prime}\right)\right|t^\prime\left(t^{\prime}-t\right)}+\frac{\operatorname{Im} g\left(t_{\mathrm{m}}\right)}{\left|\bar{\Omega}\!\left(t_{\mathrm{m}}\right)\right|t_\mathrm{m}\!\left(t_{\mathrm{m}}-t\right)^{x}} I_{+}(t)\ ,
\end{align}
where $\tau \to 0^+$ and
\begin{align}
    I_{+}(t)=\int_{0}^{\tilde{\tau}(t)} \frac{\mathrm{d} v}{v^{x}(1 + v)}=\frac{\tilde{\tau}(t)^{1-x}}{1-x} - \int_{0}^{\tilde{\tau}(t)} \mathrm{d} v \frac{v^{1-x}}{1 + v}\ , \quad \tilde{\tau}(t)=\frac{\tau}{|t_{\mathrm{m}}-t|}>0\ .
\end{align}
Similarly, there are two cases when dealing with the integral between $t_\pi$ to $t_\mathrm{m}$. For $t_{\pi} \leq t < t_{\mathrm{m}} - \tau$, the integral is given by 
\begin{align}
    \bbint_{t_{\pi}}^{t_{\mathrm{m}}} \mathrm{d} t^{\prime} \frac{\Delta\left(t^{\prime}\right) \sin \phi\left(t^{\prime}\right)}{\left|\Omega\left(t^{\prime}\right)\right|t^\prime\left(t^{\prime}-t\right)}=& \int_{t_{\pi}}^{t_{\mathrm{m}}-\tau} \frac{\mathrm{d} t^{\prime}}{t^{\prime}-t}\!\left(\frac{\Delta\left(t^{\prime}\right) \sin \phi\left(t^{\prime}\right)}{t^\prime\left|\Omega\left(t^{\prime}\right)\right|}-\frac{\Delta(t) \sin \phi(t)}{t|\Omega(t)|}\right) \nonumber\\
    &+\frac{\Delta(t) \sin \phi(t)}{t|\Omega(t)|} \ln \frac{t_{\mathrm{m}}-\tau-t}{t-t_{\pi}}+\frac{\Delta\left(t_{\mathrm{m}}\right) \sin \phi\left(t_{\mathrm{m}}\right)}{\left|\bar{\Omega}\!\left(t_{\mathrm{m}}\right)\right|t_\mathrm{m}\!\left(t_{\mathrm{m}}-t\right)^{x}} I_{-}(t)\ ,
\end{align}
while for $t_{\mathrm{m}}-\tau \leq t \leq t_{\mathrm{m}}$, it reads 
\begin{align}
    \bbint_{t_{\pi}}^{t_{\mathrm{m}}} \mathrm{d} t^{\prime} \frac{\Delta\left(t^{\prime}\right) \sin \phi\left(t^{\prime}\right)}{\left|\Omega\left(t^{\prime}\right)\right|t^\prime\left(t^{\prime}-t\right)}=\int_{t_{\pi}}^{t_{\mathrm{m}}-\tau} \frac{\mathrm{d} t^{\prime}}{t^{\prime}-t} \frac{\Delta\left(t^{\prime}\right) \sin \phi\left(t^{\prime}\right)}{t^\prime\left|\Omega\left(t^{\prime}\right)\right|}+\frac{\Delta\left(t_{\mathrm{m}}\right) \sin \phi\left(t_{\mathrm{m}}\right)}{\left|\bar{\Omega}\!\left(t_{\mathrm{m}}\right)\right|t_\mathrm{m}\!\left(t_{\mathrm{m}}-t\right)^{x}} \tilde{I}_{-}(t)\ ,
\end{align}
where 
\begin{align}
    &I_{-}(t)=\int_{0}^{\tilde{\tau}(t)} \frac{\mathrm{d} v}{v^{x}(1 - v)}=\frac{\tilde{\tau}(t)^{1-x}}{1-x} + \int_{0}^{\tilde{\tau}(t)} \mathrm{d} v \frac{v^{1-x}}{1 - v}\ ,\\
    &\tilde{I}_{-}(t)=\bbint_{0}^{\tilde{\tau}(t)} \frac{\md v}{v^{x}(1-v)}=-\ln |\tilde{\tau}(t)-1|+\frac{\tilde{\tau}(t)^{1-x}}{1-x}+\int_{0}^{\tilde{\tau}(t)} \mathrm{d} v \frac{v^{1-x}-1}{1-v}\ .
\end{align}
We assume $0 < x < 1$, so the integrals defined above are convergent. 
The singularity of the integrand at $v=0$, which leads to the endpoint singularity at the matching point, has been separated analytically. 
This allows for a stable numerical computation of the MO integral. 

The above technique is also applicable when $t_\text{m} < t < t_c$, where $t_c$ is an arbitrary value greater than $t_\text{m}$ (an upper limit is established for numerical integration). 
For $t_\text{m} < t < t_\text{m} + \tau$, the integral is 
\begin{align}
    \bbint^{t_c}_{t_\text{m}}\md t^\prime~\frac{\operatorname{Im}g(t^\prime)}{|\Omega(t^\prime)|t^\prime(t^\prime-t)}=\int^{t_c}_{t_\text{m}+\tau}\md t^\prime~\frac{\operatorname{Im}g(t^\prime)}{|\Omega(t^\prime)|t^\prime(t^\prime-t)}-\frac{\operatorname{Im}g(t_\text{m})}{|\overline{\Omega}(t_\text{m})|t_\text{m}}\frac{\tilde{I}_-(t)}{(t-t_\text{m})^x}\ ,
\end{align}
while for $t_\text{m}+\tau<t<t_c$, it is 
\begin{align}
    \bbint^{t_c}_{t_\text{m}}\md t^\prime~\frac{\operatorname{Im}g(t^\prime)}{|\Omega(t^\prime)|t^\prime(t^\prime-t)}= &-\frac{\operatorname{Im}g(t_\text{m})I_-(t)}{|\Omega(t^\prime)|t_\text{m}(t-t_\text{m})}+\int^{t_c}_{t_\text{m}}\frac{\md t}{t^\prime-t}\!\left(\frac{\operatorname{Im}g(t^\prime)}{t^\prime|\Omega|(t^\prime)}-\frac{\operatorname{Im}g(t)}{t|\Omega(t)|}\right) \nonumber\\
    &+\frac{\operatorname{Im}g(t)}{t|\Omega(t)|}\ln\left(\frac{t_c-t}{t-t_\text{m}-\tau}\right) ,
\end{align}
and
\begin{align}
    \int_{t_{\pi}}^{t_{\mathrm{m}}} \mathrm{d} t^{\prime} \frac{\Delta\left(t^{\prime}\right) \sin \phi\left(t^{\prime}\right)}{\left|\Omega\left(t^{\prime}\right)\right|t^\prime\left(t^{\prime}-t\right)}=\int_{t_{\pi}}^{t_{\mathrm{m}}-\tau} \mathrm{d} t^{\prime} \frac{\Delta\left(t^{\prime}\right) \sin \phi\left(t^{\prime}\right)}{\left|\Omega\left(t^{\prime}\right)\right|t^\prime\left(t^{\prime}-t\right)}-\frac{\Delta(t_{\mathrm{m}})\sin\phi(t_{\mathrm{m}})}{|\overline{\Omega}(t_{\mathrm{m}})|t_{\mathrm{m}}}\frac{I_+(t)}{(t-t_{\mathrm{m}})^x}\ .
\end{align}

Additionally, this method has the potential advantage of ensuring the continuity of the numerical solution at the matching point~\cite{Buettiker:2003pp, Ditsche:2012fv}. 
This requirement is highly non-trivial; we refer to Ref.~\cite{Hoferichter:2012wf} for a detailed discussion.

%%%%%%%%%%%%%%%%%%%%%%%%%%%%%%%%%%%%%%%%%%%%%%%%%%%%%%%%%%%%%%%%%%%%%%%%%%%%

\section{Solutions of Roy-Steiner-type equations}\label{sec:s-channel}

In Section~\ref{sec:t-channel}, we reviewed the construction of the solution to the MO problem with a finite matching point. 
Once the DT of Eq.~\eqref{eq:RS_t} has been determined, we can directly obtain the low-energy $t$-channel $S$- and $P$-wave amplitudes using Eq.~\eqref{eq:RS_MO}. 
Given a set of $t$-channel partial-wave amplitudes, the $s$-channel RS-type equations~\eqref{eq:RS_s} can be recast into a closed set of Roy-like equations. 

Unfortunately, Roy or RS-type equations generally have multiple solutions or even no solutions, but the physical solution should always be unique. 
This indicates that the physical solution imposes certain requirements on the DTs of the equations, i.e., not all DTs satisfy such physical restrictions. 
For a modern discussion of this mathematical problem, see Refs.~\cite{Gasser:1999hz, Wanders:2000mn}. 
In this section, we first briefly discuss how to determine the unique solution to the RS-type equations for the $\pi K$ system at $m_\pi=391~$MeV, then provide the corresponding numerical strategy to solve them based on the restrictions of the unique solution, and finally present the complete solution of the low-energy, low partial-wave amplitudes including both $t$- and $s$-channels.

\subsection{Uniqueness for the $s$-channel solutions}

The $s$-channel RS-type equations, as a set of infinitely coupled infinite-dimensional equations, can be truncated up to $P$-wave for numerical calculations. 
In this approach, the $I=3/2$ $P$-wave is treated as part of the DT due to its negligible effects. 
As a result, in total three partial waves need to be solved. We take the $s$-channel matching point $s_\text{m}=\left(1.3~\mathrm{GeV}\right)^2$, which is slightly above the $\eta K$ threshold at $1.138~$GeV but below the three-body $\pi\pi K$ threshold at $1.331$~GeV. 
Thus, our goal is to solve the partial-wave amplitudes numerically from the $\pi K$ threshold $m_+^2$ to the matching point $s_\text{m}$. 
The number of independent solutions to the RS-type equations depends on the input phase shifts at the matching point, which are derived from LQCD phase shifts~\cite{Wilson:2014cna},
\begin{align}
    \delta^{1/2}_0(s_\mathrm{m})=(62\sim 82)^\circ\ ,\quad \delta^{3/2}_0(s_\mathrm{m})=-(18\pm 1)^\circ\ ,\quad \delta^{1/2}_1(s_\mathrm{m})=-(4\pm 1)^\circ\ .
\end{align}
According to the mathematical discussions in Refs.~\cite{Gasser:1999hz, Wanders:2000mn}, the multiplicity index of the RS-type equations is $m=0-1-1=-2$. 
This results in an overdetermined system of equations, which is typically inconsistent and admits no exact solution.

The aforementioned mathematical discussion does not involve the intrinsic parameters in Eq.~\eqref{eq:RS_s}, namely scattering lengths $a^{{1}/{2},{3}/{2}}_0$ and the positions and residues of the two poles $K^*(892)$ and $\sigma$. 
In fact, these parameters cannot be simply determined from LQCD analyses as inputs to the RS-type equations. 
The scattering lengths $a^{{1}/{2},{3}/{2}}_0$ cannot be determined precisely through a simple phase shift analysis alone, so it is convenient to treat them as outputs of the RS-type equations rather than predetermined inputs. 
For the position and residue of the BS $\sigma$ pole, i.e., $s_\sigma$ and $g_{\sigma \pi\pi}$, we use the results from a recent Roy equation analysis~\cite{Cao:2023ntr} as inputs. 
In contrast, the sign of $g_{\sigma K\bar{K}} = \pm(0 \sim 900)~\text{MeV}$~\cite{Briceno:2017qmb} cannot even be definitively determined.\footnote{The sign here is relative to $g_{\sigma \pi\pi}$. Once the sign of $g_{\sigma \pi\pi}$ is determined, the sign of $g_{\sigma K\bar{K}}$ has a clear physical meaning.} 
In practice, the no-cusp condition of $g^0_0$ at the matching point $t_\mathrm{m}$ can be used to determine it, so the coupling constant $g_{\sigma K\bar{K}}$ is also treated as an input parameter when dealing with the $s$-channel RS-type equations. 
The pole position of $K^*$ is already quite precise with an accuracy of $0.2\%$ and thus can be treated as an input parameter here. 
However, the uncertainty of the residue $g_{K^* \pi K}$ is relatively large, so we do not treat it as an input parameter. 
Thus, the $s$-channel RS-type equations have three free parameters: $a^{{1}/{2},{3}/{2}}_0$ and $g_{K^* \pi K}$. 
These parameters can be fine-tuned to ensure that the equations have a (physical) solution.
Mathematically, this transforms the system from $m = -2$ to $-2 + 3 = 1$, thereby changing it from an overdetermined system to an underdetermined one. 
This implies the existence of a one-parameter family of solutions. This one-parameter family of solutions can be physically constrained using the no-cusp condition of the $\left(I,J\right)=\left({1}/{2}, 0\right)$ channel at the matching point $t_\text{m}$, thereby ensuring a unique solution for the system. 
At this point, the other two channels are also nearly smooth at the matching point. 
For convenience, we also apply the no-cusp condition to the other two channels, which does not introduce any visible numerical impact.

\subsection{Numerical strategy to solve Roy-Steiner-type equations}

Given the constraints corresponding to the physical solution of the equations, we can outline the iterative strategy for obtaining the numerical solution of the system:
\begin{itemize}
    \item[$(1)$] To begin with, we examine the $t$-channel problem. In the initial iteration, the input $s$-channel partial-wave amplitudes are derived from LQCD analysis~\cite{Wilson:2014cna} (in subsequent iterations, the partial waves are obtained from the solution in step $(2)$ below). 
    The initial parameter values are: $\sqrt{s_\sigma}=759~\text{MeV}$~\cite{Cao:2023ntr}, $\sqrt{s_{K^*}}=934~\text{MeV}$~\cite{Wilson:2019wfr}, $|g_{\sigma\pi\pi}|=493~\text{MeV}$~\cite{Cao:2023ntr}, $g_{K^* \pi K}=51~\text{MeV}$~\cite{Wilson:2019wfr}, $m_\pi a^{1/2}_0=1.00$~\cite{Wilson:2014cna}, $m_\pi a^{3/2}_0=-0.278$~\cite{Wilson:2014cna}. Additionally, the value of $g_{\sigma K\bar{K}}$ is determined by the no-cusp condition of $g^0_0$.
    \item[$(2)$] Once the $t$-channel amplitudes from step $(1)$ are obtained, the next step is to solve the $s$-channel equations to determine the three types of low-energy phase shifts in the $s$-channel. 
    At this stage, the two scattering lengths $a^{{1}/{2},{3}/{2}}_0$ and the residue $g_{K^* \pi K}$ are provided as outputs along with the solution.\footnote{Note that $g_{\sigma K\bar{K}}$ remains unchanged. Since the cut due to the $\sigma$ pole is far from the $\pi K$ threshold (see the right panel in Fig.~\ref{fig:singularity} and the values in Eq.~\eqref{eq:values}), its direct influence on the $s$-channel physical region is very weak. The precise value cannot be determined from the information available in the physical region of $s$-channel $\pi K$ scattering. In contrast, since the $\sigma$ pole is close to the $\pi\pi$ threshold, its coupling has a significant influence on the $t$-channel, especially the possible cusp at the matching point.}
    \item[$(3)$] Repeat steps $(1)$ and $(2)$ until convergence to obtain the solution of the entire system.
\end{itemize}

The most critical step is solving the $s$-channel equations in step $(2)$. Detailed numerical methods can be found in Refs.~\cite{Ananthanarayan:2000ht, Buettiker:2003pp, Hoferichter:2015hva}. 
The key idea is to employ a suitable phase shift parameterization to transform the nonlinear integral equations of the $s$-channel into a constrained optimization problem.

\subsection{Parameterizations of the $s$-channel partial waves}

The $s$-channel RS-type equation~\eqref{eq:RS_s} can be rewritten in terms of phase shifts as
\begin{align}\label{eq:RSv2}
    \operatorname{Re}f^I_J[\delta^I_J]=\sum_{J^\prime, I^\prime}\mathcal{F}[\operatorname{Im}f^{I^\prime}_{J^\prime}[\delta^{I^\prime}_{J^\prime}]]\ ,
\end{align}
where the symbol $\mathcal{F}\left[\operatorname{Im}f_J^I\right]$ denotes the right-hand side of the RS-type equations~\eqref{eq:RS_s}.
Our goal is to find phase shift functions that minimize the difference between the left and right sides of this equation. 
The analysis must systematically explore the complete allowed parameter space of phase shifts.
However, for practical reasons, we restrict the solution to a class of simple parameterizations. Here, we use a Schenk-like parameterization~\cite{Schenk:1991xe} for all three partial waves:
\begin{align}\label{eq:schenk}
    \tan \delta_{J}^{I}=|q|^{2 J+1}\!\left(A_{J}^{I}+B_{J}^{I} q^2+C_{J}^{I} q^4+D_{J}^{I} q^6\right) \frac{m_{+}^2-s_{J}^{I}}{s-s_{J}^{I}}\ ,
\end{align}
with $A^I_J=a^I_J$. Additionally, as discussed above, the physical solution requires a smooth transition at the matching point $s_\text{m}$. 
We require that Eq.~\eqref{eq:schenk} satisfies the no-cusp condition. 
The simplest approach is to impose continuity and derivative continuity at $s_\mathrm{m}$ for each phase shift explicitly~\cite{Hoferichter:2015hva}. Therefore, we can define
\begin{align}
    \delta_{\mathrm{m}, J}^{I}=\lim _{\epsilon \rightarrow 0} \delta_{J}^{I}\!\left(s_{\mathrm{m}}+\epsilon\right) , \quad \delta_{\mathrm{m}, J}^{\prime I}=\left.\lim _{\epsilon \rightarrow 0} \frac{\mathrm{d} \delta_{J}^{I}(s+\epsilon)}{\mathrm{d} s}\right|_{s=s_{\mathrm{m}}}\ ,
\end{align}
and the last two parameters in Eq.~\eqref{eq:schenk} can then be expressed as functions of $s_\mathrm{m}$, $\delta_{\mathrm{m}, J}^{I}$, and the derivative $\delta_{\mathrm{m}, J}^{\prime I}$. 
This allows us to constrain the function space so that the matching conditions are automatically satisfied. Consequently, we have 
\begin{align}
    & C_{J}^{I}=-\frac{1}{q_{\mathrm{m}}^4}\left\{3 A_{J}^{I}+2 B_{J}^{I} q_{\mathrm{m}}^2+\frac{s_{\mathrm{m}}-s_{J}^{I}}{2 q_{\mathrm{m}}^{2J} q_{\mathrm{m}}^{\prime}\!\left(m_{+}^2-s_{J}^{I}\right)}\!\left(\frac{\delta_{\mathrm{m}, J}^{\prime I}}{\cos ^2 \delta_{\mathrm{m}, J}^{I}}+\frac{\tan \delta_{\mathrm{m}, J}^{I}}{s_{\mathrm{m}}-s_{J}^{I}}-\frac{(7+2J) q_{\mathrm{m}}^{\prime}}{q_{\mathrm{m}}} \tan \delta_{\mathrm{m}, J}^{I}\right)\right\}\ , \nonumber\\
    & D_{J}^{I}=\frac{1}{q_{\mathrm{m}}^6}\left\{2 A_{J}^{I}+B_{J}^{I} q_{\mathrm{m}}^2+\frac{s_{\mathrm{m}}-s_{J}^{I}}{2 q_{\mathrm{m}}^{2J} q_{\mathrm{m}}^{\prime}\!\left(m_{+}^2-s_{J}^{I}\right)}\!\left(\frac{\delta_{\mathrm{m}, J}^{\prime I}}{\cos ^2 \delta_{\mathrm{m}, J}^{I}}+\frac{\tan \delta_{\mathrm{m}, J}^{I}}{s_{\mathrm{m}}-s_{J}^{I}}-\frac{(5+2J) q_{\mathrm{m}}^{\prime}}{q_{\mathrm{m}}} \tan \delta_{\mathrm{m}, J}^{I}\right)\right\}\ ,
\end{align}
where $q_{\mathrm{m}}=q\left(s_{\mathrm{m}}\right)$ and $q_{\mathrm{m}}^{\prime}=\frac{\mathrm{d} q(s)}{\mathrm{d} s}|_{s=s_{\mathrm{m}}}$. 
Each parameterization contains 3 parameters (including the $S$-wave scattering lengths $a^{{1}/{2},{3}/{2}}_0$ and $B^I_J$), resulting in a total of $4 \times 3 = 12$ parameters. 
Including $g_{K^* \pi K}$, the $s$-channel system has 13 parameters. Additionally, there is one parameter $g_{\sigma K\bar{K}}$ in the $t$-channel, yielding a total of 14 parameters in the full $\pi K$ RS-type equation system.

\subsection{Numerical determination of the solutions and error estimations}

\begin{figure}[t]
    \centering
    \includegraphics[width=.5\textwidth,angle=-0]{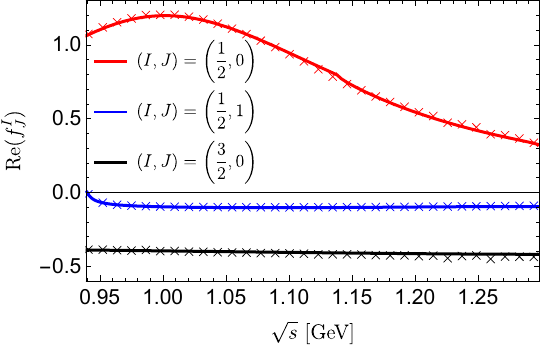} 
    \caption{Comparison of the left-hand sides (lines) and right-hand sides (crosses) of the RS-type equations after the $\chi^2$ minimization.} \label{fig:accuracy}
\end{figure} 
Based on the previous discussion, the crucial inputs for the numerical solution of the $\pi K$ RS-type equations are the phase shifts and their derivatives at the matching point $s_\text{m}$. 
These values can be obtained from LQCD~\cite{Wilson:2014cna}: for the $\left(I,J\right)=\left(\frac{1}{2},0\right)$ channel, 
$$\delta_0^{1/2}\!\left(s_{\mathrm{m}}\right)=\left.\delta_0^{1/2}\!\left(s_{\mathrm{m}}+0^{+}\right)\right|_{\text{input}}=76^{\circ}\ ,\quad \frac{\mathrm{d} \delta_0^{1/2}\!\left(s_{\mathrm{m}}\right)}{\mathrm{d} s}=\left.\frac{\mathrm{d} \delta_0^{1/2}\!\left(s_{\mathrm{m}}+0^{+}\right)}{\mathrm{d} s}\right|_{\text{input}}=73^{\circ}~\mathrm{GeV}^{-2};$$ 
for the $\left(I,J\right)=\left(\frac{1}{2},1\right)$ channel, 
$$\delta_1^{1/2}\!\left(s_{\mathrm{m}}\right)=\left.\delta_1^{1/2}\!\left(s_{\mathrm{m}}+0^{+}\right)\right|_{\text{input}}=-4^{\circ}\ ,\quad \frac{\mathrm{d} \delta_1^{1/2}\!\left(s_{\mathrm{m}}\right)}{\mathrm{d} s}=\left.\frac{\mathrm{d} \delta_1^{1/2}\!\left(s_{\mathrm{m}}+0^{+}\right)}{\mathrm{d} s}\right|_{\text{input}}=-1.5^{\circ}~\mathrm{GeV}^{-2};$$ 
and for the $\left(I,J\right)=\left(\frac{3}{2},0\right)$ channel, 
$$\delta_0^{3/2}\!\left(s_{\mathrm{m}}\right)=\left.\delta_0^{3/2}\!\left(s_{\mathrm{m}}+0^{+}\right)\right|_{\text{input}}=-18^{\circ}\ ,\quad \frac{\mathrm{d} \delta_0^{3/2}\!\left(s_{\mathrm{m}}\right)}{\mathrm{d} s}=\left.\frac{\mathrm{d} \delta_0^{3/2}\!\left(s_{\mathrm{m}}+0^{+}\right)}{\mathrm{d} s}\right|_{\text{input}}=-7^{\circ}~\mathrm{GeV}^{-2}.$$

\begin{figure}[tb]
    \centering
    \includegraphics[width=1\textwidth,angle=-0]{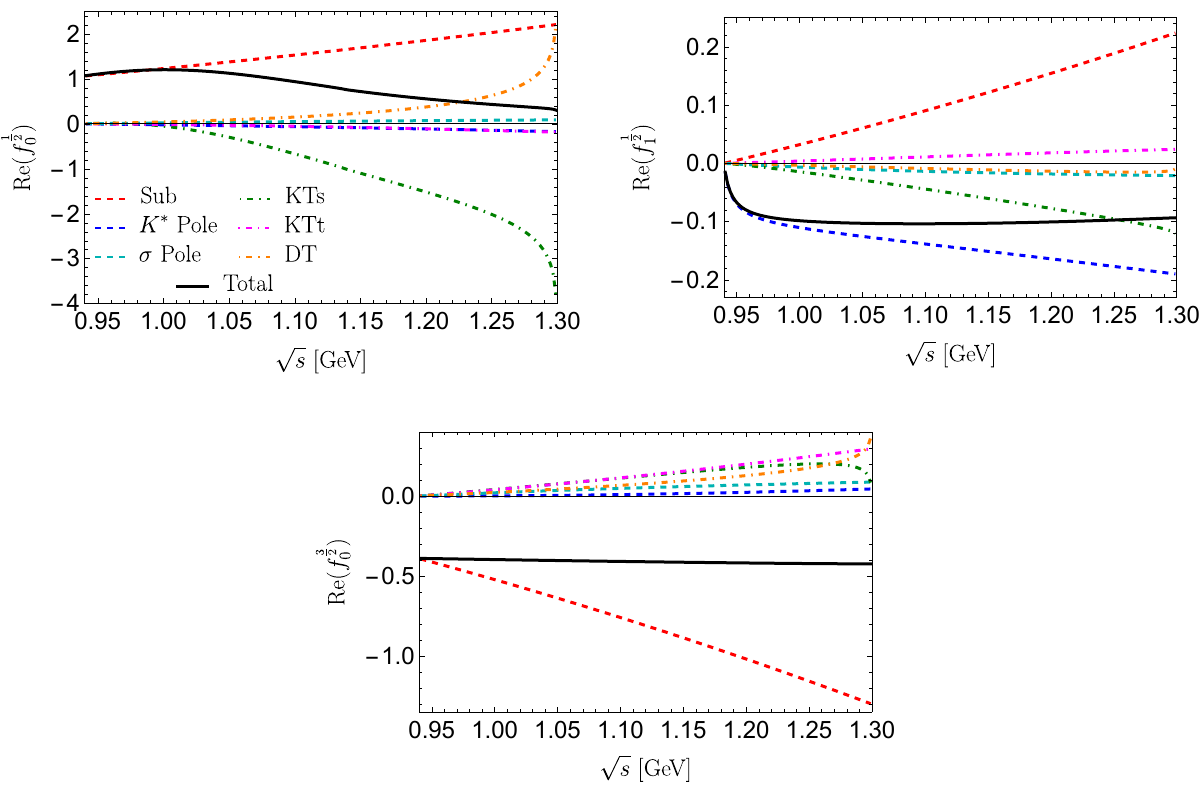} 
    \caption{Decomposition of the right-hand sides of RS-type equations into different contributions. Black solid lines correspond to the sum of all contributions. Red dashed lines denote the subtraction contribution (``Sub''), whereas the blue and cyan dashed lines refer to the bound $K^*(892)$ and $\sigma$ pole terms (``$K^*$ Pole'' and ``$\sigma$ Pole''). The $s$- and $t$-channel kernel terms (``KTs'' and ``KTt'') are given by the green and magenta dot-dashed lines, and the driving terms (``DT'') are described by the orange dot-dashed lines.} \label{fig:decomposition}
\end{figure}

All parameters are determined through an optimization procedure by minimizing a $\chi^2$-like function~\cite{Buettiker:2003pp, Ditsche:2012fv}
\begin{align}
    \chi^2=\sum_{I, J} \sum_{i=1}^N\left\{\frac{\operatorname{Re} f_J^I\left(s_i\right)-\mathcal{F}\left[\operatorname{Im}f_J^I\right]\left(s_i\right)}{\operatorname{Re} f_J^I\left(s_i\right)}\right\}^2\ .
\end{align}
We have verified the stability of the solution with respect to the choice of the number of uniformly distributed grid points from the $\pi K$ threshold to the matching point $s_\text{m}$, $N$, which is varied between $20$ and $30$, and finally $N=25$ is adopted. 
A $\chi^2$ value of $10^{-3}$ indicates that our numerical solution provides a valid approximation.
The precision of our results is quantified in Fig.~\ref{fig:accuracy}.
In Fig.~\ref{fig:decomposition}, we show the effects of different parts on the right-hand sides of the RS-type equations~\eqref{eq:RS_s}.
While the pole term of the $P$-wave $K^*$ makes a significant contribution in the $(I,J)=(1/2,1)$ channel, the other pole terms have minor impact in the physical region. 
However, the LHCs generated by these poles through crossing symmetry are essential for the analytic behavior of the partial-wave amplitudes in the subthreshold region.

It should be noted that the subtraction terms in Fig.~\ref{fig:decomposition} contribute significantly in each partial wave and largely cancel out the other contributions in the energy region near the matching point. 
Analogous to the $\pi\pi$ scattering analysis at unphysical quark masses~\cite{Cao:2023ntr}, the $\pi K$ scattering analysis also employs a twice-subtracted DR. 
However, the asymptotic behavior of the $s$-channel kernels, $K_{J, J^\prime}^{I, I^\prime}(s,s^\prime)\sim {1}/{s^{\prime 2}}$ as $s^\prime\to\infty$ (unlike the conventional twice-subtracted behavior ${1}/{s^{\prime 3}}$), results directly from our use of hyperbolic DRs.
As shown in Fig.~\ref{fig:DT_Decomposition}, the Regge contributions, which have some model dependence, are quite small except for the $P$-wave and can be almost neglected below $s_\mathrm{m}$. 
The results in the crucial intermediate energy region, namely $1.3~$GeV to $1.8~$GeV, are compatible with more reliable LQCD data. 
Therefore, our analysis based on the RS-type equations and LQCD data can be regarded as a model-independent study.

\begin{figure}[t]
    \centering
    \includegraphics[width=1\textwidth,angle=-0]{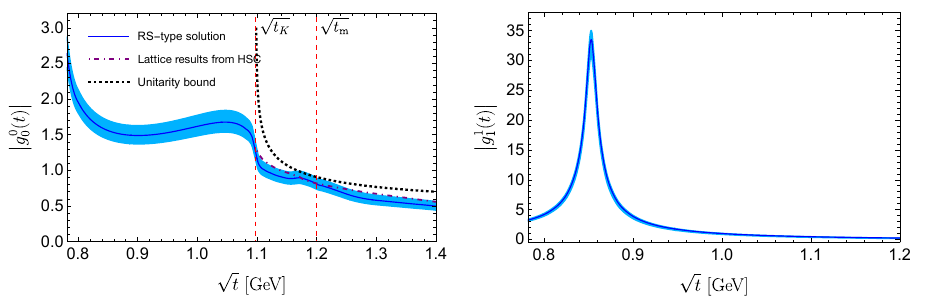} 
    \caption{Comparison of the absolute values of $g_0^0$ and $g_1^1$ obtained from solving the RS-type equations and the corresponding LQCD result from HSC~\cite{Briceno:2017qmb}.} \label{fig:t_PW}
\end{figure}

The procedure for evaluating uncertainties involves varying various inputs, which include the inputs at the matching point $s_\mathrm{m}$ and the DTs, including $\delta_0^{1/2}\!\left(s_{\mathrm{m}}\right)=(67 \sim 82)^{\circ}$, $\delta_0^{3/2}\!\left(s_{\mathrm{m}}\right)=-(18 \pm 1)^{\circ}$, $\delta_1^{1/2}\!\left(s_{\mathrm{m}}\right)=-(4 \pm 1)^{\circ}$, $s_{\mathrm{m}}=\left(1.3~\mathrm{GeV}\right)^2$, $\sqrt{s_{K^*}}=(934\pm 2)~\text{MeV}$~\cite{Wilson:2019wfr}, $\sqrt{s_\sigma}=759^{+~7}_{-16}~\text{MeV}$~\cite{Cao:2023ntr}, $g_{\sigma\pi\pi}=493^{+27}_{-46}~\text{MeV}$~\cite{Cao:2023ntr}, $\lambda=10^{+10}_{-~5}$, the LQCD data for $\pi K$ scattering above $s_{\mathrm{m}}$~\cite{Wilson:2014cna,Wilson:2019wfr}, and the crossed $t$-channel results obtained in Ref.~\cite{Cao:2023ntr}. 
As one of the key inputs, the LQCD data above $s_{\mathrm{m}}$ still carry non-negligible uncertainties, preventing us from predicting the phase shifts between $m_+^2$ and $s_\mathrm{m}$ via the RS-type equations with the same precision as in the physical quark mass scenario. 
Based on these input variations, the uncertainties of the phase shifts in the three channels and the pole positions discussed below are obtained using the bootstrap approach. 
For a detailed error estimation for the RS-type equations, see Refs.~\cite{Ananthanarayan:2000ht, Buettiker:2003pp, Hoferichter:2015hva}.
The final $t$-channel partial-wave amplitudes are illustrated in Fig.~\ref{fig:t_PW}, where we compare the moduli of $g^0_0$ computed from the RS-type equations with the LQCD inputs.

Note that the $t$-channel matching point for the $P$-wave has been chosen to be $\sqrt{t_{\mathrm{m}}}=1.7$~GeV. 
For $m_\pi=391$~MeV, there is a single extremely narrow resonance below 1~GeV, which is the $\rho$ meson with a pole located at $\sqrt{s_\rho}=\left(853.3_{-1.1}^{+1.1}\right)-i\left(6.7_{-0.7}^{+0.2}\right)$~MeV~\cite{Cao:2023ntr}. 
We have verified that varying $\sqrt{t_{\mathrm{m}}}$ from 1.5~GeV to 1.9~GeV yields rather stable $P$-wave amplitude results, and the impact on the $\kappa$ pole is negligibly small (at the keV level), as they primarily affect high-energy $t$-channel contributions, which are strongly suppressed in the dispersive integral.

For the $S$-wave, we find that the agreement for $g_0^0(t)$ is reasonably good. Importantly, in the range $t \geq t_K$, the unitarity bound, i.e., Eq.~\eqref{eq:UB}, is always satisfied. Adopting a larger value for the $S$-wave matching point $t_{\mathrm{m}}$ will improve the agreement between input and output in the region spanning $1.2\sim 1.4~\rm{GeV}$ but will lead to a violation of unitarity for $t<t_{\mathrm{m}}$~\cite{Buettiker:2003pp}. 
We find that only when the matching point $\sqrt{t_{\mathrm{m}}}$ is near 1.2~GeV (approximately between 1.18~GeV and 1.22~GeV) can the $S$-wave $\pi\pi \to K \bar{K}$ amplitude satisfy the unitarity constraint $\left|g_0^{1/2}(t)\right|<{\sqrt{t}}/{\left[4\left(q_\pi q_K\right)^{1 / 2}\right]}$.
Slight variations of this matching point while maintaining unitarity, the effect on the $\kappa$ pole result is also negligible.

\begin{figure}[tb]
    \centering
    \includegraphics[width=1\textwidth,angle=-0]{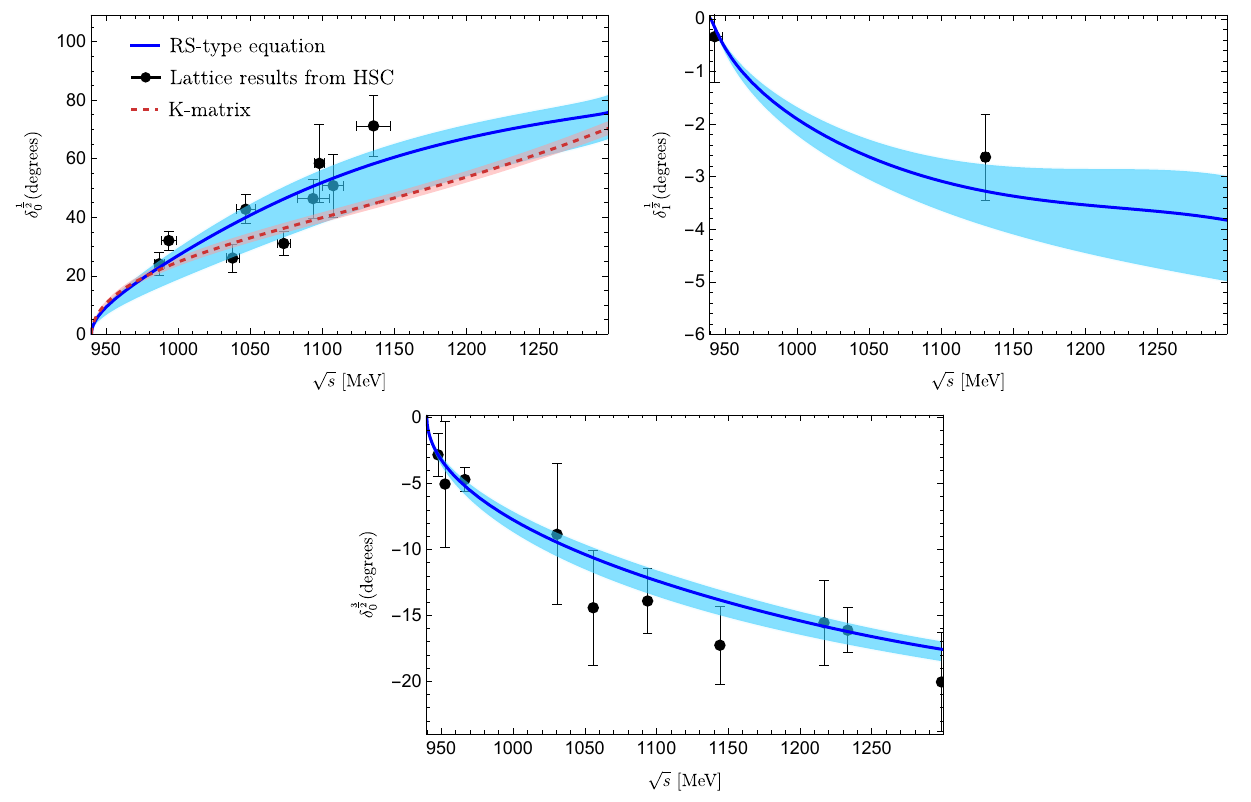} 
    \caption{Phase shifts for the $(I,J)=(1/2,0)$, $(1/2,1)$, and $(3/2,0)$ channels obtained by solving the RS-type equations. The LQCD results, represented by data points with error bars, are taken from HSC~\cite{Wilson:2014cna}. We also show as a red dashed curve the $(1/2,0)$ phase shift obtained from the preferred $K$-matrix parametrization fit in Ref.~\cite{Wilson:2014cna}.} \label{fig:s_PW}
\end{figure}

The corresponding parameter $g_{\sigma K\bar K}$ is $g_{\sigma K\bar K}=-\left(296^{+45}_{-44}\right)~$MeV.\footnote{We assume that the coupling satisfies $g_{\sigma \pi\pi}>0$, i.e., $g_{\sigma \pi\pi} g_{\sigma K \bar K}<0$.}
For the $P$-wave, a pronounced narrow $\rho$ peak is evident in $g_1^1(t)$. 
The three $S$- and $P$-wave phase shifts predicted by the RS-type equations are shown in Fig.~\ref{fig:s_PW}, where the LQCD data from HSC~\cite{Wilson:2014cna} are also displayed.
Given that the most relevant channel for extracting the $\kappa$ pole comes from the $S_0^{1 / 2}$ wave, we also compare the phase shift from the $K$-matrix parametrization fit presented in Ref.~\cite{Wilson:2014cna}. For clarity, we display only the one with $\pi K$-$\eta K$ coupled channels, with a single pole coupled to both channels plus a constant background matrix, which is the preferred one in Ref.~\cite{Wilson:2014cna}. 
While our dispersive solution tends to show slightly higher values than the $K$-matrix result above 1~GeV, the HSC results fall within the uncertainties of our solution.
Our phase shift results are valuable for future LQCD simulations and phenomenological studies to compare with. 
We quote our results for the scattering lengths and the residue of the $K^*$ pole~\cite{Cao:2024zuy}:
\begin{align}
    m_\pi a_0^{1/2} & =0.92^{+0.06}_{-0.28}\ ,\quad m_\pi a_0^{3/2}=-\left(0.32^{+0.05}_{-0.02}\right) ,\nonumber\\
    \sqrt{s_{K^*}} & =934\pm 2~\mathrm{MeV}~\text{\cite{Wilson:2019wfr}}\ (\text{input})\ ,\quad \left|g_{K^* \pi K}\right|=35^{+5}_{-7}~\mathrm{MeV}\ ,
\end{align}
which are obtained at $m_\pi=391$~MeV. 

\begin{figure}[t]
    \centering
    \includegraphics[width=.5\textwidth,angle=-0]{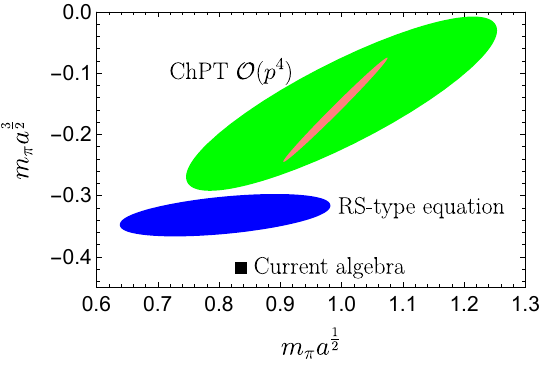} 
    \caption{Comparison of the scattering lengths between determinations from solving the RS-type equations and the ChPT calculations at NLO, i.e., $\mathcal{O}(p^4)$, as well as the current algebra result. The green and pink ellipses correspond to the error estimates for the NLO ChPT results from Set-\textrm{I} and Set-\textrm{II} scenarios, respectively. See the main text for detailed explanations.} \label{fig:SL_plot}
\end{figure}

There is a moderate correlation between these two scattering lengths, as shown in Fig.~\ref{fig:SL_plot}. 
The predictions of $\pi K$ scattering from ChPT at  $\mathcal{O}(p^4)$~\cite{Bernard:1990kw, GomezNicola:2001as} for the scattering lengths are also plotted. 
For the mesonic $\mathcal{O}(p^4)$ low-energy constants (LECs) $L_i^r(\mu)$ at the renormalization scale $\mu=770~\mathrm{MeV}$, we adopt the values fitted at $\mathcal{O}(p^4)$ given in Table~6 of Ref.~\cite{Bijnens:2014lea}, which are $L_1^r=1.0(1) \times 10^{-3}, L_2^r=1.6(2) \times 10^{-3}, L_3^r=$ $-3.8(3) \times 10^{-3}, L_4^r=0.0(3) \times 10^{-3}, L_5^r=1.2(1) \times 10^{-3}, L_6^r=0.0(4) \times 10^{-3}, L_7^r=-0.3(2)\times 10^{-3}$ and $L_8^r=0.5(2) \times 10^{-3}$. 
To estimate the uncertainty of the $\pi K \rightarrow \pi K$ amplitude, we generate a large number of parameter sets by randomly sampling the LECs with the assumption that they follow a Gaussian distribution and are uncorrelated with each other. 
Since the latter assumption clearly overestimates the uncertainty, in an alternative scenario we simply neglect the uncertainties of the LECs of $L_{4,6,7}^r$, which are suppressed in the $1/N_c$ expansion~\cite{Gasser:1984gg}. 
The uncertainty analysis that includes the error bars of all LECs in the random sampling is referred to as Set-\textrm{I}, which corresponds to the green ellipse in Fig.~\ref{fig:SL_plot}. 
The analysis that only includes the error bars of $L_{1,2,3,5,8}^r$ in the random sampling procedure is denoted as Set-\textrm{II}, corresponding to the smaller pink ellipse in Fig.~\ref{fig:SL_plot}. 

Interestingly, the uncertainty ellipse representing the chiral $O(p^4)$ results does not overlap with the RS-type equation solutions, as shown in Fig.~\ref{fig:SL_plot}. Given the current pion mass of $m_\pi = 391$~MeV, the convergence of ChPT could be problematic, as evidenced by the sizable difference between the current algebra results and the magnitude of the ChPT $O(p^4)$ corrections. 
In particular, if the $\kappa$ pole lies close to the $\pi K$ threshold, the perturbative result for $a^{1/2}_0$ will inevitably fail. In the next section, we focus specifically on the $\kappa$ pole.

%%%%%%%%%%%%%%%%%%%%%%%%%%%%%%%%%%%%%%%%%%%%%%%%%%%%%%%%%%%%%%%%%%%%%%%%%%%%

\section{Dispersive determination of $\kappa$}\label{sec:kappa}

Analyses of $\pi K \rightarrow \pi K$ scattering phase shifts have been performed using simple $K$-matrix approaches applied to LQCD data, as reported by HSC~\cite{Dudek:2014qha, Wilson:2014cna, Wilson:2019wfr}. 
One of the most important conclusions is that the $\kappa$ appears to be a deep VS pole at $m_\pi = 391$~MeV. 
In fact, the $I=1/2$ $S$-wave $\pi K$ phase shift exhibits atypical resonant behavior at low energies. 
Definitive conclusions cannot be drawn from these phase shift data alone, since they constrain only information on the real energy axis. A complete determination of the $\kappa$ properties requires analytic continuation to the complex energy plane to access the resonance pole position. 
The RS-type equation representation provides a reliable and rigorous approach in which analyticity, unitarity, and crossing symmetry are satisfied. Below, we focus on the determination of the $\kappa$ pole and relegate comprehensive discussions of other intriguing structures, including subthreshold poles and the Adler zero from chiral symmetry, to Appendix~\ref{app:VS & Adler0}.

In order to determine the position and residue of the $\kappa$ state, one should recast the partial-wave $S$-matrix as
\begin{align}
    S_J^I(s)=1-2 \frac{\sqrt{\left(m_{+}^2-s\right)\left(s-m_{-}^2\right)}}{s} f_J^I(s)\ .
\end{align}
The values of the $S$-matrix element on the second Riemann sheet can be related to those on the first sheet via the unitarity relation,
\begin{align}
    S^I_J(s)^\textrm{II}=\frac{1}{S^I_J(s)^\textrm{I}}\ ,
\end{align}
where the upper indices I and II denote the first and second Riemann sheets, respectively. Thus, a pole on the second sheet of the $S$-matrix $S^I_J(s)^\textrm{II}$ corresponds to a zero of $S^I_J(s)^\textrm{I}$ on the first sheet. 
Therefore, our task is to evaluate Eq.~\eqref{eq:RS_s} for real and complex values of $s$ within the validity domain and identify the possible zero of the $S$-matrix, corresponding to the $\kappa$ pole, in the $\left(I,J\right)=\left(1/2,0\right)$ channel. 
In our analysis of the scattering amplitude using the RS-type equation, we find that $S^{1/2}_0$ indeed contains a complex zero, i.e., $S^{1/2}_0(s_\kappa) = 0$ in the vicinity of the $\pi K$ threshold with~\cite{Cao:2024zuy},   
\begin{align}\label{eq.kappapole}
    \sqrt{s_\kappa} & =966_{-24}^{+41}-i 198_{-17}^{+38}~\text{MeV}\ ,\\
    g_{\kappa \pi K} & =759^{+63}_{-25}e^{-i (1.05^{+0.14}_{-0.06})}~\text{MeV}\ ,
\end{align}
where the corresponding coupling $g_{\kappa \pi K}$ is extracted from the residue of the amplitude at the $\kappa$ pole, i.e., $g_{\kappa}^2=\lim _{s \rightarrow s_\kappa}\!\left(s_\kappa-s\right) f_0^{1/2}(s)$.
Additionally, we also identify a pair of resonance poles above 1.5~GeV on the second Riemann sheet, which might correspond to the poles of the $K_0^*(1430)$ state proposed by HSC on the second Riemann sheet when $m_\pi=391~$MeV. 
We do not address the related discussions here due to both the complexity arising from many coupled channels and significant uncertainties in this inelastic high-energy region. 

According to the results in Eq.~\eqref{eq.kappapole}, our study using RS-type equations reveals that the scalar $\kappa$ state remains a broad resonance at $m_\pi=391$~MeV in the complex energy plane, which is clearly different from the findings in Refs.~\cite{Dudek:2014qha, Wilson:2014cna} that conclude the $\kappa$ to be a VS pole. 
We note that the $K$-matrix analyses in Refs.~\cite{Dudek:2014qha, Wilson:2014cna} almost completely neglect the influence of LHCs, especially the one induced by the cross-channel $\sigma$ exchange, resulting in an inaccurate description. 
We therefore establish that proper treatment of LHC contributions is essential for the accurate determination of atypical resonance poles, such as the $\kappa$.

%%%%%%%%%%%%%%%%%%%%%%%%%%%%%%%%%%%%%%%%%%%%%%%%%%%%%%%%%%%%%%%%%%%%%%%%%%%%

\section{Summary and outlook}\label{sec:summary}

In this work, we have provided a comprehensive analysis of Roy-Steiner-type equations for $\pi K$ scattering at $m_\pi = 391$~MeV and presented details that led to the conclusions in Ref.~\cite{Cao:2024zuy}. 
The analysis includes solutions of the $s$- and $t$-channel systems, a comprehensive examination of $S$-wave scattering lengths, and extraction of the $\kappa$ pole. 
The main results are summarized as follows:
\begin{itemize}
    \item[$1.$] We established a framework to construct the non-minimal subtracted Roy-Steiner-type equations for $\pi K$ scattering at unphysical large pion masses when $\sigma/f_0(500)$ and $K^*(892)$ both become bound state poles. 
    We derived the kernel functions for the equations with $a \neq 0$, many of which are presented for the first time in this work. In addition, we introduced a generic method to derive the complex validity domain of the Roy-Steiner-type equations for any hyperbolic parameter $a$.
    
    \item[$2.$] In the $t$-channel system, unitarity, as implemented through Muskhelishvili–Omn\`es techniques, provides the necessary connection at least up to the onset of inelastic contributions, and even beyond if inelasticities are sufficiently well constrained. 
    Consequently, the $t$-channel equations can be solved, which in turn reformulates the $s$-channel equations into Roy-like equations. 
    Utilizing LQCD data and Regge theory as driving terms, we have provided a full solution for the entire system via an iterative process. The resulting $S$-wave scattering lengths sizably differ from the results of chiral perturbation theory up to $\mathcal{O}(p^4)$. 
    We also find that the $S$-wave scattering lengths exhibit a moderate correlation, analogous to the situation at the physical pion mass~\cite{Buettiker:2003pp}.
    
    \item[$3.$] The most important finding is that the $\kappa$ state at $m_\pi=391~$MeV is not a deep virtual-state pole below the $\pi K$ threshold~\cite{Dudek:2014qha}, but rather remains a broad resonance~\cite{Cao:2024zuy}. 
    We have demonstrated that the restrictions from crossing symmetry play a crucial role in $\pi K$ scattering at large pion masses, especially in the presence of bound state poles.
\end{itemize}
Anticipated improvements in the precision of LQCD calculations will undoubtedly increase the need for rigorously extracting resonance information. 
In this work, we have demonstrated that the sophisticated dispersive Roy-Steiner-type equations can serve as a powerful and model-independent tool for analyzing lattice data. 
It would be interesting to extend similar analyses to other systems such as $\pi D$, $\pi N$, and even $KN$, $\bar{K}N$ scatterings at both physical and unphysical quark masses. Such analyses can provide valuable insights into low-lying resonances in these systems and the realization of chiral symmetry in low-energy QCD.

\bigskip 

\begin{acknowledgements}
We thank Zhi-Guang Xiao and Han-Qing Zheng for useful discussions.
XHC thanks the kind hospitality of Sichuan University during a research visit where part of this work was done.
This work is supported in part by National Natural Science Foundation of China (NSFC) under Grants No.~12347120, No.~12475078, No.~12150013, No.~12125507, No.~12447101 and No.~12335002; by the Postdoctoral Fellowship Program of China Postdoctoral Science Foundation under Grant No.~GZC20232773 and No.~2023M74360; by Chinese Academy of Sciences under Grant No.~YSBR-101. 
ZHG is also partially supported by the Science Foundation of Hebei Normal University with Contract No.~L2023B09.
\end{acknowledgements}

\appendix

%%%%%%%%%%%%%%%%%%%%%%%%%%%%%%%%%%%%%%%%%%%%%%%%%%%%%%%%%%%%%%%%%%%%%%%%%%%%

\section{Integral kernels}\label{app:kernel}

In this appendix, we provide details about the different contributions to the $s$- and $t$-channels of the RS-type equation systems. The relevant results are dispersed across multiple studies~\cite{Buettiker:2003pp, Pelaez:2020gnd}, with most analyses limited to simplified scenarios (e.g., the $a=0$ case) or the minimal subtraction scheme.
Here, our goal is to provide a comprehensive derivation, including subtraction terms, pole terms, and integral kernels when $a\neq 0$.

\subsection{Partial-wave projection for the $t$-channel amplitudes}\label{app:kernel_t}

We start by enumerating all kernels that appear in the equation of $g_1^1$. According to Eq.~\eqref{eq:t_PW}, the partial-wave projection of the $t$-channel for $F^-$ (we only need the $P$-wave) can be written as
\begin{align}
    g_1^1(t)=\frac{\sqrt{2}}{16 \pi q_\pi q_K} \int_0^1 \md z_t~ z_t F^{-}(s, t)=\frac{\sqrt{2}}{4 \pi} \int_0^1 \mathrm{d} z_t~ z_t^2 \frac{F^{-}\!\left(s, t\right)}{s-u}\ .
\end{align}
First, we perform the $t$-channel $P$-wave partial-wave projection using the above equation on both sides of Eq.~\eqref{eq:F-fin}. 
The left-hand side of Eq.~\eqref{eq:F-fin} becomes $g^1_1(t)$, and the first subtraction term becomes
\begin{align}
    \text{ST}^1_1(t)=\frac{\sqrt{2}}{4 \pi} \int_0^1 \mathrm{d} z_t~ z_t^2\frac{8 \pi m_{+} a_0^{-}}{m_{+}^2-m_{-}^2}=\frac{2 \sqrt{2} m_{+} a_0^{-}}{3\left(m_{+}^2-m_{-}^2\right)}\ ,
\end{align}
which is consistent with the results in Refs.~\cite{Buettiker:2003pp,Pelaez:2020gnd}. The contribution from the pole term yields
\begin{align}
    \text{PT}^1_1(t)= &\, \mathbf{
    \frac{g_{K^* \pi K}^2}{12\sqrt{2}}\left[\frac{32}{(s_{K^*}-m_+^2)(s_{K^*}-m_-^2)}\right.} \nonumber\\
    &\mathbf{\left.-\frac{3\left(2 s_{K^*} t+\lambda_{s_{K^*}}\right)\left(2(2s_{K^*}+t-2\Sigma)A(t,s_{K^*})-8q_K(t)q_\pi(t)\right)}{\left(q_K(t)q_\pi(t)\right)^3 \lambda_{s_{K^*}}}\right] .}
\end{align}
where we have defined
\begin{align}
    A\left(t, s\right)=&\operatorname{arctanh}\!\left(\frac{4 q_K(t) q_\pi(t)}{2 s+t-2 \Sigma}\right) .
\end{align}

The contributions from the partial waves can be decomposed into the $t$- and $s$-channel parts via dispersion integrals. 
The relevant integral contribution from the $t$-channel partial waves requires performing a partial-wave expansion on $\operatorname{Im}\left[\frac{G^1\left(t^{\prime}, s_b^{\prime}\right)}{\left(s_b^{\prime}-u_b^{\prime}\right)}\right]$:
\begin{align}
    & \frac{\sqrt{2}}{4 \pi} \int_0^1 \mathrm{d} z_t~ z_t^2\frac{t}{\pi}\int_{t_\pi}^\infty \frac{\md t^\prime}{t^\prime(t^\prime-t)}\frac{\operatorname{Im}G^1(t^\prime,s^\prime)}{2(s^\prime-u^\prime)} \nonumber\\
    =&\, \frac{t}{\pi}\int_{t_\pi}^\infty \frac{\md t^\prime}{t^\prime} \sum_{J}\underset{G^{1}_{1,J}(t,t^\prime)}{\underbrace{\left[\frac{(2J+1)\left(q_\pi(t^\prime)q_K(t^\prime)\right)^{J-1}}{(t^\prime-t)}\int_0^1 \mathrm{d} z_t~ \frac{z_t^2 P_J(z_t^\prime)}{z_t^\prime} \right]}}\operatorname{Im}g^1_J(t^\prime)\ .
\end{align}
It is important to note that $z_t^\prime $ is a function of $t, t^\prime$ and $z_t$, i.e., $z_t^\prime = z_t^\prime(t, t^\prime; z_t)$, because the hyperbolic DR fulfills the hyperbolic constraint such that $(s-a)(2\Sigma-s-t-a) = (s^\prime-a)(2\Sigma-s^\prime-t^\prime-a)$. 
Using $z_t = \frac{2s+t-2\Sigma}{4 q_\pi(t) q_K(t)}$ and $z_t^\prime = \frac{2s^\prime+t^\prime-2\Sigma}{4 q_\pi(t^\prime) q_K(t^\prime)}$, we obtain~\cite{Ditsche:2012fv}
\begin{align}
    z_t^{\prime\, 2}\!\left(t, t^{\prime};z_t\right)=\frac{q_\pi^2(t) q_K^2(t)}{q_\pi^2(t^\prime) q_K^2(t^\prime)} z_t^2+\frac{t^{\prime}-t}{16 q_\pi^2(t^\prime) q_K^2(t^\prime)}\!\left(t+t^{\prime}-4 \Sigma+4 a\right) .
\end{align}
The general results for the integrals $\int_0^1 \mathrm{d} z_t~ \frac{z_t^2 P_J(z_t^\prime)}{z_t^\prime}$ can be found in Appendix~B of Ref.~\cite{Ditsche:2012fv}. 
Here, we provide the results for the first few kernels $G^1_{1,1}$ and $G^1_{1,3}$:
\begin{align}
    G^1_{1,1}(t,t^\prime)& =\frac{1}{t^\prime-t}\ ,\quad G_{1,3}^1(t, t^{\prime})=\frac{7}{48}\!\left(t^{\prime}+t-4 \Sigma+10 a\right) ,
\end{align}
which are consistent with Refs.~\cite{Pelaez:2018qny,Pelaez:2020gnd}. 
The remaining terms involve the contributions of the $s$-channel partial waves to the dispersion integrals. 
Several relations are needed~\cite{Ditsche:2012fv}:
\begin{align}
    s(t,z_t)&=\frac{1}{2}\!\left(4q_\pi(t)q_K(t)z_t-t+2\Sigma\right) ,\quad u(t,z_t)=\frac{1}{2}\!\left(-4q_\pi(t)q_K(t)z_t-t+2\Sigma\right) , \nonumber\\
    z_s^{\prime}\!\left(t, s^{\prime} ; z_t\right)&=\frac{2 q_\pi^2(t) q_K^2(t)}{q^2(s^\prime)\left(s^{\prime}-a\right)}\!\left(z_t^2-\frac{(t-2\Sigma+2 a)^2-4\left(s^{\prime}-a\right)\left(2 q^2(s^\prime)+2\Sigma-s^{\prime}-a\right)}{16 q_\pi^2(t) q_K^2(t)}\right) .
\end{align}
Then, the integral contribution of the $S$-wave part is written as
\begin{align}
    \frac{1}{\pi} \sum_{J} \int_{m_{+}^2}^{\infty} \md s^{\prime} G_{1, J}^{-}\!\left(t, s^{\prime}\right) \operatorname{Im} f_{J}^{-}\!\left(s^{\prime}\right) ,
\end{align}
where
\begin{align}
    G_{1, J}^{-}\!\left(t, s^{\prime}\right)=&\, 4\sqrt{2}(2J+1)\int^1_0\md z_t~z_t^2 \nonumber\\
    &\times\left\{\left[\frac{1}{s^{\prime 2}-2 \Sigma s^{\prime}+s(t,z_t) u(t,z_t)-a\left(s(t,z_t)+u(t,z_t)-2\Sigma\right)}-\frac{1}{s^{\prime 2}-2 \Sigma s^{\prime}+\Delta^2}\right]\right. \nonumber\\
    & +\left[\frac{1}{s^{\prime 2}-2 \Sigma s^{\prime}+s(t,z_t) u(t,z_t)-a\left(s(t,z_t)+u(t,z_t)-2 \Sigma\right)+s^\prime t-at}\right. \nonumber\\
    & \left.\left.-\frac{1}{s^{\prime 2}-2 \Sigma s^{\prime}+s(t,z_t) u(t,z_t)-a\left(s(t,z_t)+u(t,z_t)-2\Sigma\right)}\right]P_J\!\left(z_s^\prime(t,s^\prime;z_t)\right)\right\}\ .
\end{align}
The explicit expressions for the first few kernels are
\begin{align}
    & G_{1,0}^{-}\!\left(t, s^{\prime}\right)=4 \sqrt{2}\left[\frac{\left(2s^{\prime}-2\Sigma+t\right)A\left(t, s^{\prime}\right)-4 q_K(t) q_\pi(t)}{16\left(q_K(t) q_\pi(t)\right)^3}-\frac{1}{3 \lambda_{s^{\prime}}}\right] , \\
    & G_{1,1}^{-}\!\left(t, s^{\prime}\right)=12 \sqrt{2}\left[P_1\left(z_{s}(s^\prime,t)\right) \frac{\left(2s^{\prime}-2\Sigma+t\right)A\left(t, s^{\prime}\right)-4 q_K(t) q_\pi(t)}{16\left(q_K(t) q_\pi(t)\right)^3}-\frac{1}{3 \lambda_{s^{\prime}}}\right] , \\
    & G_{1,2}^{-}\!\left(t, s^{\prime}\right)=20 \sqrt{2}\left[P_2\left(z_{s}(s^\prime,t)\right) \frac{\left(2s^{\prime}-2\Sigma+t\right)A\left(t, s^{\prime}\right)-4 q_K(t) q_\pi(t)}{16\left(q_K(t) q_\pi(t)\right)^3}-\frac{1}{3 \lambda_{s^{\prime}}}-\frac{2 s^{\prime 2} t}{\lambda_{s^{\prime}}^2\left(s^{\prime}-a\right)}\right] .
\end{align}
Note that the results of $G_{1, J}^{-}\left(t, s^{\prime}\right)$ in Refs.~\cite{Buettiker:2003pp,Pelaez:2020gnd} (denoted as $\hat{G}_{1, J}^{-}\left(t, s^{\prime}\right)$ in~\cite{Pelaez:2020gnd}) contain some typos\footnote{For example, the expression $\hat{G}_{1, 0}^{-}\left(t, s^{\prime}\right)=\left(t, s^{\prime}\right)=4 \sqrt{2}\left[\left(s^{\prime}-\Sigma+t/2\right)\frac{A\left(t, s^{\prime}\right)-4 q_K(t) q_\pi(t)}{16\left(q_K(t) q_\pi(t)\right)^3}-\frac{1}{3 \lambda_{s^{\prime}}}\right]$ of Eq.~(E.4) in Ref.~\cite{Pelaez:2020gnd} should be written as $4 \sqrt{2}\left[\frac{\left(2s^{\prime}-2\Sigma+t\right)A\left(t, s^{\prime}\right)-4 q_K(t) q_\pi(t)}{16\left(q_K(t) q_\pi(t)\right)^3}-\frac{1}{3 \lambda_{s^{\prime}}}\right]$, and the rest are similar. The expression $G_{1 l^{\prime}}^{-}\left(t, s^{\prime}\right)=4 \sqrt{2}\left(2 l^{\prime}+1\right)\left\{\frac{F(x)}{\left(s^{\prime}-\Sigma+t / 2\right)^2}-\frac{1}{3 \lambda_{s^{\prime}}}+C_{l^{\prime}}\left(t, s^{\prime}\right)\right\}$ in Ref.~\cite{Buettiker:2003pp} should be written as $4 \sqrt{2}\left(2 l^{\prime}+1\right)\left\{\frac{F(x)}{\left(s^{\prime}-\Sigma+t / 2\right)^2}P_{l^\prime}\left(z_{s^\prime}\right)-\frac{1}{3 \lambda_{s^{\prime}}}+C_{l^{\prime}}\left(t, s^{\prime}\right)\right\}$, where $z_{s^\prime}=1+\frac{2s^\prime t}{\lambda_{s^\prime}}$.}. 

Next, we present the partial-wave projection of the $t$-channel part for $F^+$ (although only the $S$-wave is needed, the expressions for the $D$-wave are also provided for comparison). 
The first contribution from the subtraction term in Eq.~\eqref{eq:F+_fin} is~\cite{Buettiker:2003pp}
\begin{align}
    \text{ST}^0_J(t)=\delta_{J, 0}\frac{\sqrt{3}m_+}{2}\!\left(a_0^{+}+t \frac{a_0^{-}}{m_{+}^2-m_{-}^2}\right) .
\end{align}
The pole term can be written as
\begin{align}
    \text{PT}^0_J(t)=&\, \mathbf{\frac{\sqrt{3}}{16 \pi\left(q_\pi q_K\right)^J} \int_0^1 \md z_t~P_J(z_t) \left[16\pi g_{K^* \pi K}^2\frac{2s_{K^*}+t-m_+^2-m_-^2}{\left(s_{K^*}-m_+^2\right)\left(s_{K^*}-m_-^2\right)}\right.} \nonumber\\
    & \mathbf{\left.-16\pi g_{K^* \pi K}^2 P_1\left(z_s(s_{K^*},t)\right) \left(\frac{1}{s_{K^*}-s(t,z_t)}+\frac{1}{s_{K^*}-u(t,z_t)} \right)+\frac{16 \pi  g_{\sigma \pi \pi} g_{\sigma K \bar{K}} t}{\sqrt{3}s_\sigma\left(s_\sigma-t\right)}\right]}\ ,
\end{align}
where the explicit expression of $\text{PT}^0_0(t)$ is
\begin{align}
    \text{PT}^0_0(t)=&\, \mathbf{\sqrt{3}g_{K^* \pi K}^2\left(\frac{2s_{K^*}+t-m_+^2-m_-^2}{\left(s_{K^*}-m_+^2\right)\left(s_{K^*}-m_-^2\right)}-\frac{(2s_{K^*} t+\lambda_{s_{K^*}})A(t,s_{K^*})}{q_K(t)q_\pi(t)\lambda_{s_{K^*}}}\right)+\frac{g_{\sigma \pi \pi} g_{\sigma K \bar{K}} t}{s_\sigma\left(s_\sigma-t\right)}}\ .
\end{align}

The contribution of the $t$-channel partial waves to the integral is somewhat complicated, as there are contributions not only from $G^0$ but also from $G^1$. 
It can be written as
\begin{align}
    & \frac{\sqrt{3}}{16 \pi\left(q_\pi(t) q_K(t)\right)^J} \int_0^1 \md z_t~P_J(z_t) \frac{t}{\pi} \int_{t_\pi}^{\infty} \frac{\mathrm{d} t^{\prime}}{t^{\prime}} \operatorname{Im}\left[\frac{ G^0\!\left(t^{\prime}, s^{\prime}\right)}{\sqrt{6} \left(t^\prime-t\right)}-\frac{G^1\!\left(t^{\prime}, s^{\prime}\right)}{2\left(s^{\prime}-u^{\prime}\right)}\right] \nonumber\\
    \equiv&\, \sum_{J^\prime}\frac{t}{\pi} \int_{t_\pi}^{\infty} \frac{\mathrm{d} t^{\prime}}{t^{\prime}}\left[G_{J,J^\prime}^0\!\left(t, t^{\prime}\right) \operatorname{Im} g_{2 J^{\prime}-2}^0\!\left(t^{\prime}\right)+G_{J,J^\prime}^1\!\left(t, t^{\prime}\right) \operatorname{Im} g_{2 J^{\prime}-1}^1\!\left(t^{\prime}\right)\right] ,
\end{align}
where
\begin{align}
    G_{J,J^\prime}^0= & \frac{1}{\left(q_\pi(t) q_K(t)\right)^J} \int_0^1 \md z_t~P_J(z_t) \frac{(2 J^\prime+1)\left(q_\pi\left(t^{\prime}\right) q_K\left(t^{\prime}\right)\right)^{J^\prime} P_{J^\prime}\!\left(z_t^{\prime}(t,t^\prime;z_t)\right)}{t^\prime-t}\ ,\\
    G_{J,J^\prime}^1= & -\frac{1}{\left(q_\pi(t) q_K(t)\right)^J} \int_0^1 \md z_t~P_J(z_t) \frac{\sqrt{6}(2 J^\prime+1)\left(q_\pi\left(t^{\prime}\right) q_K\left(t^{\prime}\right)\right)^{J^\prime} P_{J^\prime}\!\left(z_t^{\prime}(t,t^\prime;z_t)\right)}{8 z_{t}^{\prime}(t,t^\prime;z_t) q_\pi\left(t^{\prime}\right) q_K\left(t^{\prime}\right)}\ .
\end{align}
The explicit expressions for the first few non-vanishing kernels are 
\begin{align}
    G_{0, 0}^{0}\!\left(t, s^{\prime}\right)& =\frac{1}{t^\prime-t}\ ,\\
    G_{0, 2}^{0}\!\left(t, s^{\prime}\right)& =\frac{5}{16}\!\left(t+t^{\prime}-4 \Sigma+6 a\right),\\
    G_{2, 2}^{0}\!\left(t, s^{\prime}\right)& =\frac{1}{t^\prime-t} \ ,\\
    G_{2, 4}^{0}\!\left(t, s^{\prime}\right)& =\frac{3}{8}\!\left(t+t^{\prime}-4 \Sigma+7 a\right) ,\\
    G_{0, 1}^{1}\!\left(t, s^{\prime}\right)& =-\frac{3\sqrt{6}}{8}\ ,\\
    G_{2, 3}^{1}\!\left(t, s^{\prime}\right)& =-\frac{7\sqrt{6}}{24}\ ,
\end{align}
which are consistent with the results of Refs.~\cite{Buettiker:2003pp,Pelaez:2020gnd}.

Finally, the contribution of the $s$-channel partial waves to the integral is given by
\begin{align}
    \sum_{J^\prime}\frac { 1 } { \pi } \int _ { m _ { + } ^ { 2 } } ^ { \infty } \mathrm { d } s ^ { \prime } \left[G_{J,J^\prime}^{+}\!\left(t, s^{\prime}\right) \operatorname{Im} f_{J^\prime}^{+}\!\left(s^{\prime}\right)+t G_{J,J^\prime}^{-}\!\left(t, s^{\prime}\right) \operatorname{Im} f_{J^\prime}^{-}\!\left(s^{\prime}\right)\right] ,
\end{align}
where
\begin{align}
    G_{J,J^\prime}^{+}\!\left(t, s^{\prime}\right)=&\, \frac{\sqrt{3}}{\left(q_\pi(t)q_K(t)\right)^J}\int^1_0 \md z_t~P_J(z_t)(2J^\prime+1) \nonumber\\
    & \times\left\{\frac{1}{s^{\prime 2}}\left[\frac{s^{\prime}(2 \Sigma)^2-2 \left[s(t,z_t) u(t,z_t)-a(s(t,z_t)+u(t,z_t)-2 \Sigma)\right](s^\prime+\Sigma)}{s^{\prime 2}-2 \Sigma s^{\prime}+s(t,z_t) u(t,z_t)-a(s(t,z_t)+u(t,z_t)-2 \Sigma)}\right.\right.\nonumber\\
    & \left.-\frac{s^{\prime}\!\left((2 \Sigma)^2-2 \Delta^2\right)-2 \Sigma \Delta^2}{s^{\prime 2}-2 \Sigma s^{\prime}+\Delta^2}\right] \nonumber\\
    & +\left[\frac{2 s^{\prime}-2 \Sigma+t}{s^{\prime 2}-2 \Sigma s^{\prime}+s(t,z_t) u(t,z_t)-a(s(t,z_t)+u(t,z_t)-2 \Sigma)+s^{\prime} t-a t}\right.\nonumber\\
    &\left.\left.-\frac{2 s^{\prime}-2 \Sigma}{s^{\prime 2}-2 \Sigma s^{\prime}+s(t,z_t) u(t,z_t)-a(s(t,z_t)+u(t,z_t)-2 \Sigma)}\right]P_{J^\prime}(z_s^\prime(t,s^\prime;z_t))\right\}\ ,\\
    G_{J,J^\prime}^{-}\!\left(t, s^{\prime}\right)=&\, \frac{\sqrt{3}}{\left(q_\pi(t)q_K(t)\right)^J}\int^1_0 \md z_t~P_J(z_t)(2J^\prime+1) \nonumber\\
    & \times\left\{\left[\frac{1}{s^{\prime 2}-2 \Sigma s^{\prime}+s(t,z_t) u(t,z_t)-a(s(t,z_t)+u(t,z_t)-2 \Sigma)}-\frac{1}{s^{\prime 2}-2 \Sigma s^{\prime}+\Delta^2}\right] \right.\nonumber\\
    & \left.-\left[\frac{1}{s^{\prime 2}-2 \Sigma s^{\prime}+s(t,z_t) u(t,z_t)-a(s(t,z_t)+u(t,z_t)-2 \Sigma)}\right]P_{J^\prime}(z_s^\prime(t,s^\prime;z_t)) \right\}\ .
\end{align}
The explicit expressions for the first few non-vanishing kernels are given by
\begin{align}
    G_{0,0}^{+}\!\left(t, s^{\prime}\right)=&\, \sqrt{3}\left[\frac{A\left(t, s^{\prime}\right)}{q_K(t) q_\pi(t)}+\frac{2\left(\Sigma-s^{\prime}\right)}{\lambda_{s^{\prime}}}\right] , \\
    G_{0,1}^{+}\!\left(t, s^{\prime}\right)=&\, 3 \sqrt{3}\left[\frac{A\left(t, s^{\prime}\right)}{q_K(t) q_\pi(t)} P_1\left(z_s\left(s^{\prime}, t\right)\right)-\frac{\left(2 s^{\prime}+2 t-2 \Sigma\right)}{\lambda_{s^{\prime}}}-\frac{2 a t}{\left(s^{\prime}-a\right) \lambda_{s^{\prime}}}\right] , \\
    G_{0,2}^{+}\!\left(t, s^{\prime}\right)=&\, 5 \sqrt{3}\left[\frac{A\left(t, s^{\prime}\right)}{q_K(t) q_\pi(t)} P_2\left(z_s\left(s^{\prime}, t\right)\right)-\frac{2 s^\prime-2 \Sigma}{\lambda_{s^{\prime}}}-\frac{6 s^\prime t\left(\Delta^2+s^{\prime}\!\left(3 s^{\prime}+2 t-4 \Sigma\right)\right)}{\left(s^{\prime}-a\right) \lambda_{s^{\prime}}^2}\right. \nonumber\\
    & \left.+\frac{3 s^{\prime 2} t\left(2 s^{\prime}+t-2 \Sigma\right)^2}{2\left(s^{\prime}-a\right)^2 \lambda_s^{\prime 2}}-\frac{8 s^{\prime 2} t\left(q_K(t) q_\pi(t)\right)^2}{\left(s^{\prime}-a\right)^2 \lambda_{s^{\prime}}^2}\right] ,\\
    G_{2,0}^{+}\!\left(t, s^{\prime}\right)=&\, \frac{\sqrt{3}\!\left(2 s^{\prime}+t-2 \Sigma\right)^2}{32 (q_K(t) q_\pi(t))^5}\left[\left(3-\left(\frac{4 q_K(t) q_\pi(t)}{2 s^{\prime}+t-2 \Sigma}\right)^2\right) A\left(t, s^{\prime}\right)-3 \left(\frac{4 q_K(t) q_\pi(t)}{2 s^{\prime}+t-2 \Sigma}\right)\right] , \\
    G_{2,1}^{+}\!\left(t, s^{\prime}\right)=&\, \frac{3 \sqrt{3}\!\left(2 s^{\prime}+t-2 \Sigma\right)^2}{32 (q_K(t) q_\pi(t))^5} P_1\left(z_s\left(s^{\prime}, t\right)\right)\left[\left(3-\left(\frac{4 q_K(t) q_\pi(t)}{2 s^{\prime}+t-2 \Sigma}\right)^2\right) A\left(t, s^{\prime}\right)\right.\nonumber\\
    &\left.-3 \left(\frac{4 q_K(t) q_\pi(t)}{2 s^{\prime}+t-2 \Sigma}\right)\right] ,\\
    G_{2,2}^{+}\!\left(t, s^{\prime}\right)=&\, 5 \sqrt{3}\left[\frac{\left(2 s^{\prime}+t-2 \Sigma\right)^2}{32 (q_K(t) q_\pi(t))^5} P_2\left(z_s(s^\prime,t)\right)\left(\left(3-\left(\frac{4 q_K(t) q_\pi(t)}{2 s^{\prime}+t-2 \Sigma}\right)^2\right) A\left(t, s^{\prime}\right)\right.\right.\nonumber\\
    &\left.\left.-3 \left(\frac{4 q_K(t) q_\pi(t)}{2 s^{\prime}+t-2 \Sigma}\right)\right)-\frac{16 s^{\prime 2} t}{5\left(s^{\prime}-a\right)^2 \lambda_{s^{\prime}}^2}\right] ,\\
    G^-_{0,0}(t,s^\prime)=&\, -\frac{\sqrt{3}}{\lambda_{s^\prime}}\ ,\\
    G^-_{0,1}(t,s^\prime)=&\, \frac{3\sqrt{3}(s^\prime+a)}{\lambda_{s^\prime} \left(s^\prime-a\right)}\ ,\\
    G^-_{0,2}(t,s^\prime)=&\, \frac{5\sqrt{3}\left[s^{\prime 2}\!\left(3\left(4t(s^\prime-a)-(2s^\prime+t-2\Sigma)^2\right)+16\left(q_K(t)q_\pi(t)\right)^2\right)+2(s^\prime-a)(5s^\prime+a)\lambda_{s^\prime}\right]}{2(s^\prime-a)^2 \lambda_{s^\prime}^2}\ ,\\
    G^-_{2,2}(t,s^\prime)=&\, \frac{16\sqrt{3}s^{\prime 2}}{(s^\prime-a)^2 \lambda_{s^\prime}^2}\ ,
\end{align}
which are in agreement with the results of Ref.~\cite{Buettiker:2003pp} in the $a\to 0$ limit. 
And the expressions of $G^+_{J,J^\prime}$ are also consistent with those from~\cite{Pelaez:2020gnd}\footnote{The expression $G_{2,2}^+$ of Eq.~(E.5) in Ref.~\cite{Pelaez:2020gnd} has a typo and $(\cdots)\frac{16 s^2 t}{5\left(s^{\prime}-a\right)^2 \lambda_{s^{\prime}}^2}$ should be written as $(\cdots)-\frac{16 s^2 t}{5\left(s^{\prime}-a\right)^2 \lambda_{s^{\prime}}^2}$.}. 

\subsection{Partial-wave projection for the $s$-channel amplitudes}\label{app:kernel_s}

We will continue to start with $F^-$. In Eq.~\eqref{eq:F-fin} (where both sides of the expression are previously multiplied by the factor $s-u$), the first subtraction terms read~\cite{Pelaez:2020gnd}
\begin{align}
    \text{ST}^-_J(s)=\delta_{J,0}\frac{m_{+} a_0^{-}}{2(m_+^2-m_-^2)} \frac{3 s^2-2\Sigma s-\Delta^2}{2 s}+\delta_{J,1}\frac{m_{+} a_0^{-}}{6(m_+^2-m_-^2)} \frac{\lambda_s}{2 s}\ .
\end{align}
The pole terms are given by
\begin{align}
    \text{PT}^-_J(s)=&\, \mathbf{\frac{1}{32 \pi} \int_{-1}^1 \md z_s~P_{J}\!\left(z_s\right)\left(s-u(s,z_s)\right)}\nonumber\\
    &\mathbf{\times\left[\frac{16\pi g_{K^* \pi K}^2}{(s_{K^*}-m_+^2)(s_{K^*}-m_-^2)}-\frac{16\pi g_{K^* \pi K}^2 P_1\left(z_s(s_{K^*},t(s,z_s))\right)}{\left(s_{K^*}-s\right)\left(s_{K^*}-u(s,z_s)\right)}\right]}\ , 
\end{align}
and $\text{PT}^-_0(s)$ and $\text{PT}^-_1(s)$ are expressed as
\begin{align}
    \text{PT}^-_0(s)= &\, \mathbf{g_{K^* \pi K}^2\left[-\frac{1}{s_{K^*}-s}+\frac{2(s-\Sigma)}{(s_{K^*}-m_+^2)(s_{K^*}-m_-^2)}+\frac{2s_{K^*}}{\lambda_{s_{K^*}}}+C(s_{K^*},s)B(s,s_{K^*})\right.}\nonumber\\
    & \mathbf{\left.-\frac{\lambda_s}{s}\!\left(\frac{1}{2(s_{K^*}-m_+^2)(s_{K^*}-m_-^2)}-\frac{s_{K^*}}{(s_{K^*}-s)\lambda_{s_{K^*}}}\right)\right]}\ ,\label{eq:PT-0}\\
    \text{PT}^-_1(s)= &\, \mathbf{g_{K^* \pi K}^2\left[C(s_{K^*},s)\left(C(s,s_{K^*})B(s,s_{K^*})+\frac{2s}{\lambda_s}\right)\right.}\nonumber\\
    &\mathbf{\left.-\frac{\lambda_s}{6s\lambda_{s_{K^*}}}\!\left(\frac{2s_{K^*}}{s_{K^*}-s}-\frac{\lambda_{s_{K^*}}}{(s_{K^*}-m_+^2)(s_{K^*}-m_-^2)}\right)\right]}\ ,\label{eq:PT-1}
\end{align}
where
\begin{align}
    \begin{aligned}
        & B\left(s, s^{\prime}\right)=\frac{s}{\lambda_s}\left[\ln \left(s^{\prime}+s-2 \Sigma\right)-\ln \left(s^{\prime}-\frac{\Delta^2}{s}\right)\right] , \\
        & C\left(s, s^{\prime}\right)=1-\frac{2 s\left(s^{\prime}+s-2 \Sigma\right)}{\lambda_s}\ .
    \end{aligned}
\end{align}
Moreover, there is the $t$-channel partial-wave contribution through the dispersion integral, which can be expressed as
\begin{align}
    \frac{1}{\pi} \sum_{J^\prime} \int_{t_\pi}^{\infty} \md t^{\prime} K_{J, J^\prime}^{-,1}\!\left(s, t^{\prime}\right) \operatorname{Im} g_{J^\prime}^1\left(t^{\prime}\right) ,
\end{align}
where 
\begin{align}
    K_{J, J^\prime}^{-,1}\!\left(s, t^{\prime}\right)=\frac{\sqrt{2}}{16}\int_{-1}^1 \md z_s~P_{J}\!\left(z_s\right) \frac{t(s,z_s)(s-u(s,z_s))}{t^\prime\left(t^\prime-t(s,z_s)\right)}\frac{\left(2 J^{\prime}+1\right)\left(q_\pi\left(t^{\prime}\right) q_K\left(t^{\prime}\right)\right)^{J^{\prime}} P_{J^{\prime}}\!\left(z_t^{\prime}(s,t^\prime;z_s)\right)}{z_t^\prime(s,t^\prime;z_s) q_\pi(t^\prime)q_K(t^\prime)}\ .
\end{align}
To calculate the above integral, we use~\cite{Ditsche:2012fv}
\begin{align}
    z_t^{\prime 2}\!\left(s, t^{\prime} ; z_s\right)=\frac{q^2(s)(s-a)}{2 q_\pi^{2}(t^\prime) q_K^{2}(t^\prime)} z_s+\frac{\left(t^{\prime}-2\Sigma+2 a\right)^2-4(s-a)\left(2 q(s)^2+2\Sigma-s-a\right)}{16 q_\pi^{2}(t^\prime) q_K^{2}(t^\prime)}\ .
\end{align}
The explicit expressions for the first few kernels are as follows
\begin{align}
    & K_{0,1}^{-,1}\!\left(s, t^{\prime}\right)=\frac{3}{4 \sqrt{2}} \frac{s\left(t^{\prime}+2 s-2 \Sigma\right)}{\lambda_s}\left[\ln \left(1+\frac{\lambda_s}{s t^{\prime}}\right)-\frac{\lambda_s}{t^{\prime} s}+\frac{\lambda_s^2}{2 t^{\prime} s^2\left(t^{\prime}+2 s-2 \Sigma\right)}\right] , \\
    & K_{1,1}^{-,1}\!\left(s, t^{\prime}\right)=\frac{3}{4 \sqrt{2}} \frac{s\left(t^{\prime}+2 s-2 \Sigma\right)}{\lambda_s}\left[P_1\left(z_s(s,t^\prime)\right) \ln \left(1+\frac{\lambda_s}{s t^{\prime}}\right)-2\right]-\frac{\lambda_s}{8 \sqrt{2} t^{\prime} s}\ .
\end{align}
We note that the formula for $K_{1,1}^{-,1}$ in Ref.~\cite{Pelaez:2020gnd} (denoted as $\hat{K}_{1,1}^1$ therein) is consistent with our result. 
However, some coefficients for $K_{0,1}^{-,1}$ in Ref.~\cite{Pelaez:2020gnd} (denoted as $\hat{K}_{0,1}^1$ therein) contain typographical errors\footnote{The expression $\hat{K}_{0,1}^{1}\!\left(s, t^{\prime}\right)=\frac{3}{4 \sqrt{2}} \frac{s\left(t^{\prime}+2 s-2 \Sigma\right)}{\lambda_s}\left[\ln \left(1+\frac{\lambda_s}{s t^{\prime}}\right)-\frac{2\lambda_s}{t^{\prime} s}+\frac{\lambda_s^2}{t^{\prime} s^2\left(t^{\prime}+2 s-2 \Sigma\right)}\right]$ of Eq.~(E.10) in Ref.~\cite{Pelaez:2020gnd} should be written as $\frac{3}{4 \sqrt{2}} \frac{s\left(t^{\prime}+2 s-2 \Sigma\right)}{\lambda_s}\left[\ln \left(1+\frac{\lambda_s}{s t^{\prime}}\right)-\frac{\lambda_s}{t^{\prime} s}+\frac{\lambda_s^2}{2 t^{\prime} s^2\left(t^{\prime}+2 s-2 \Sigma\right)}\right].$}.
The remaining comes from the $s$-channel partial-wave contribution to the dispersion integral,
\begin{align}
    \frac{1}{\pi} \sum_{J^\prime} \int_{m_{+}^2}^{\infty} \md s^{\prime} K_{J, J^\prime}^{-,-}\!\left(s, s^{\prime}\right) \operatorname{Im} f_{J^\prime}^{-}\!\left(s^{\prime}\right) ,
\end{align}
where 
\begin{align}
    K_{J, J^\prime}^{-,-}\!\left(s, s^{\prime}\right)=&\, \frac{1}{2} \int_{-1}^1 \md z_s~P_{J}\!\left(z_s\right) \left(s-u(s,z_s)\right)\left(2J^\prime+1\right)\nonumber\\
    & \times\left\{\left[\frac{1}{s^2-2 \Sigma s^{\prime}+s u(s,z_s)-a(s+u(s,z_s)-2 \Sigma)}-\frac{1}{s^2-2 \Sigma s^{\prime}+\Delta^2}\right]\right.\nonumber\\
    & +\left[\frac{1}{s^{\prime 2}-2 \Sigma s^{\prime}+s u(s,z_s)-a(s+u(s,z_s)-2 \Sigma)+s^{\prime} t(s,z_s)-a t(s,z_s)}\right. \nonumber\\
    & \left.\left.-\frac{1}{s^2-2 \Sigma s^{\prime}+s u(s,z_s)-a(s+u(s,z_s)-2 \Sigma)}\right]P_{J^\prime}(z_s^\prime\left(s,s^\prime;z_s\right))\right\}\ .
\end{align}
In particular, the first several explicit expressions are
\begin{align}
    & K_{0,0}^{-,-}\!\left(s, s^{\prime}\right)=\frac{1}{s^{\prime}-s}-B\left(s, s^{\prime}\right)+\frac{\lambda_s}{2 s \lambda_{s^{\prime}}}-2 \frac{(s-\Sigma)}{\lambda_{s^{\prime}}}\ , \\
    & K_{0,1}^{-,-}\!\left(s, s^{\prime}\right)=3\left[\frac{1}{s^{\prime}-s}-2 \frac{\left(s^{\prime}+s-\Sigma\right)}{\lambda_{s^{\prime}}}-\frac{\lambda_s\left(s^{\prime}+s\right)}{2\left(s^{\prime}-s\right) s \lambda_{s^{\prime}}}-C\left(s^{\prime}, s\right) B\left(s, s^{\prime}\right)\right] , \\
    & K_{0,2}^{-,-}\!\left(s, s^{\prime}\right)=\frac{5}{4s^2(s^\prime-a)(s-s^\prime)\lambda_{s^\prime}^2}\bigg[-8(s-a)s^{\prime 2}\lambda_s^2 \nonumber\\
    &\qquad -2s\lambda_s\left(-6(s-s^\prime)s^{\prime 2}(2s+s^\prime-a-2\Sigma)-(s^\prime-a)(s+5s^\prime)\lambda_{s^\prime}\right) \nonumber\\
    &\qquad +4s^2(s^\prime-a)\left(6(s-s^\prime)s^{\prime 2}(s+s^\prime-2\Sigma)-2(s-s^\prime)(s+3s^\prime-\Sigma)\lambda_{s^\prime}-\lambda_{s^\prime}^2\right) \nonumber\\
    &\qquad -4s^2(s^\prime-a)(s-s^\prime)\left(6s^{\prime 2}(s+s^\prime-2\Sigma)^2-6s^\prime(s+s^\prime-2\Sigma)\lambda_{s^\prime}+\lambda_{s^\prime}^2\right)B(s,s^\prime)\bigg]\ , \\
    & K_{1,0}^{-,-}\!\left(s, s^{\prime}\right)=-C\left(s, s^{\prime}\right) B\left(s, s^{\prime}\right)-\frac{2 s}{\lambda_s}-\frac{\lambda_s}{6 s \lambda_{s^{\prime}}}\ , \\
    & K_{1,1}^{-,-}\!\left(s, s^{\prime}\right)=-3 C\left(s^{\prime}, s\right)\left[C\left(s, s^{\prime}\right) B\left(s, s^{\prime}\right)+\frac{2 s}{\lambda_s}\right]+\frac{\left(s^{\prime}+s\right) \lambda_s}{2 s\left(s^{\prime}-s\right) \lambda_{s^{\prime}}}\ ,\\
    & K_{1,2}^{-,-}\!\left(s, s^{\prime}\right)=\frac{5}{6s^2 \lambda_s^2 \lambda_{s^\prime}^2}\bigg[\frac{6(s-a)s^{\prime 2}\lambda_s^4}{(s^\prime-a)(s-s^\prime)}+s\lambda_s^3\left(\frac{6s^{\prime 2}(2s+s^\prime-a-2\Sigma)}{s^\prime-a}-\frac{(s+5s^\prime)\lambda_{s^\prime}}{s-s^\prime}\right)-\nonumber\\
    &\qquad -6s^2\lambda_s^2\left(6s^{\prime 2}(s+s^\prime-2\Sigma)^2-6(s+s^\prime-2\Sigma)\lambda_{s^\prime}+\lambda_{s^\prime}^2\right)C(s,s^\prime)B(s,s^\prime)\bigg]\ ,
\end{align}
which are consistent with the results in Ref.~\cite{Pelaez:2020gnd}. Additionally, we also present the formulas for the kernels $K^{-,-}_{J,2}$.

Next, we deal with $F^+$ following the same procedure. Starting from the subtraction term on the right-hand side of Eq.~\eqref{eq:F+_fin}, we have 
\begin{align}
    \text{ST}^+_J(s)&= \frac{1}{32 \pi} \int_{-1}^1 \md z_s P_{J}\!\left(z_s\right) 8 \pi m_{+}\!\left(a_0^{+}+t(s,z_s) \frac{a_0^{-}}{m_{+}^2-m_{-}^2}\right) \nonumber\\
    &= \frac{m_+}{2}\!\left(a_0^+-\frac{a_0^- \lambda_s}{2s\left(m_+^2-m_-^2\right)}\right)\delta_{J,0}+\frac{m_+}{6}\frac{a_0^- \lambda_s}{2s\left(m_+^2-m_-^2\right)}\delta_{J,1}\ .
\end{align}
The pole term can be written as
\begin{align}
    \text{PT}^+_J(s)=&\, \mathbf{\frac{1}{32 \pi} \int_{-1}^1 \md z_s~P_{J}\!\left(z_s\right)\left[16\pi g_{K^* \pi K}^2\frac{2s_{K^*}+t(s,z_s)-m_+^2-m_-^2}{\left(s_{K^*}-m_+^2\right)\left(s_{K^*}-m_-^2\right)}\right.} \nonumber\\
    & \mathbf{\left.-16\pi g_{K^* \pi K}^2 P_1\left(z_s(s_{K^*},t(s,z_s))\right) \left(\frac{1}{s_{K^*}-s}+\frac{1}{s_{K^*}-u(s,z_s)} \right)+\frac{16 \pi  g_{\sigma \pi \pi} g_{\sigma K \bar{K}} t(s,z_s)}{\sqrt{3}s_\sigma\left(s_\sigma-t(s,z_s)\right)}\right]}\ ,
\end{align}
where the expressions for $\text{PT}^+_0(s)$ and $\text{PT}^+_1(s)$ read
\begin{align}
    \text{PT}^+_0(s)=&\, \mathbf{g_{K^* \pi K}^2\left[-\frac{1}{s_{K^*}-s}+\frac{1}{s_{K^*}-m_+^2}+\frac{1}{s_{K^*}-m_-^2}-\frac{2s_{K^*}}{\lambda_{s_{K^*}}}\right.}\nonumber\\
    & \mathbf{-\frac{\lambda_s}{s}\!\left(\frac{1}{2(s_{K^*}-m_+^2)(s_{K^*}-m_-^2)}-\frac{s_{K^*}}{(s_{K^*}-s)\lambda_{s_{K^*}}}\right)+C(s_{K^*},s)B(s,s_{K^*})\bigg ]} \nonumber\\
    &\mathbf{+g_{\sigma \pi\pi}g_{\sigma K \bar{K}}\frac{1}{s_\sigma}\!\left(\frac{s_\sigma s}{\lambda_s}\ln\left(1+\frac{\lambda_s}{s_\sigma s}\right)-1\right)}\ ,\label{eq:PT+0}\\
    \text{PT}^+_1(s)=&\, \mathbf{-g_{K^* \pi K}^2\left[C(s_{K^*},s)\left(C(s,s_{K^*})B(s,s_{K^*})+\frac{2s}{\lambda_s}\right)\right.}\nonumber\\
    &\mathbf{\left.+\frac{\lambda_s}{6s\lambda_{s_{K^*}}}\!\left(\frac{2s_{K^*}}{s_{K^*}-s}-\frac{\lambda_{s_{K^*}}}{(s_{K^*}-m_+^2)(s_{K^*}-m_-^2)}\right)\right]}\nonumber\\
    & \mathbf{+g_{\sigma \pi\pi}g_{\sigma K \bar{K}}\frac{s}{\lambda_s}\!\left(\left(2\frac{s_\sigma s}{\lambda_s}+1\right)\ln\left(1+\frac{\lambda_s}{s_\sigma s}\right)-2\right)}\ .\label{eq:PT+1}
\end{align}
The contribution corresponding to the $t$-channel partial waves in Eq.~\eqref{eq:F+_fin} is given by
\begin{align}
    & \frac{1}{32 \pi} \int_{-1}^1 \md z_s P_{J}\!\left(z_s\right) \frac{t(s,z_s)}{\pi} \int_{t_\pi}^{\infty} \frac{\mathrm{d} t^{\prime}}{t^{\prime}}\operatorname{Im} \left[\frac{G^0\left(t^{\prime}, s_b^{\prime}\right)}{\sqrt{6}\!\left(t^{\prime}-t(s,z_s)\right)}-\frac{G^1\left(t^{\prime}, s_b^{\prime}\right)}{2\left(s_b^{\prime}-u_b^{\prime}\right)}\right] \nonumber\\
    \equiv& \frac{1}{\pi} \sum_{J^\prime} \int_{t_\pi}^{\infty} \md t^{\prime} K_{J, J^\prime}^{+,0}\!\left(s, t^{\prime}\right) \operatorname{Im} g_{J^\prime}^0\left(t^{\prime}\right)+ \frac{1}{\pi} \sum_{J^\prime} \int_{t_\pi}^{\infty} \md t^{\prime} K_{J, J^\prime}^{+,1}\!\left(s, t^{\prime}\right) \operatorname{Im} g_{J^\prime}^1\left(t^{\prime}\right) ,
\end{align}
where
\begin{align}
    K_{J, J^\prime}^{+,0}\!\left(s, t^{\prime}\right)& =\frac{\sqrt{3}}{6} \int_{-1}^1 \md z_s P_{J}\!\left(z_s\right) \frac{\left(2 J^{\prime}+1\right)t(s,z_s)\left(q_\pi\left(t^{\prime}\right) q_K\left(t^{\prime}\right)\right)^{J^{\prime}} P_{J^{\prime}}\!\left(z_t^{\prime}(s,t^\prime;z_s) \right)}{t^\prime \left(t^{\prime}-t(s,z_s)\right)}\ ,\\
    K_{J, J^\prime}^{+,1}\!\left(s, t^{\prime}\right)& =-\frac{\sqrt{2}}{16} \int_{-1}^1 \md z_s P_{J}\!\left(z_s\right) \frac{\left(2 J^{\prime}+1\right)t(s,z_s)\left(q_\pi\left(t^{\prime}\right) q_K\left(t^{\prime}\right)\right)^{J^{\prime}} P_{J^{\prime}}\!\left(z_t^{\prime}(s,t^\prime;z_s) \right)}{t^\prime z_t^{\prime}(s,t^\prime;z_s) q_\pi\left(t^{\prime}\right) q_K\left(t^{\prime}\right)}\ .
\end{align}
The explicit expressions for the first few non-vanishing kernels are
\begin{align}
    K_{0,0}^{+,0}\!\left(s, t^{\prime}\right)= &\,\frac{1}{\sqrt{3}} \frac{s}{\lambda_s}\left[\ln \left(1+\frac{\lambda_s}{s t^{\prime}}\right)-\frac{\lambda_s}{s t^{\prime}}\right] , \\
    K_{0,2}^{+,0}\!\left(s, t^{\prime}\right)= &\,\frac{5}{32\sqrt{3}} \frac{s}{\lambda_s}\!\left(3\left(2s+t^\prime-2\Sigma\right)^2-16\left(q_\pi\left(t^{\prime}\right) q_K\left(t^{\prime}\right)\right)^2 \right)\left[\ln \left(1+\frac{\lambda_s}{s t^{\prime}}\right)-\frac{\lambda_s}{s t^{\prime}}\right] \nonumber\\
    &+\frac{15 \lambda_s(s-a)}{16\sqrt{3} s t^{\prime}}\ , \\
    K_{1,0}^{+,0}\!\left(s, t^{\prime}\right)= &\,\frac{1}{\sqrt{3}} \frac{s}{\lambda_s}\left[P_1\left(z_s(s,t^\prime)\right) \ln \left(1+\frac{\lambda_s}{s t^{\prime}}\right)-2\right] ,\\
    K_{1,2}^{+,0}\!\left(s, t^{\prime}\right)= &\,\frac{5}{32\sqrt{3}} \frac{s}{\lambda_s}\!\left(3\left(2s+t^\prime-2\Sigma\right)^2-16\left(q_\pi\left(t^{\prime}\right) q_K\left(t^{\prime}\right)\right)^2 \right)\left[P_1(z_s(s,t^\prime))\ln \left(1+\frac{\lambda_s}{s t^{\prime}}\right)-2\right] \nonumber\\
    &-\frac{5 \lambda_s(s-a)}{16 \sqrt{3} s t^{\prime}}\ ,\\
    K_{0,1}^{+,1}\!\left(s, t^{\prime}\right)= &\,\frac{3\lambda_s}{8\sqrt{2}st^\prime}\ ,\\
    K_{1,1}^{+,1}\!\left(s, t^{\prime}\right)= &\,-\frac{\lambda_s}{8\sqrt{2}st^\prime}\ .
\end{align}
$K_{0,0}^{+,0}$ and $K_{1,0}^{+,0}$ are consistent with the results in Ref.~\cite{Pelaez:2020gnd} (denoted as $K^{0}_{J,J^\prime}$ therein).

The remaining contributions from the $s$-channel partial waves are 
\begin{align}
    \frac{1}{\pi} \sum_{J^{\prime}} \int_{m_{+}^2}^{\infty} \mathrm{d} s^{\prime}~ K_{J, J^{\prime}}^{+,+}\!\left(s, s^{\prime}\right) \operatorname{Im} f_{J^{\prime}}^{+}\!\left(s^{\prime}\right)+\frac{1}{\pi} \sum_{J^{\prime}} \int_{m_{+}^2}^{\infty} \mathrm{d} s^{\prime}~ K_{J, J^{\prime}}^{+,-}\!\left(s, s^{\prime}\right) \operatorname{Im} f_{J^{\prime}}^{-}\!\left(s^{\prime}\right) ,
\end{align}
where
\begin{align}
    K_{J, J^{\prime}}^{+,+}\!\left(s, s^{\prime}\right)=&\, \frac{1}{2}\int^1_{-1}\md z_s~P_J(z_s)\left(2J^\prime+1\right)\left\{\frac{1}{s^{\prime 2}}\left[\frac{s^{\prime}(2 \Sigma)^2-2 \left[s u(s,z_s)-a(s+u(s,z_s)-2 \Sigma)\right](s^\prime+\Sigma)}{s^{\prime 2}-2\Sigma s^\prime+s u(s,z_s)-a(s+u(s,z_s)-2 \Sigma)}\right.\right.\nonumber\\
    &\left.\left.-\frac{s^\prime\left((2\Sigma)^2-2\Delta^2\right)-2\Sigma \Delta^2}{s^{\prime 2}-2\Sigma s^\prime+\Delta^2}\right]\right. \nonumber\\
    & +\left[\frac{2 s^{\prime}-2 \Sigma+t(s,z_s)}{s^{\prime 2}-2 \Sigma s^{\prime}+s u(s,z_s)-a(s+u(s,z_s)-2 \Sigma)+s^{\prime} t(s,z_s)-a t(s,z_s)}\right. \nonumber\\
    & \left.\left.-\frac{2 s^{\prime}-2 \Sigma}{s^{\prime 2}-2 \Sigma s^{\prime}+s u(s,z_s)-a(s+u(s,z_s)-2 \Sigma)}\right]P_{J^{\prime}}\!\left(z_s^{\prime}\!\left(s, s^{\prime} ; z_s\right)\right) \right\}\ ,\\
    K_{J, J^{\prime}}^{+,-}\!\left(s, s^{\prime}\right)=&\, \frac{1}{2}\int^1_{-1}\md z_s~P_J(z_s)\left(2J^\prime+1\right)t(s,z_s)\\
    &\times\left\{\frac{1}{s^{\prime 2}-2 \Sigma s^{\prime}+s u(s,z_s)-a(s+u(s,z_s)-2 \Sigma)}-\frac{1}{s^2-2 \Sigma s^{\prime}+\Delta^2}\right.\nonumber\\
    &\left.-\left[\frac{1}{s^{\prime 2}-2 \Sigma s^{\prime}+s u(s,z_s)-a(s+u(s,z_s)-2 \Sigma)}\right] P_{J^{\prime}}\!\left(z_s^{\prime}\!\left(s, s^{\prime} ; z_s\right)\right)\right\}\ .
\end{align}
The explicit expressions for the first few non-vanishing kernels are
\begin{align}
    & K_{0,0}^{+,+}\!\left(s, s^{\prime}\right)=\frac{2\left(\left(s^{\prime}+\Sigma\right) \Delta^2-2 s^{\prime} \Sigma^2\right)}{s^{\prime 2} \lambda_{s^{\prime}}}-\frac{s^{\prime 2}+2 s^{\prime}(\Sigma-s)-2 s \Sigma}{s^{\prime 2}\!\left(s^{\prime}-s\right)}+B\left(s, s^{\prime}\right) , \\
    & K_{0,1}^{+,+}\!\left(s, s^{\prime}\right)=3\left[\frac{s\left(s^{\prime}+2 \Sigma\right)-\Delta^2}{\lambda_{s^{\prime}} s}+\frac{s^{\prime} \lambda_s}{s\lambda_{s^{\prime}}\!\left(s^{\prime}-a\right)}+C\left(s^{\prime}, s\right) B\left(s, s^{\prime}\right)\right] , \\
    & K_{0,2}^{+,+}(s,s^\prime)=\frac{5\left(s^{\prime 2}+2(\Sigma-s)s^\prime-2s\Sigma\right)}{s^{\prime 2}(s-s^\prime)}+\frac{5}{\lambda_{s^\prime}^2}\bigg[-6s^{\prime 2}(s+s^\prime-2\Sigma)-\frac{2(s-a)^2s^{\prime 2}\lambda_s^2}{s^2(s^\prime-a)^2(s-s^\prime)} \nonumber\\
    & +\frac{2(3 s^{\prime 3}+(\Delta^2-2\Sigma^2)s^\prime+\Delta^2\Sigma)\lambda_{s^\prime}}{s^{\prime 2}}+\frac{3(s-a)s^\prime \lambda_s\left((s-s^\prime)s^\prime(s+a-2\Sigma)+(s^\prime-a)\lambda_{s^\prime}\right)}{s(s^\prime-a)^2(s-s^\prime)} \nonumber\\
    & +B(s,s^\prime)\left(6 s^{\prime 2}(s+s^\prime-2\Sigma)^2-6 s^{\prime}(s+s^\prime-2\Sigma)\lambda_{s^\prime}+\lambda_{s^\prime}^2\right)\bigg]\ , \\
    & K_{1,0}^{+,+}\!\left(s, s^{\prime}\right)=C\left(s, s^{\prime}\right) B\left(s, s^{\prime}\right)+2 \frac{s}{\lambda_s}\ , \\
    & K_{1,1}^{+,+}\!\left(s, s^{\prime}\right)=3 C\left(s^{\prime}, s\right)\left[C\left(s, s^{\prime}\right) B\left(s, s^{\prime}\right)+2 \frac{s}{\lambda_s}\right]+\frac{s^{\prime}(s-a) \lambda_s}{ s\left(s^{\prime}-a\right) \lambda_{s^{\prime}}\!\left(s^{\prime}-s\right)}\ ,\\
    & K_{1,2}^{+,+}(s,s^\prime)=\frac{5}{s^2\lambda_s^2\lambda_{s^\prime}^2}\bigg\{\frac{(s-a)^2 s^{\prime 2}\lambda_s^4}{(s^\prime-a)^2(s^\prime-s)}-\frac{(s-a)ss^\prime \lambda_s^3\left((s-s^\prime)s^\prime(s+a-2\Sigma)+(s^\prime-a)\lambda_{s^\prime}\right)}{(s^\prime-a)^2(s-s^\prime)} \nonumber\\
    & +s^2\lambda_s^2\left[C(s,s^\prime)B(s,s^\prime)+2\frac{s}{\lambda_s}\right]\bigg\}\ ,\\
    & K_{0,0}^{+,-}(s,s^\prime)=\frac{\lambda_s}{2s\lambda_{s^\prime}}\ ,\\
    & K_{0,1}^{+,-}(s,s^\prime)=-\frac{3(s^\prime+a)\lambda_s}{2s(s^\prime-a)\lambda_{s^\prime}}\ ,\\
    & K_{0,2}^{+,-}(s,s^\prime)=-\frac{5\lambda_{s}\!\left(2s^{\prime 2}\!\left(3s (s-s^\prime)(s+s^\prime-2\Sigma)+2(s-a)\lambda_s\right)+s(s^\prime-a)(5s^\prime+a)\lambda_{s^\prime}\right)}{2s^2(s^\prime-a)^2\lambda_{s^\prime}^2}\ , \\
    & K_{1,0}^{+,-}(s,s^\prime)=-\frac{\lambda_s}{6s\lambda_{s^\prime}}\ ,\\
    & K_{1,1}^{+,-}(s,s^\prime)=\frac{(s^\prime+a)\lambda_s}{2s(s^\prime-a)\lambda_{s^\prime}}\ ,\\
    & K_{1,2}^{+,-}(s,s^\prime)=\frac{5\lambda_{s}\!\left(6s^{\prime 2}\!\left(s (s-s^\prime)(s+s^\prime-2\Sigma)-(s-a)\lambda_s\right)-s(s^\prime-a)(5s^\prime+a)\lambda_{s^\prime}\right)}{6s^2(s^\prime-a)^2\lambda_{s^\prime}^2}\ ,
\end{align}
which are consistent with the results of Ref.~\cite{Pelaez:2020gnd}.
Additionally, we provide the explicit expressions for the $D$-wave kernels $K^{+,\pm}_{0(1),2}(s,s^\prime)$.

%%%%%%%%%%%%%%%%%%%%%%%%%%%%%%%%%%%%%%%%%%%%%%%%%%%%%%%%%%%%%%%%%%%%%%%%%%%%

\section{Subthreshold singularities: virtual state poles and Adler zeros}\label{app:VS & Adler0}

In this appendix, we provide a comprehensive study of the subthreshold singularities of the $\pi K$ scattering amplitudes at $m_\pi=391~$MeV.

\subsection{$\left(I,J\right)=\left({1}/{2},0\right)$ channel}

First, let us recall the prediction of ChPT. In the $\left(I,J\right)=\left({1}/{2},0\right)$ channel, there are two Adler zeros~\cite{Adler:1964um}, which at leading order (LO) of ChPT are given by
\begin{align}
    s_{A\pm}^{{1}/{2}}=\frac{1}{5}\!\left(m_K^2+m_\pi^2 \pm 2 \sqrt{4 m_K^4-7 m_K^2 m_\pi^2+4 m_\pi^4}\right) .
\end{align}
Note that only one of them is close to the $\pi K$ threshold. The next-to-leading order (NLO) ChPT prediction of $\sqrt{s^{1/2}_{A+}}$ is $427^{+21}_{-24}~$MeV for $m_\pi=391$~MeV. In the RS-type equation analysis, this Adler zero is found at $719^{+20}_{-19}~$MeV, and its position is significantly shifted toward the threshold due to the influence of the LHCs induced by the cross-channel pole exchanges.

\begin{figure}[t]
    \centering
    \includegraphics[width=.6\textwidth,angle=-0]{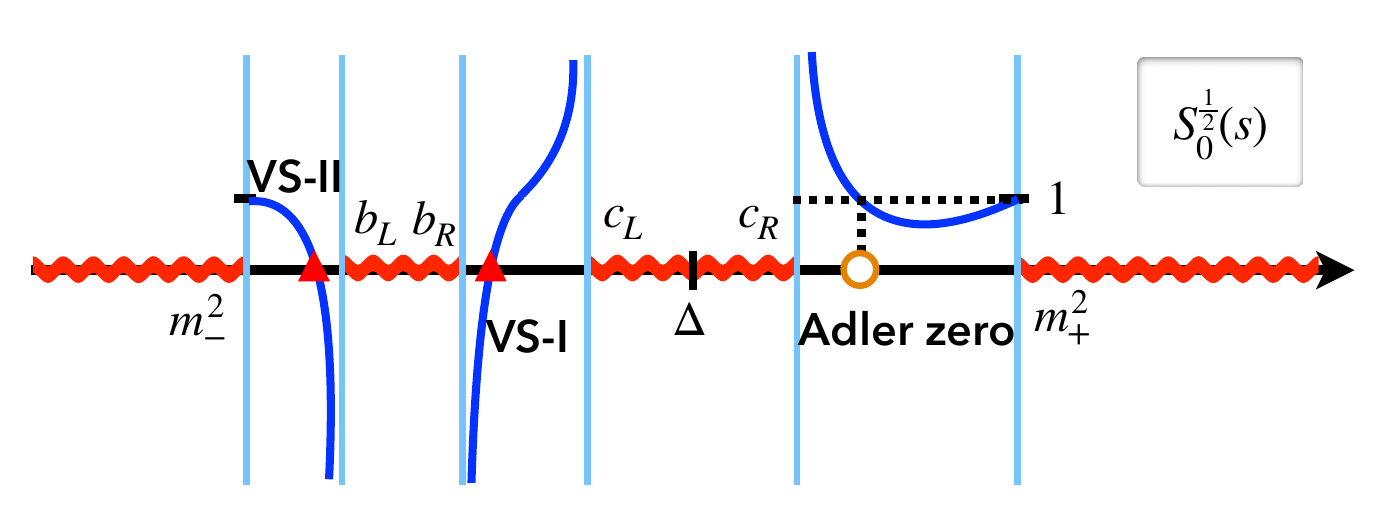}    \caption{The $I=1/2$ $S$-wave $\pi K$ partial-wave $S$-matrix $S_0^{1/2}(s)$ below the $\pi K$ threshold. The abscissa represents the values of $s$. Note that the $S$-matrix is real in the region $[m_-^2,b_L)\cup(b_R,c_L)\cup(c_R,m_+^2]$.
    The intersection points where the $S$-matrix, shown as blue curves, crosses the $s$ axis correspond to the VS locations.} \label{fig:S1_sub}
\end{figure}

Furthermore, the presence of two new LHCs on the real axis below threshold leads to the appearance of new VS poles. Recall that, due to unitarity, the zeros of the partial-wave $S$-matrix on the first sheet correspond to poles on the second sheet. Therefore, we focus on the zero behavior of the $S$-matrix on the first sheet. Using the expression of the pole terms in Appendix~\ref{app:kernel_s}, we can determine the behavior of the corresponding $S$-matrix near the branch points of the LHCs, as shown in Fig.~\ref{fig:S1_sub}. Clearly, there is (at least) one zero on the first sheet between $m_-^2$ and $b_L$ as well as between $b_R$ and $c_L$. The numerical results of the RS-type equation also support this picture.
This VS phenomenon was first discussed in Ref.~\cite{PhysRev.123.692} (discussed lately in $\pi\pi$ scattering in Refs.~\cite{Zhou:2004ms, Dai:2019zao, Cao:2023ntr, Lyu:2024lzr, Lyu:2024elz} and in $\pi N$ scattering~\cite{Li:2021oou, Cao:2022zhn}). 

\subsection{$\left(I,J\right)=\left({1}/{2},1\right)$ channel}

\begin{figure}[t]
    \centering
    \includegraphics[width=.6\textwidth,angle=-0]{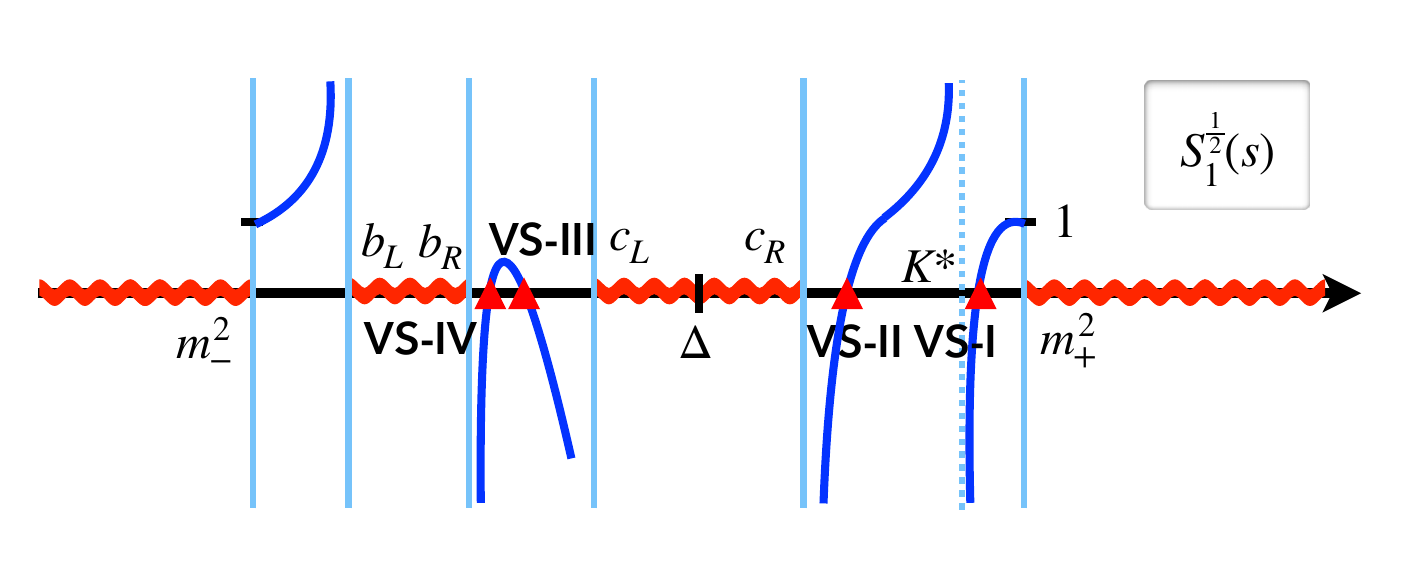}    \caption{Same as Fig.~\ref{fig:S1_sub} but showing the $I=1/2$ $P$-wave $\pi K$ partial-wave $S$-matrix $S^{1/2}_1(s)$.} \label{fig:P1_sub}
\end{figure}

In the $\left(I,J\right)=\left({1}/{2},1\right)$ channel, there is a $P$-wave BS below the threshold, which is always accompanied by a VS pole at nearly the same position on the second sheet. 
Furthermore, as illustrated in Fig.~\ref{fig:P1_sub}, the RS-type equation analysis reveals three additional VS poles. 
This is also attributed to the LHCs.

\subsection{$\left(I,J\right)=\left({3}/{2},0\right)$ channel}

\begin{figure}[t]
    \centering
    \includegraphics[width=.6\textwidth,angle=-0]{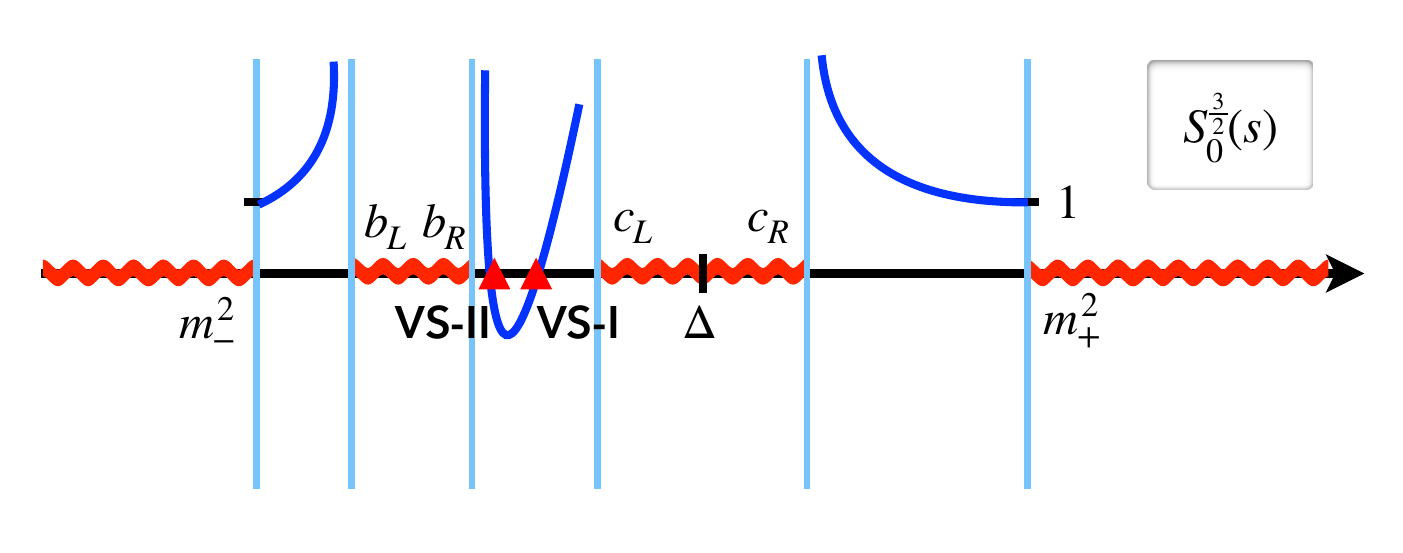}    \caption{Same as Fig.~\ref{fig:S1_sub} but showing the $I=3/2$ $S$-wave $\pi K$ partial-wave $S$-matrix $S^{3/2}_0(s)$.} \label{fig:S3_sub}
\end{figure}

The singularities of the $\left(I,J\right)=\left({3}/{2},0\right)$ channel may seem trivial compared to those of the $\left(I,J\right)=\left({1}/{2},0\right)$ channel. However, this is not the case.
As shown in Fig.~\ref{fig:S3_sub}, there appear to be only two VS poles. 
In addition, in the $\left(I,J\right)=\left({3}/{2},0\right)$ channel, there is an Adler zero, which at LO of ChPT is given by
\begin{align}
    s^{{3}/{2}}_A=m_\pi^2+m_K^2\ .
\end{align}
The NLO ChPT prediction yields $\sqrt{s^{{3}/{2}}_A} = 667^{+31}_{-26}~$MeV at $m_\pi=391$~MeV, which lies closer to the threshold than $\sqrt{s^{1/2}_{A+}}$. 
Nevertheless, the RS-type equation analysis does not find this zero between $c_R$ and the threshold $m_+^2$. 
Moreover, it is worth noting that a pair of conjugate resonance poles, $\sqrt{s^{3/2}_\text{pole}} = 252^{+36}_{-33} \pm i329^{+14}_{-2}~\text{MeV}$, and a pair of conjugate zeros of $f^{3/2}_0$, $\sqrt{s^{3/2}_\text{zero}} = 337^{+48}_{-37} \pm i398^{+26}_{-31}~\text{MeV}$, can be found in the complex $s$ plane outside the circular cut. 
How can these complex poles and zeros be understood?

These subthreshold structures may be understood by varying the pion mass. 
Actually, Ref.~\cite{Zheng:2003rw} uses the inverse amplitude method (IAM)~\cite{Dobado:1996ps} to unitarize the ChPT $\mathcal{O}(p^4)$ amplitude of $\pi K$ scattering and discovers a pair of resonance poles inside the circular cut, $371\pm i61~$MeV.\footnote{The poles must exist in conjugate pair as a consequence of the Schwarz reflection principle. The pole position differs slightly from the result of Ref.~\cite{Zheng:2003rw} due to the use of updated LECs~\cite{Bijnens:2014lea}.} 
However, since this resonance pole is far from the physical region, it was never thought that such a pole would have any physical impact. 

\begin{figure}[ht]
    \centering
    \includegraphics[width=.6\textwidth,angle=-0]{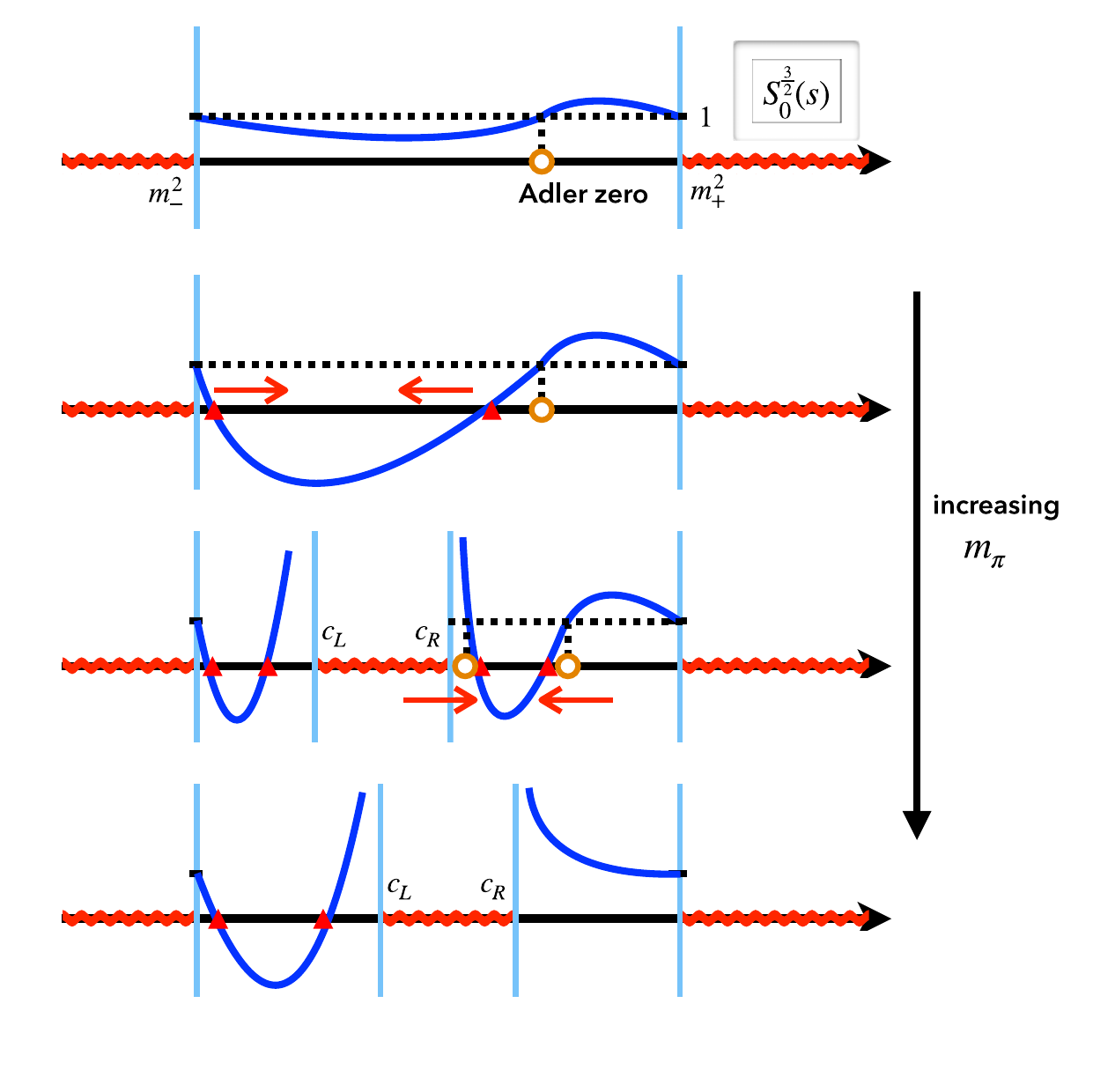}    \caption{Qualitative evolution of $S^{3/2}_0$ with increasing pion mass. } \label{fig:S3_mass}
\end{figure}

Nevertheless, as the light quark mass increases, the situation could change.
\begin{figure}[t]
    \centering
    \includegraphics[width=.5\textwidth,angle=-0]{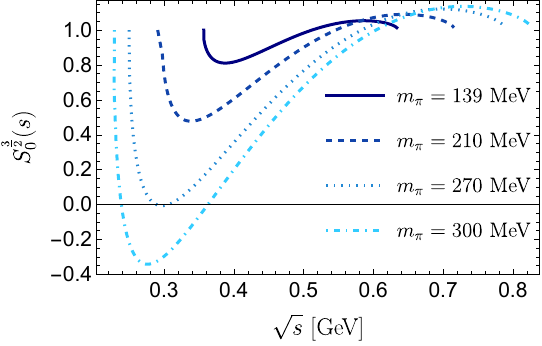}    \caption{The NLO ChPT predictions of the $I=3/2$  $S$-wave $S$-matrix $S_0^{3/2}(s)$ below the $\pi K$ threshold for several pion masses.} \label{fig:S3_chiral}
\end{figure}
As shown in Fig.~\ref{fig:S3_chiral}, the NLO ChPT calculation predicts that this pair of $S$-matrix zeros gradually approaches the real axis. 
For a fully unitary $S$-matrix, a zero of the $S$-matrix on the first Riemann sheet corresponds to a pole on the second Riemann sheet. Although the NLO ChPT $S$-matrix satisfies unitarity only perturbatively, one can expect the $S$-matrix pole on the second Riemann sheet to be located at roughly the same positions.\footnote{We have checked that the whole picture shown in Fig.~\ref{fig:S3_chiral} remains if instead using the IAM unitarization of the NLO ChPT amplitude.} 
The zero reaches the real axis around $m_\pi\sim270~$MeV, becoming a pair of zeros and thus a pair of VS poles on the second Riemann sheet. 
As $m_\pi$ increases further, the two VS poles move in opposite directions, as shown in Fig.~\ref{fig:S3_mass}.
When $m_\pi\sim 300~$MeV, the $\sigma$ pole becomes a BS, accompanied by a LHC from $c_L$ to $c_R$. Since the $S$-matrix $S^{3/2}_0$ tends to $+\infty$ near $c_R$, which is a logarithmic branch point, a new VS and a zero\footnote{A zero of the partial-wave amplitude corresponds to the partial-wave $S$-matrix being equal to unity.} of the partial-wave amplitude $f^{3/2}_0$ are generated near $c_R$. 
This VS and the previously upward-moving VS meet and then leave the real $s$-axis, becoming a pair of subthreshold resonance poles. As $m_\pi$ increases further, the Adler zero and the new zero also meet and become a pair of conjugate complex zeros.
This explains the pair of resonance poles and the pair of partial-wave amplitude zeros observed in the RS-type equation analysis, as well as the ``missing'' Adler zero.
In summary, the intriguing structures observed in the $\left(I,J\right)=\left({3}/{2},0\right)$ channel is a consequence of intricate interplay between chiral symmetry and crossing symmetry.

%%%%%%%%%%%%%%%%%%%%%%%%%%%%%%%%%%%%%%%%%%%%%%%%%%%%%%%%%%%%%%%%%%%%%%%%%%%%

% \bibliographystyle{unsrt}
\bibliography{refs}
\end{document}